\documentclass[useAMS,usenatbib,useaastex,usletter]{mn2e}
\usepackage{natbib}
\usepackage{graphicx}
\usepackage{amssymb}
\usepackage{amsmath}
\usepackage{mathrsfs}
\usepackage{color}

\def\lesssim{\mathrel{\hbox{\rlap{\hbox{\lower4pt\hbox{$\sim$}}}\hbox{$<$}}}}
\def\gtrsim{\mathrel{\hbox{\rlap{\hbox{\lower4pt\hbox{$\sim$}}}\hbox{$>$}}}}

\title{Where is the Radiation Edge in Magnetized Black Hole Accretion discs?}

\author[Beckwith et al.]{Kris Beckwith$^{1,2}$, John F. Hawley$^1$ and 
Julian H. Krolik$^3$ \\$^1$Astronomy Department, University of 
Virginia, P.O. Box 400325, Charlottesville, VA 22904-4325 
\\$^{2}$Institute of Astronomy, University of Cambridge, 
Madingley Road, Cambridge. CB3 0HA, United Kingdom
\\$^3$Department of Physics and Astronomy, Johns Hopkins 
University, Baltimore, MD 21218}

\pagerange{\pageref{firstpage}--\pageref{lastpage}} \pubyear{2004}

\def\LaTeX{L\kern-.36em\raise.3ex\hbox{a}\kern-.15em
T\kern-.1667em\lower.7ex\hbox{E}\kern-.125emX}

\newcommand{\aap}   {A\&A}

\newcommand{\apj}   {ApJ}
\newcommand{\apjl}  {ApJ}

\newcommand{\mnras} {MNRAS}

\newcommand{\pasj}  {PASJ}

\newcommand{\physrep}  {Phys. Rev.}

\begin{document}

\label{firstpage}

\maketitle

\begin{abstract} General Relativistic (GR) Magnetohydrodynamic (MHD)
simulations of black hole accretion find significant magnetic stresses
near and inside the innermost stable circular orbit (ISCO), suggesting
that such flows could radiate in a manner noticeably different from
the prediction of the standard model, which assumes that there are no
stresses in that region.  We provide estimates of how phenomenologically
interesting parameters like the ``radiation edge", the innermost ring of
the disc from which substantial thermal radiation escapes to infinity,
may be altered by stresses near the ISCO.  These estimates are based on
data from a large number of three-dimensional GRMHD simulations combined
with GR ray-tracing.  For slowly spinning black holes ($a/M<0.9$),
the radiation edge lies well inside where the standard model predicts,
particularly when the system is viewed at high inclination.  For more
rapidly spinning black holes, the contrast is smaller.  At fixed total luminosity,
the characteristic temperature of the accretion flow increases between 
a factor of $1.2-2.4$ over that predicted by the standard model, whilst at fixed 
mass accretion rate, there is a corresponding enhancement of the accretion luminosity
which may be anywhere from tens of percent to order unity.  When all these considerations are combined, we find that, for fixed black hole mass, luminosity, and inclination angle, our uncertainty in
the characteristic temperature of the radiation reaching distant observers
due to uncertainty in dissipation profile (around a factor of 3) is {\it greater} 
than the uncertainty due to a complete lack of knowledge 
of the black hole's spin (around a factor of 2) and furthermore that spin estimates based on the stress-free inner boundary condition provide an upper limit to $a/M$.
\end{abstract}


\section{Introduction}\label{intro}

Recent years have seen a rapid growth in the specificity and detail with
which we attempt to describe accretion onto black holes.  In the early
days of the subject three decades ago, we were content with the simplest
of models \cite[the `standard' model, ][]{Novikov:1973}, hereafter NT,
in which the accretion flow was assumed to be time-steady, axisymmetric,
geometrically thin, and following circular Keplerian orbits at all
radii outside the innermost stable circular orbit (the ISCO, located
at radius $r_{ms}$).  Relying on heuristic arguments framed in a purely
hydrodynamic context \citep{Page:1974}, it was further assumed that all
stresses ceased inside the ISCO, so that the inner edge of the disc could
be described as falling precisely at that radius.  Dimensional analysis
was the basis for any link between the inter-ring stresses essential to
accretion and local physical conditions \citep{Shakura:1973}.

Today, there is tremendous effort to obtain and interpret 
direct observational diagnostics of the innermost parts of accretion 
flows onto black holes.  Apparently relativistically-broadened Fe 
K$\alpha$ profiles can be discerned in numerous examples 
\citep{Reynolds:2003}. Extensive efforts are made to fit detailed disc 
spectral models to observed continuum spectra 
\citep{Gierlinski:2004,Davis:2005,Shafee:2006a,Shafee:2006b,Hui:2007}. These spectral 
fits can be used to infer the mass of the central black hole; both 
methods may be used to constrain the spin of the black hole 
\citep{Makishima:2000,Miller:2004,Miniutti:2004,Brenneman:2006,Shafee:2006a, 
Shafee:2006b}. Some hope to use relativistic fluid dynamics to define 
normal modes of disc oscillation that could then also be used to 
constrain the central black hole's spin 
\citep{Wagoner:2001,Rezzolla:2003,KatoS:2004}.

At the same time, there has also been much progress on the theoretical
side. A strong consensus now supports the idea that accretion stresses
are the result of correlated MHD turbulence, driven by a pervasive
magneto-rotational instability \citep{Balbus:1998}.  
Numerical simulations developing this idea have given us a detailed
and quantitative description of the vertical profiles of pressure
and density inside discs \citep{Miller:2000,Hirose:2005,Hirose:2007},
as well as provided us with detailed pictures of how their properties
vary smoothly across the ISCO \citep{Krolik:2005}.  These simulations have
vindicated the prescient remark made by \citet{Thorne:1974}:

\noindent ``In the words of my referee, James M. Bardeen (which
echo verbal warnings that I have received from Ya. B. Zel'dovich and
V.F. Schwartzman), `It seems quite possible that magnetic stresses
could cause large deviations from circular orbits in the very inner
part of the accretion disk and change the energy-angular-momentum
balance of the accreting matter by an amount of order unity'."

Indeed, when the black hole rotates, significant magnetic stresses can
be found throughout the accretion flow, all the way to the edge of the
event horizon \citep{Krolik:2005}.

Work on specific dynamical pictures of accretion has stimulated a
reexamination of the simple picture that discs have sharp edges, within
which little of interest happens.  Although it is true that there are
qualitative changes across the ISCO, they do not happen discontinuously.
As argued by \cite{Krolik:2002}, what one means by ``edge" depends on the
question asked.  For example, the ``reflection edge", the edge outside of
which most of the observed Fe K$\alpha$ and Compton reflection photons are
created, is likely to lie near, but possibly either inside or outside, the
ISCO \citep{Reynolds:1997,Krolik:2002,Brenneman:2006,Reynolds:2008}---its
exact position depends on the density and optical depth of the gas
in that region, and on the intensity of ionizing radiation striking
it. Similarly, the ``radiation edge" (the edge inside of which
little of the total luminosity emitted by the flow escapes to infinity)
should be near the ISCO, but is not necessarily identical to it.
Its position depends both on the profile of dissipation and
on the ability of photons
to escape to infinity, and to do so without excessive loss of energy to
gravitational redshift.

In this paper we examine the influence of magnetic stresses at
and inside the ISCO on the apparent size, luminosity and characteristic
temperature of black hole accretion discs.  This effort is important to
black hole phenomenology because so many observational diagnostics depend
upon these three parameters.  A prime example is provided by attempts to
use spectral fits to constrain black hole mass and spin.  This
program rests upon the idea that the observed luminosity,
$L$, is essentially thermal, so that it may be characterized
by an effective radiating area and a characteristic temperature:
$L = A(a/M,\theta) T^{4}_{char}$.  Because both $T_{char}$ and $L$
can be measured, $A$ can be inferred through this relation.  With
a theoretically-supported connection between $A$ and $a/M$ (and some
other constraint on $\theta$), the inference of $A$ leads to an
inference of $a/M$.  Current efforts connect $T_{char}$ to $M$, $\dot M$, and
$a/M$ using the \cite{Novikov:1973} model for the radial dissipation profile.
This model depends in an essential way on the assumption that
all stresses cease at and within the ISCO, whose location (at
Boyer-Lindquist radial coordinate $r/M=r_{ms}$) is a function only of
the black hole mass and spin.  This temperature is further adjusted by
an appropriate correction
for gravitational and Doppler energy shifts and a ``color temperature
correction'' due to opacity effects \cite[see e.g.][]{Done:2008}.  The
apparent size of the disc, $A$, is likewise found from
the dissipation profile of the Novikov-Thorne model.  As first noted by
\cite{Page:1974},
significant magnetic forces undercut the rationale for the zero-stress
boundary condition; the simulation data we discuss here shows quantitatively
how these magnetic forces alter the connections between both $A$ and $T_{char}$
and the black hole spin parameter.  Although more work is needed to explore
fully these new effects, in this paper we begin the discussion of
how they can influence these inferences.

The simulations reported here employ full general relativity
and three-dimensional MHD, so that whilst their treatment of angular momentum
flow and inflow dynamics is quite accurate, they do not directly
track dissipation.  
In an accretion disc, energy is extracted from
orbital motion and transformed into kinetic and magnetic energy on the
largest scales of the turbulence.  Subsequently, this energy cascades
down to a dissipation scale (either viscous or resistive) where it is
finally thermalized.  Current simulations can describe well the first
stages of this process, but can mimic only indirectly the last step:
grid-level effects intervene at lengthscales far larger than the physical
scale of dissipation.  In fact, the simulation code whose data we will
use solves only the internal energy equation, and makes no attempt to
follow dissipative energy losses except those associated with shocks.

We proceed by instead making a plausible {\it ansatz} for heating within
the disc that can be determined \textit{a posteriori } from simulation
data.  As we shall see, our results depend primarily on the qualitative
fact that dissipation continues smoothly across the ISCO, and on the
nature of photon trajectories deep in a relativistic potential; for
this reason, we believe that a non-rigorous, but physically-motivated,
{\it ansatz} will not be misleading.  We opt for the simplest choice:
a connection between the heating rate and the stress that follows
from the standard model for energy conservation in an accretion disc
\citep{Page:1974,Balbus:1994,Hubeny:1998}.

We can also investigate the importance of enhanced stress in
a semi-analytic fashion that bypasses most of the limitations of the
current simulations.  We employ the model formulated by \cite{Agol:2000}
(hereafter the AK model) that allows for non-zero stress at the ISCO
but otherwise computes the dissipation profile using the same approach
as the standard model.  The AK model cannot be extrapolated into the
plunging region, and has in common with the NT model a
fixed disc size.  Its unique feature is
the enhanced total dissipation due to the nonzero stress at the ISCO;
simulation data provides the single parameter needed to calibrate the model.
The AK model, therefore, provides an important link between the
standard model and the full simulation results and allows us to gauge
the appropriateness of the radiation {\it ansatz} employed for the
latter.

\begin{table*}
\caption{Simulation Parameters}
\label{sims}
\begin{tabular}{@{}lccccccc}
\hline
Name          &
$a/M$         &
$r_{+}/M$ &
$r_{\rm in}/M$ &
$r_{\rm ms}/M$  &
Field &
$T_{avg}\times10^{3}M$ &
Originally Presented in \\
\hline
KD0b & 0.0   & 2.00 & 2.104  & 6.00 & Dipole & 6--8 & \cite{De-Villiers:2003b}  \\
KD0c & 0.0   & 2.00 & 2.104  & 6.00 & Dipole & 8--10 & \cite{Hawley:2006}  \\
QD0d & 0.0   & 2.00 & 2.104  & 6.00 & Quadrupole & 8--10 & This work  \\
KDIa & 0.5   & 1.86 & 1.904  & 4.23 & Dipole & 6--8  & \cite{De-Villiers:2003b}    \\
KDIb & 0.5   & 1.86 & 1.904  & 4.23 & Dipole & 8--10  & \cite{Hawley:2006}   \\
KDPd & 0.9   & 1.44 & 1.503  & 2.32 & Dipole & 6--10  & \cite{De-Villiers:2003b}  \\
KDPg & 0.9   & 1.44 & 1.503  & 2.32 & Dipole & 8--10  & \cite{Hawley:2006}  \\
QDPa & 0.9   & 1.44 & 1.503  & 2.32 & Quadrupole & 8--10 & \cite{Beckwith:2008a}  \\
TDPa & 0.9   & 1.44 & 1.503  & 2.32 & Toroidal & 20--22 & \cite{Beckwith:2008a}  \\
KDG  & 0.93  & 1.37 & 1.458  & 2.10 & Dipole & 8--10  & \cite{Hawley:2006}  \\
KDH  & 0.95  & 1.31 & 1.403  & 1.94 & Dipole & 8--10  & \cite{Hawley:2006}  \\
KDJd & 0.99  & 1.14 & 1.203  & 1.45 & Dipole & 8--10  & This work  \\
KDEa & 0.998  & 1.084 & 1.175  & 1.235 & Dipole & 6--8  & \cite{De-Villiers:2003b} \\
KDEb & 0.998  & 1.084 & 1.175  & 1.235 & Dipole & 8--10  & This work \\
QDEb & 0.998  & 1.084 & 1.175  & 1.235 & Quadrupole & 8--10  & This work \\
\hline
\end{tabular}

\medskip Here $a/M$ is the spin parameter of the black hole, $r_{+}$ is 
the horizon radius, $r_{\rm in}$ is the innermost radius in the 
computational grid, field is the initial field topology in the torus and 
$T_{avg}\times10^{3}$ is the time-interval over which simulation data 
was averaged. For reference, we note where individual simulations were 
originally presented.
\end{table*}

To relate dissipation rates to radiation received at infinity, we must
perform one additional calculation using three further assumptions:
that the dissipated heat is efficiently converted to photons, that
these photons emerge from the accretion flow very near where they are
created, and that they are radiated isotropically in the fluid frame.
The first two assumptions are equivalent to requiring the timescales
for dissipation, radiation, and photon diffusion to be shorter than the
inflow timescale for all fluid elements. The third, while not strictly
justified, is the simplest guess we can make.  Given those assumptions,
we use a general relativistic ray-tracing code to relate the luminosity
radiated by each fluid element to the luminosity received by observers
located at different polar angles far from the black hole.

A further result of this calculation is a new estimate of the radiative 
efficiency of accretion.  The traditional calculation of this quantity 
follows directly from a primary assumption that there are no forces 
inside the ISCO and two additional assumptions, that the radiation is 
prompt and all of it reaches infinity.  We improve upon this traditional 
estimate in two ways: we allow for dissipation associated with the 
stresses we measure at and inside the ISCO; and we calculate the 
radiated energy (even within the NT model) that actually reaches distant 
observers. However, we do not regard our result as sufficiently final or 
complete to give it much weight, as it is likely to be more model- and 
parameter-dependent than our placement of the radiation edge.  One 
reason for downplaying this result is that we find it necessary to omit 
any estimate of dissipation outside the main part of the accretion flow.  
We have less confidence that our dissipation prescription is appropriate 
in the jet, or even in the disc corona, than we do when it is applied to 
the main disc body and plunging region.  Moreover, radiation from these 
lower-density regions is less likely to be thermalised and contribute to 
the radiation usually identified with the disc continuum.

The rest of this paper is structured as follows. In \S\ref{numerics} we 
briefly review the numerical scheme employed to solve the equations of 
GRMHD in the simulations whose data we use, give an overview of the 
parameters of the simulations included in this work, and describe our 
general relativistic ray-tracing code.  In \S\ref{dissmodels}, we 
contrast the dissipation rate distributions predicted by the standard 
model, its Agol \& Krolik extension, and our simulation-based {\it 
ansatz}.  In \S\ref{rad} we compute the luminosity at infinity predicted 
by each of these dissipation profiles and discuss their consequences for 
the location of the radiation edge and the characteristic temperature of the
accretion flow. Finally, in \S\ref{summ}, we draw 
specific conclusions from our results and describe their implications 
for black hole phenomenology.

\section{Numerical Details}\label{numerics}

\subsection{GRMHD Simulations}

The calculations presented here continue the analysis of a program of
black hole accretion disc simulations begun in \cite{De-Villiers:2003b}
and continued by \cite{Hirose:2004}, \cite{De-Villiers:2005},
\cite{Krolik:2005}, \cite{Hawley:2006} and \cite{Beckwith:2008a}.
The simulation code is described in \cite{De-Villiers:2003}.  This code
solves the equations of ideal non-radiative MHD in the static
Kerr metric of a rotating black hole using Boyer-Lindquist coordinates.
Values are expressed in gravitational units $(G = M = c = 1)$ with line
element $ds^{2} = g_{tt} dt^{2} + 2g_{t \phi} dt d\phi + g_{rr} dr^{2} +
g_{\theta \theta} d \theta^{2} + g_{\phi \phi} d\phi^{2}$ and signature
$(-,+,+,+)$.  The determinant of the $4$-metric is $\alpha \sqrt{\gamma}$,
where $\alpha = (-g^{tt})^{-1/2}$ is the lapse function and $\gamma$
is the determinant of the spatial $3$-metric.

The relativistic fluid at each grid zone is described by its density
$\rho$, specific internal energy $\epsilon$, $4$-velocity $u^{\mu}$, and
isotropic pressure $P$. The relativistic enthalpy is $h = 1 + \epsilon
+ P / \rho$. The pressure is related to $\rho$ and $\epsilon$ via the
equation of state for an ideal gas, $P = \rho \epsilon ( \Gamma - 1)$. The
magnetic field is described by two sets of variables.  The first is the
constrained transport magnetic field $\mathcal{B}^{i} = [ijk] F_{jk}$,
where $[ijk]$ is the completely anti-symmetric symbol, and $F_{jk}$ are
the spatial components of the electromagnetic field strength tensor.
From these are derived the magnetic field four-vector, $(4\pi)^{1/2}
b^{\mu} = {}^{*}F^{\mu \nu} u_{\nu}$, and the magnetic field scalar,
$||b^{2}|| = b^{\mu} b_{\mu}$. The electromagnetic component of the
stress-energy tensor is $T^{\mu \nu}_{\mathrm{(EM)}} = \frac{1}{2}g^{\mu
\nu} ||b||^{2} + u^{\mu} u^{\nu} ||b||^{2} - b^{\mu} b^{\nu}$.

The initial conditions for the simulations consist of an isolated
hydrostatic gas torus orbiting near the black hole, with an inner edge
at $r = 15 M$,  a pressure maximum located at $r \approx 25 M$, and a
(slightly) sub-Keplerian distribution of angular momentum throughout.
Parameters for all of the simulations analyzed in this work are summarised
in Table \ref{sims}.  We use three different initial field configurations.
Models labeled KD have an initial dipole field that lies along surfaces of
constant pressure within the torus.  The initial field in the QD models
is a quadrupolar configuration consisting of field of opposite polarity
located above and below the equatorial plane in the torus.  The TD
model's initial field is purely toroidal.  Many of these simulations
are studied in more detail in the referenced work given in the last
column of the table.  A few simulations are new to this work and these
were computed with the latest version of the GRMHD code as described
by \cite{De-Villiers:2006}. Whenever we require time-averaged
data, we average over the times $T_{avg}$ shown in Table~\ref{sims},
from which we have full three-dimensional snapshots taken every $80M$.  Thus, 26
time-samples go into each time-average.

Each simulation uses $192 \times 192 \times 64$ $(r,\theta,\phi)$
grid zones.  The radial grid extends from an inner boundary located
just outside the black hole event horizon (see Table \ref{sims} for
the precise location in each case), to the outer boundary located at
$r_{out} = 120 M$ in all cases.  The radial grid is graded according to
a hyperbolic cosine function in order to concentrate grid zones close
to the inner boundary, except in the TDPa model where a logarithmic
grid was used.  An outflow condition is applied at both the inner and
outer radial boundary.  The $\theta$-grid spans the range $0.045 \pi \le
\theta \le 0.955 \pi$ using an exponential distribution that concentrates
zones near the equator.  A reflecting boundary condition is used along
the conical cutout surrounding the coordinate axis. Further discussion
of the $r,\theta$ griding and boundary conditions can be found in
\cite{Hawley:2006}. Finally, the $\phi$-grid spans the quarter plane,
$0 \le \phi \le \pi / 2$, with periodic boundary conditions in $\phi$.
The use of this restricted angular domain significantly reduces
the computational requirements of the simulation \cite[for further
discussion of the effects of this restriction see][]{De-Villiers:2003a}.
Most of the simulations were run to either 8000 or $10^4M$,
which corresponds to approximately 10 or $12$ orbits at the initial
pressure maximum.  The toroidal simulation was run to time $29000M$
due to the longer timescale taken for accretion to commence in this
case \cite[see][]{HK:2002}. For each simulation the time step $\Delta
t$ is determined by the extremal light crossing time for a zone
on the spatial grid and remains constant for the entire simulation
\cite[]{De-Villiers:2003}.

In all the models the initial evolution is driven by the MRI and by
growing magnetic pressure due to shear amplification of poloidal
field (if present) into toroidal.  By the end of the simulation,
a quasi-steady state accretion disc extends from the hole out to
$r\sim 20M$.  Beyond this radius the net mass motion shifts to outward
flow as it absorbs angular momentum from the inner disc.  In this paper,
therefore, we focus our attention on the region inside $20M$ and on late
times after the quasi-steady disc has been established.  Whenever we
require time-averaged data, we average over the times $T_{avg}$ shown in
Table~\ref{sims}, from which we have full 3-d snapshots taken every $80M$.
Thus, 26 time-samples go into each time-average.

\subsection{Ray-Tracing}

Transformation into the reference frame of a distant observer was 
accomplished by means of a ray-tracing code. We assign to this observer 
a coordinate system (``photographic plate'') defined by the photon 
impact parameters $(\alpha,\beta)$, which can be simply related to the 
photon's constants of motion, $(\lambda,q)$ via \cite[][]{Bardeen:1972}:
\begin{equation}
\alpha = - \frac{\lambda}{\sin \theta_{o}}; \;\; \beta = \pm \sqrt{q + a^{2} cos^{2} \theta_{o} - \lambda^{2} \mathrm{cot}^{2} \theta_{o}}
\end{equation}
where $\theta_{o}$ is the co-latitude 
(inclination) of the distant observer. Once $(\lambda,q)$ are known, 
then the photon 4-momentum, $p^{\mu}$ at the observer can be 
constructed. Photon paths were then traced from a distant observer to 
the surface of the accretion disc by integration of the null-geodesic 
equation:
\begin{equation}
\label{photoneqnmotion} 
\frac{d^{2}x^{\kappa}}{ds^{2}} + \Gamma^{\kappa}_{\mu \nu} \frac{d x^{\mu}}{ds} \frac{d x^{\nu}}{ds} = 0 
\end{equation}
where $x^{\mu}$ is the spacetime coordinate of the photon, $s$ is an 
affine parameter, $\Gamma^{\kappa}_{\mu \nu}$ are the connection 
coefficients and the ${d x^{\mu}} / {ds}$ are specified initially by the 
$p^{\mu}$ at the distant observer.  Equation~\ref{photoneqnmotion} may 
be recast as a set of coupled first-order differential equations. We 
solve this set by applying the fifth-order Runge-Kutta integrator 
described by \cite{Brankin:1991} as implemented in the NAG FORTRAN 
Library (Mark 21). The local error in the integration was kept to 
$10^{-8}$. Where the geodesic passed through a radial turning point, the 
location of the turning point found by the integrator was compared with 
its analytic value \cite[see][]{Beckwith:2005} and was found to agree to 
the order of the local error in the integration. The integration of the 
geodesic was terminated when the photon intersected the disc surface, 
which was located in the $\theta = \pi / 2$ plane. Once the intersection 
with the disc surface, $x^{\mu}_{\mathrm{surf}}(\alpha,\beta)$ is known, 
the flux measured by the distant observer can be calculated.  By 
treating the emission as a line flux, a given radius within the disc can 
be associated with an observed flux by
\begin{equation}
\label{obsflux} 
F(\alpha,\beta) = g^{4}[x^{\mu}_{\mathrm{surf}}(\alpha,\beta)] Q [x^{\mu}_{\mathrm{surf}}(\alpha,\beta)] \delta (E_{o} - g E_{e}). 
\end{equation}
The term $g$ is the Doppler and gravitational energy shift of the photon 
between the disc and the distant observer, found by projecting the 
photon four-momentum onto the tetrad describing the fluid rest frame 
(see Appendix A). For the standard disc models the fluid motion is 
assumed to be purely Keplerian, that is $V^{\phi} = 1/ (a + r^{3/2})$, 
whilst for fluxes calculated from the simulations the fluid motion on 
the emission surface is obtained directly from the data.

\newpage

\section{Dissipation in the Fluid Frame}\label{dissmodels}

The radial dependence of the fluid-frame dissipation rate per unit disc 
surface area in the standard model is
\begin{equation} \label{eqn:NTdissprofile} 
Q_{\mathrm{NT}} = \frac{3 GM \dot{M}}{4 \pi r^{3}} R_{R}(r),
\end{equation} 
where $R_{R}(r)$ is a function encapsulating all of the
relativistic effects relevant for disc dynamics and the effect of the
assumed stress-free inner boundary condition at the ISCO. 
In the Newtonian limit for a Keplerian
potential, $R_{R}(r)$ takes the familiar form:  
\begin{equation}
R_{R}(r) = 1 - \left( r_{m}/{r} \right)^{1/2},
\end{equation} 
where $r_{m}$ is the radius in the disc at which the stress is zero. $Q$ 
is defined in the frame that comoves with the surfaces of the disc (the 
``disc frame").  Determining the form of this function as measured by a 
distant observer then requires a transformation into that observer's 
reference frame, accomplished by explicitly tracing photons from the 
emitter to the observer \cite[see e.g.][]{Beckwith:2004}.

The NT dissipation profile is derived from conservation laws applied
in a time-steady context.  We require the instantaneous dissipation
in a non-stationary flow.  To estimate it, we begin with the
vertically-integrated and azimuthally-averaged fluid-frame Maxwell
stress ${\cal W}^{(r)}_{(\phi)}$.  Following \cite{Krolik:2005},
we find the fluid-frame Maxwell stress from the following expression:
\begin{equation}
\label{mxwstress}
{\cal W}^{(r)}_{(\phi)}=  \frac{ \int_{\mathrm{disc}}
{e^{(r)}_{\mu} e^\nu_{(\phi)} T^\mu_{\nu \; \mathrm{(EM)}} dx^{(\theta)} dx^{(\phi)}}}
{\int \, dx^{(\phi)}(r, \theta=\pi/2)}
\end{equation}
where $e^{\mu}_{(\nu)}$ are the basis vectors describing the rest frame
of the fluid (see Appendix A) and $dx^{(\mu)} = e^{(\mu)}_{\nu} dx^{\nu}$
is the fluid-frame co-ordinate element. The subscript ``disc'' denotes
that we are including only contributions to the integral from bound
material ($-hu_{t}<1$) that lies within one density-scale height of
the midplane (we define the density scale height $\theta_h$
at a given $r$ and $\phi$ by $\rho(r,\phi,\theta_h) = \rho(r,\phi,\pi/2)/e$).
This form of the stress facilitates comparison with
the conventional representation of the vertically-integrated stress
given by \cite{Novikov:1973}; we discuss later why we restrict the
integration to matter within a single density scale height of the plane.
Plots showing the radial profile of ${\cal W}^{(r)}_{(\phi)}$ compared
to the conventional stress are given in \cite{Krolik:2005} for four
different black hole spins and in \cite{Beckwith:2008a} for several
different magnetic field topologies.  In every case examined in both
these studies, the Maxwell stress in the fluid frame extends all the
way to the event horizon whenever the black hole rotates and reaches
deep into the plunging region even when $a/M = 0$.

To determine the observational implications of this additional stress
we must compute its associated dissipation.
As discussed above, this requires adopting a model for dissipation;
we consider two.  The first is the AK model, which
extends the standard disc approach to allow for a non-zero stress at
the ISCO.  The second is a model, derived below, that determines
the local dissipation in terms of the local stress and gradients of
the fluid velocity.

\subsection{Dissipation Derived from the Agol \& Krolik Model}\label{ak2000}

\begin{figure}
\begin{center}
\leavevmode
\includegraphics[width=\columnwidth]{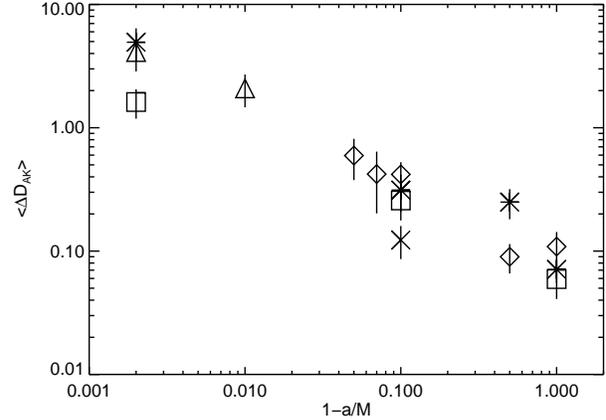}
\end{center}
\caption[]{Time-averaged fractional increase in total dissipation,
$\langle \Delta D_{\mathrm{AK}} \rangle$, derived using the
Agol-Krolik model and the simulation Maxwell
stress at the ISCO. Error bars show $\pm1$ standard deviation about the
average and reflect the intrinsic variability in the stress at the 
ISCO seen in the simulation.
The dipolar field simulations
described by \cite{De-Villiers:2003b} are shown by star symbols,
those described by \cite{Hawley:2006} are marked by diamonds and the
new high spin cases computed for this paper are marked by triangles.
Simulations with quadrupolar topologies are denoted by squares. The
toroidal field simulation ($a/M=0.9$) is denoted by a cross. Simulation
data was time-averaged over the intervals given in Table \ref{sims}.}
\label{eta} 
\end{figure}

AK showed how the inner boundary condition for the standard NT model can
be modified to allow for arbitrary stress at the ISCO.  Parameterizing
the stress at the ISCO in terms of the resultant increase in radiative
efficiency $\Delta \epsilon$ yields the following expression for the
vertically-integrated $r-\phi$ stress in the fluid frame:
\begin{equation} \label{epsdef}
-\int{dz S_{(r) (\phi)}(z)} = \frac{\dot{M} \Omega}{2 \pi}
\left[\frac{r^{3/2}_{ms} C^{1/2} (r_{ms}) \Delta \epsilon}{D(r)
r^{1/2}} + R_{T}(r) \right]
\end{equation}  
where $S_{(r) (\phi)}$ is the $r-\phi$ component of the fluid-frame 
viscous stress tensor; $C(r)$, $D(r)$ are relativistic correction 
factors and $R_{T}(r)$ is similar in nature to $R_{R}(r)$, reducing to 
an identical expression in the Newtonian limit.  The corresponding 
expression for the fluid-frame dissipation rate $Q_{AK}$ is:
\begin{equation} \label{eqn:AKdissprofile} 
Q_{\mathrm{AK}} = \frac{3 GM \dot{M}}{4 \pi
r^{3}} \left[ \frac{r^{3/2}_{ms} C^{1/2} (r_{ms}) \Delta \epsilon}{C(r)
r^{1/2}} + R_{R}(r) \right]
\end{equation} 
This semi-analytic prescription lacks only a specification of the
increase in radiative efficiency.  This is supplied from the
simulation data by setting
$-\int{dz S_{(r) (\phi)}(z)} = {\cal W}^{(r)}_{(\phi)}$
at the ISCO for each time-step in the data-set
and then inverting Equation~\ref{epsdef} to determine $\Delta \epsilon$.
Lastly, we average $\Delta \epsilon$ over the different time-steps.
It is important to recognise that the nature of the AK model is
such that stress data from one location alone---the ISCO---are
enough to determine the stress (and dissipation) everywhere else
because all locations in the disc outside $r_{ms}$ are linked through
the assumptions of time-steadiness and axisymmetry.

\newpage

It is also important to recognise that there are certain intrinsic
limitations built into the AK model, inherited from its NT parent.
These are best illustrated by returning to its basis, the relativistic
conservation of momentum-energy, expressed by the requirement that
$\nabla_\nu T^{\mu\nu} = 0$.  In both of these models, the stress
tensor was defined to have the form
\begin{equation}
T^{\mu \nu} = \rho h u^{\mu} u^{\nu} + Pg^{\mu \nu} + S^{\mu \nu} +
u^{\mu} q^{\nu} + u^{\nu}q^{\mu}
\end{equation}
where $S^{\mu \nu}$ is the part of the stress tensor responsible for 
inter-ring torques and $q^{\mu}$ is the energy flux four-vector.  Energy
flux contributes to the stress tensor in proportion to the four-velocity
because energy can be directly advected with the material.  It is 
assumed that both $S^{\mu\nu}$ and $q^\mu$ are orthogonal to the four 
velocity: $u_{\mu} S^{\mu \nu} = 0$ and $u_{\mu} q^{\mu} = 0$.  In the 
case of the stress tensor, this choice was made with classical viscosity 
in mind and also ensures that heat advection can be cleanly separated from
other kinds of stress.  In the case of the energy flux four-vector, orthogonality
with the four-velocity ensures
true stationarity.  Lastly, $S^{\mu\nu}$ is also assumed to be symmetric.  In the 
context of MHD stresses, we have
\begin{equation}
T^{\mu\nu}_{\mathrm{(EM)}} = ||b||^2 u^\mu u^\nu + (||b||^2/2)g^{\mu\nu} - b^\mu b^\nu,
\end{equation}
which is manifestly symmetric, but is {\it not} orthogonal to $u_\mu$:
\begin{equation}
u_\mu T^{\mu\nu}_{\mathrm{(EM)}} = -(||b||^2/2) u^\nu .
\end{equation}
In other words, Maxwell stresses fail to be orthogonal to the fluid 
four-velocity to the degree that field energy is carried by fluid 
motions.

Moreover, both the AK and NT models also assume that $u_r = u_\theta = 0$
in the (Boyer-Lindquist) coordinate frame.  Combined with the orthogonality
between the four-velocity and the stress tensor, this assumption places rather
special conditions on the relation between $b^\phi$ and $b^t$.  Although
$u_r$ and $u_\theta$ are very small compared to $u_\phi$ and $u_t$ in
the disc body, this assumption begins to fail as the ISCO is approached,
and becomes entirely invalid in the plunging region.

In any event, independent of the method used to find it, $Q(r)$ can be
used to compute
the instantaneous total dissipation by integrating 
\begin{equation}
D = \int^{\infty}_{r_{ms}} \int \, Q dx^{(t)} dx^{(r)} dx^{(\phi)}
\end{equation}
where $dx^{(\mu)} = e^{(\mu)}_{\nu} dx^{\nu}$ are the fluid frame differential
coordinate elements in the $\theta=\pi/2$ plane.
The time-averaged fractional increase in total energy dissipated
due to the additional stress at the ISCO is then given by
$\langle \Delta D_{\mathrm{i}} \rangle$: 
\begin{equation} \label{fracin}
\langle \Delta D_{\mathrm{i}} \rangle = 
\langle \frac{D_{\mathrm{i}} - D_{\mathrm{NT}}} {D_{\mathrm{NT}}} \rangle 
\end{equation}
Here the index $i$ could be either AK or MW, and
all 26 of our full three-dimensional snapshots are used for these time-averages.

Making use of the data from all our simulations, we plot the fractional
increases in dissipation for the AK model in Figure \ref{eta}.  This model
predicts that nonzero stress at the ISCO produces a dissipation rate that
is $\simeq 10\%$ greater than in the NT model for zero spin, and increases
steadily as the spin increases; when $a/M = 0.998$ the total dissipation
rate is several times greater than in the NT model, with the specific
factor depending on the initial magnetic topology.   Generally speaking,
non-dipolar magnetic field topologies lead to fractional increases
that are smaller than their dipolar counterparts.  This contrast
arises from a distinguishing characteristic of the dipole simulations
for high spin holes: in the region near $r_{ms}$, the stress rises more
rapidly toward small radius for this field geometry than for quadrupolar
or toroidal fields (see Fig.~4 in \cite{Beckwith:2008a}).

\subsection{Dissipation Derived from the Local Stress-Energy Tensor}\label{pyntflx}

\begin{figure*}
\begin{center}
\leavevmode
\includegraphics[width=0.24\textwidth]{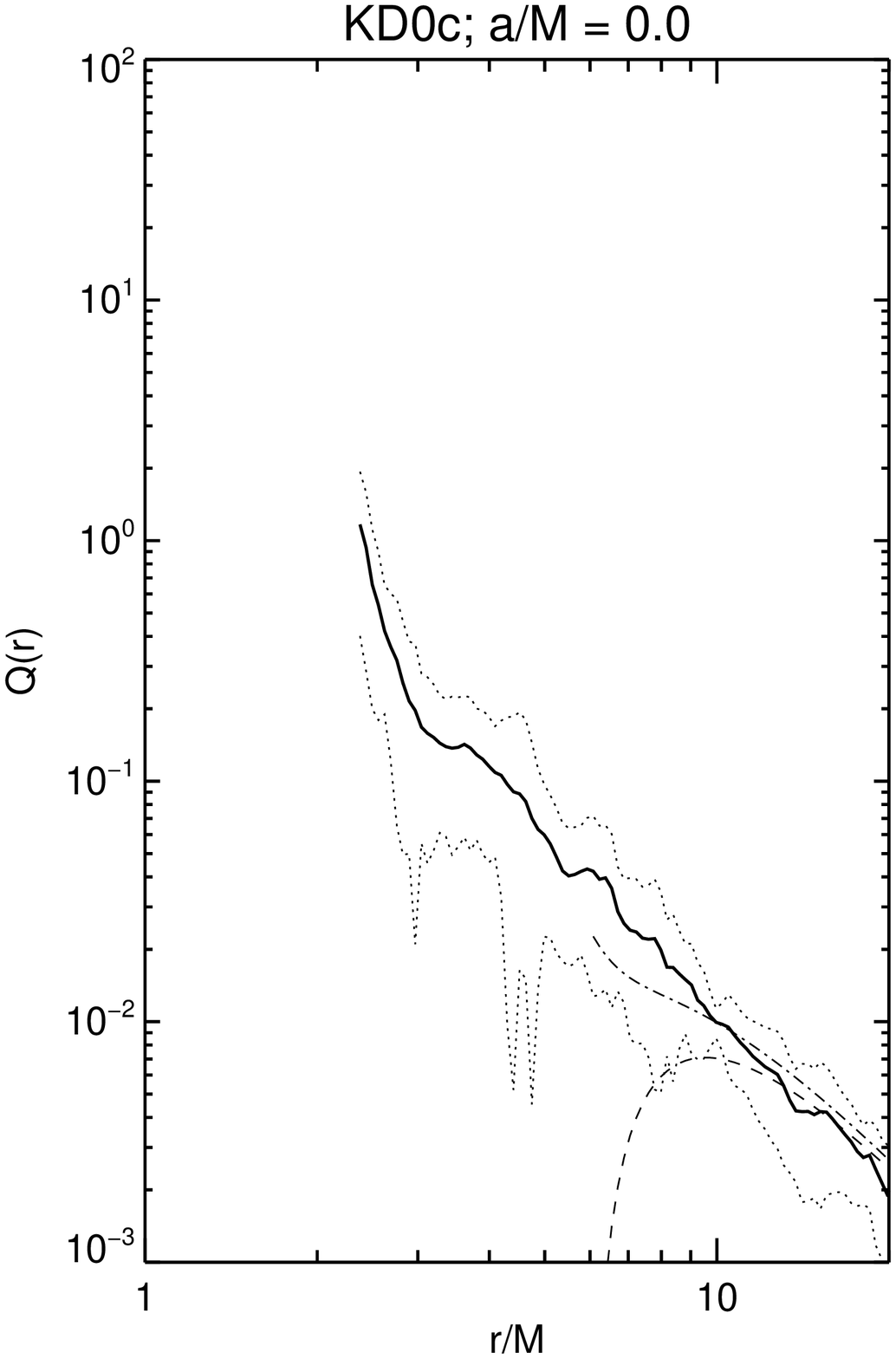}
\includegraphics[width=0.24\textwidth]{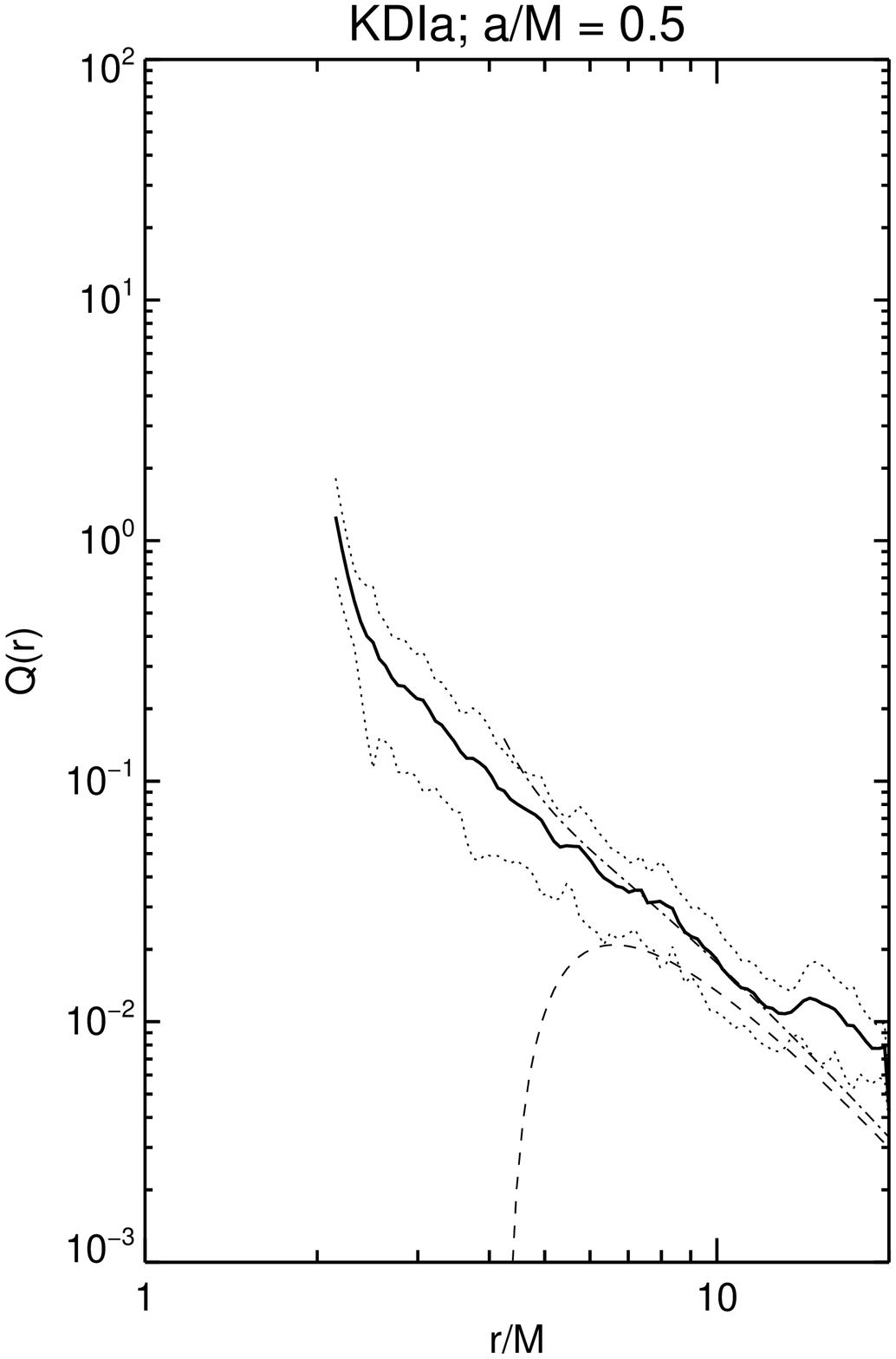}
\includegraphics[width=0.24\textwidth]{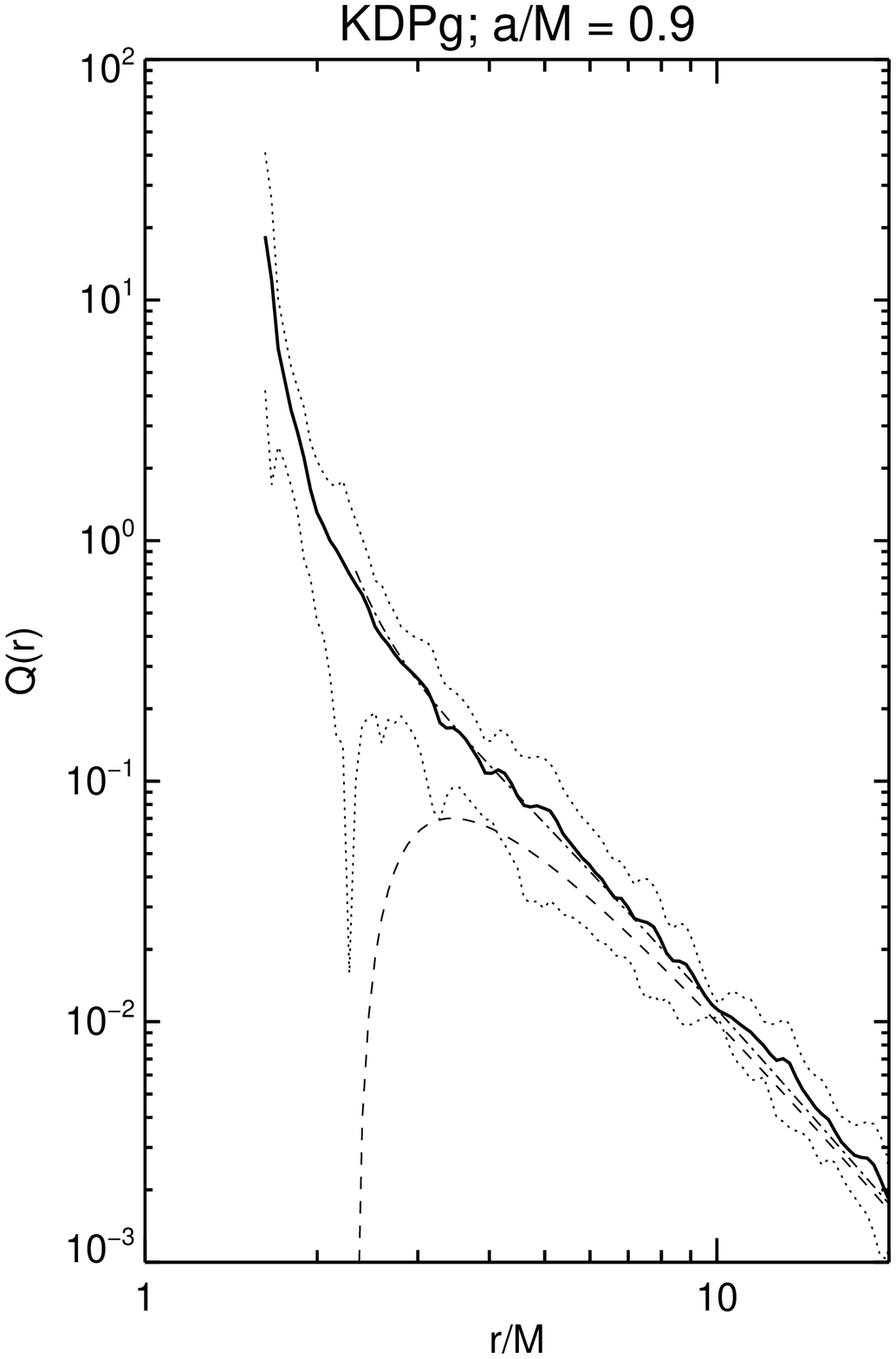}
\includegraphics[width=0.24\textwidth]{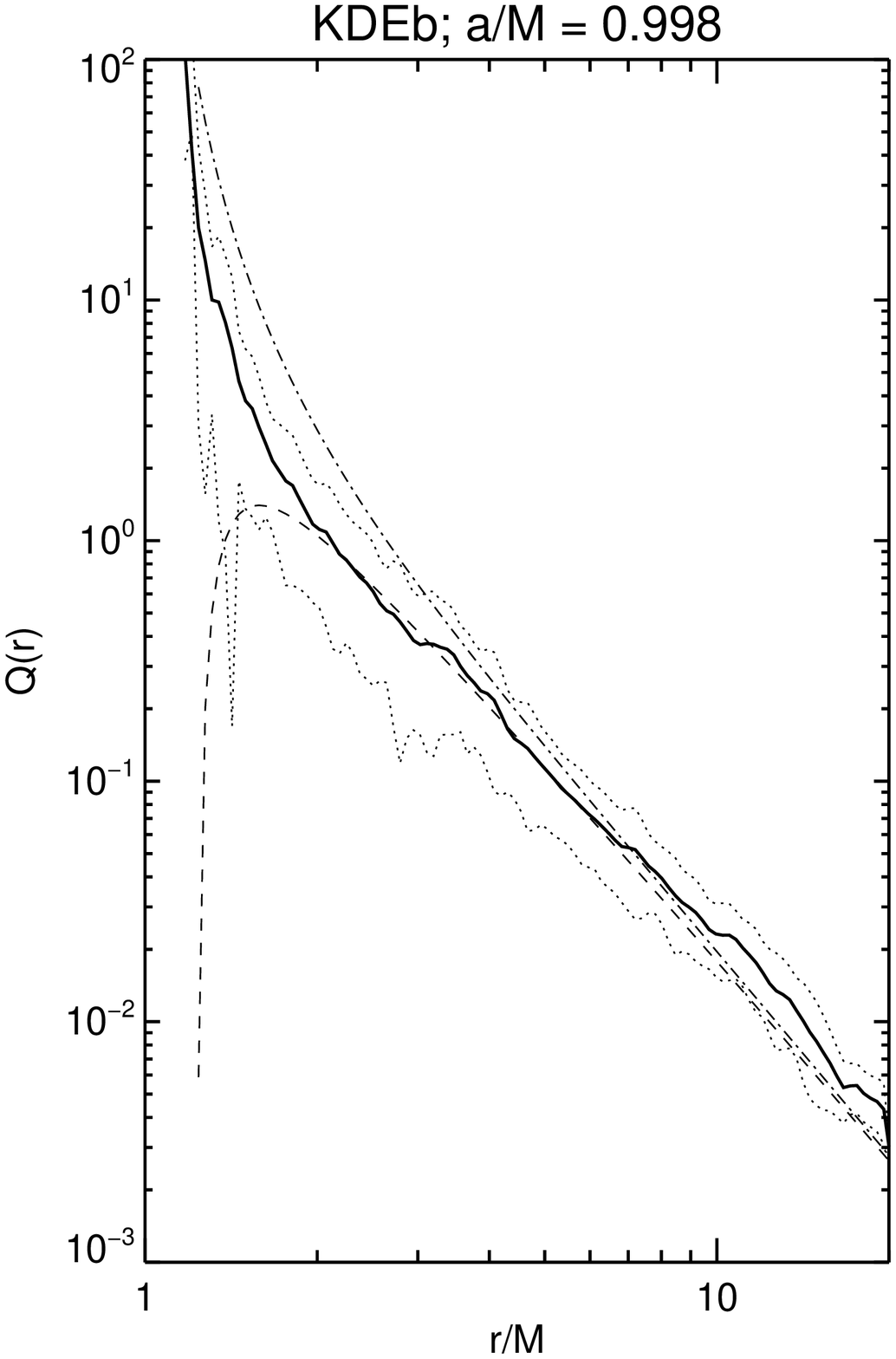}
\end{center}
\caption[]{Time-averaged radial profile of  $Q_{\mathrm{MW}}$
(solid lines) compared to $Q_{\mathrm{NT}}$  (dashed lines) and
$Q_{\mathrm{AK}}$ (dot-dash lines) for (from left to right) dipole
simulations KD0c ($a/M=0$, KDIb ($a/M=0.5$), KDPg ($a/M=0.9$) and KDEb
($a/M=0.998$). Dashed lines denoted $\pm1$ std. deviation from the
mean $Q_{\mathrm{MW}}$. Simulation data was time-averaged over the
periods given in Table \ref{sims}}
\label{spindiss} 
\end{figure*}

\begin{figure*}
\begin{center}
\leavevmode
\includegraphics[width=0.32\textwidth]{KDPhrg_fz}
\includegraphics[width=0.32\textwidth]{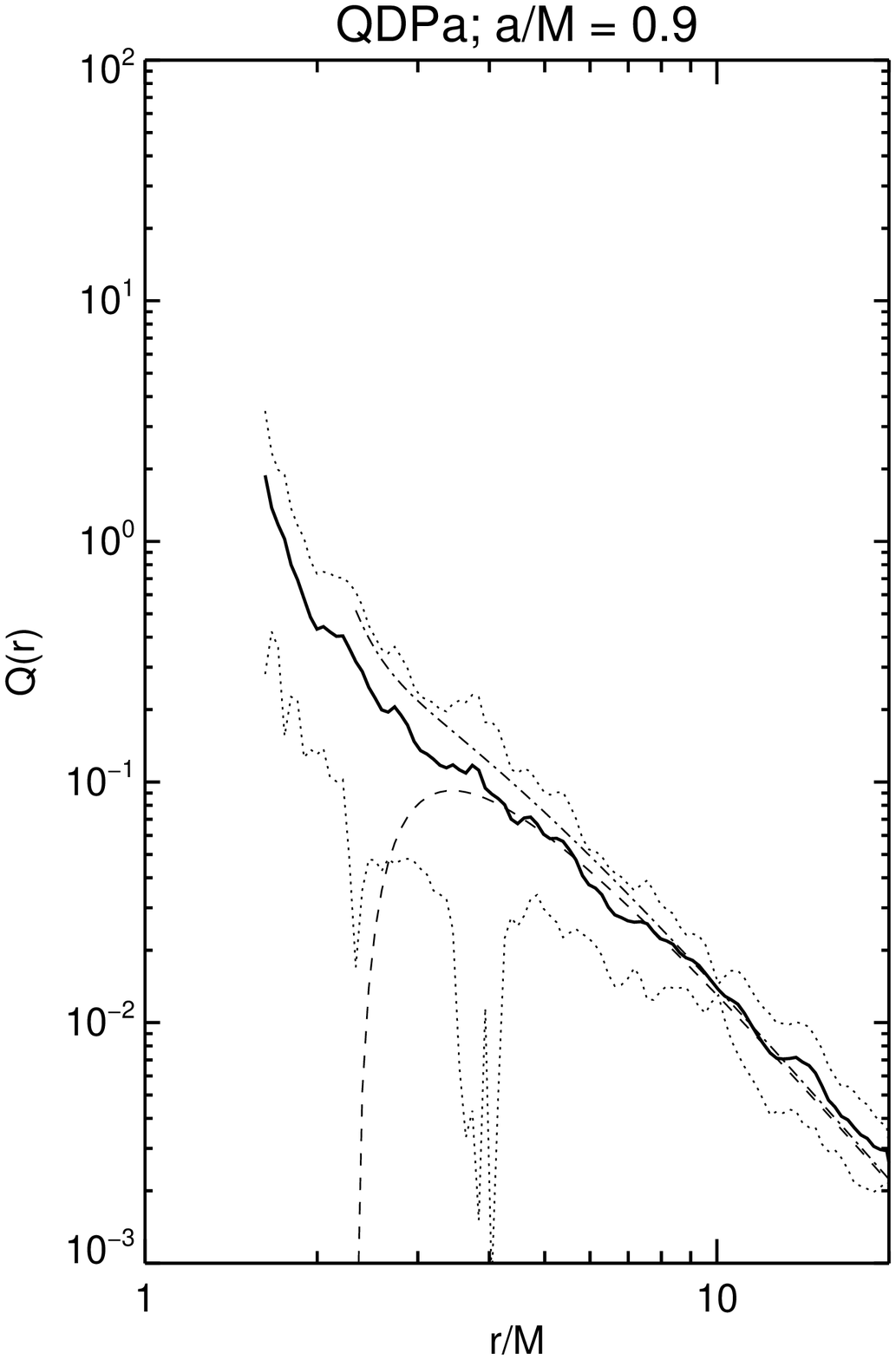}
\includegraphics[width=0.32\textwidth]{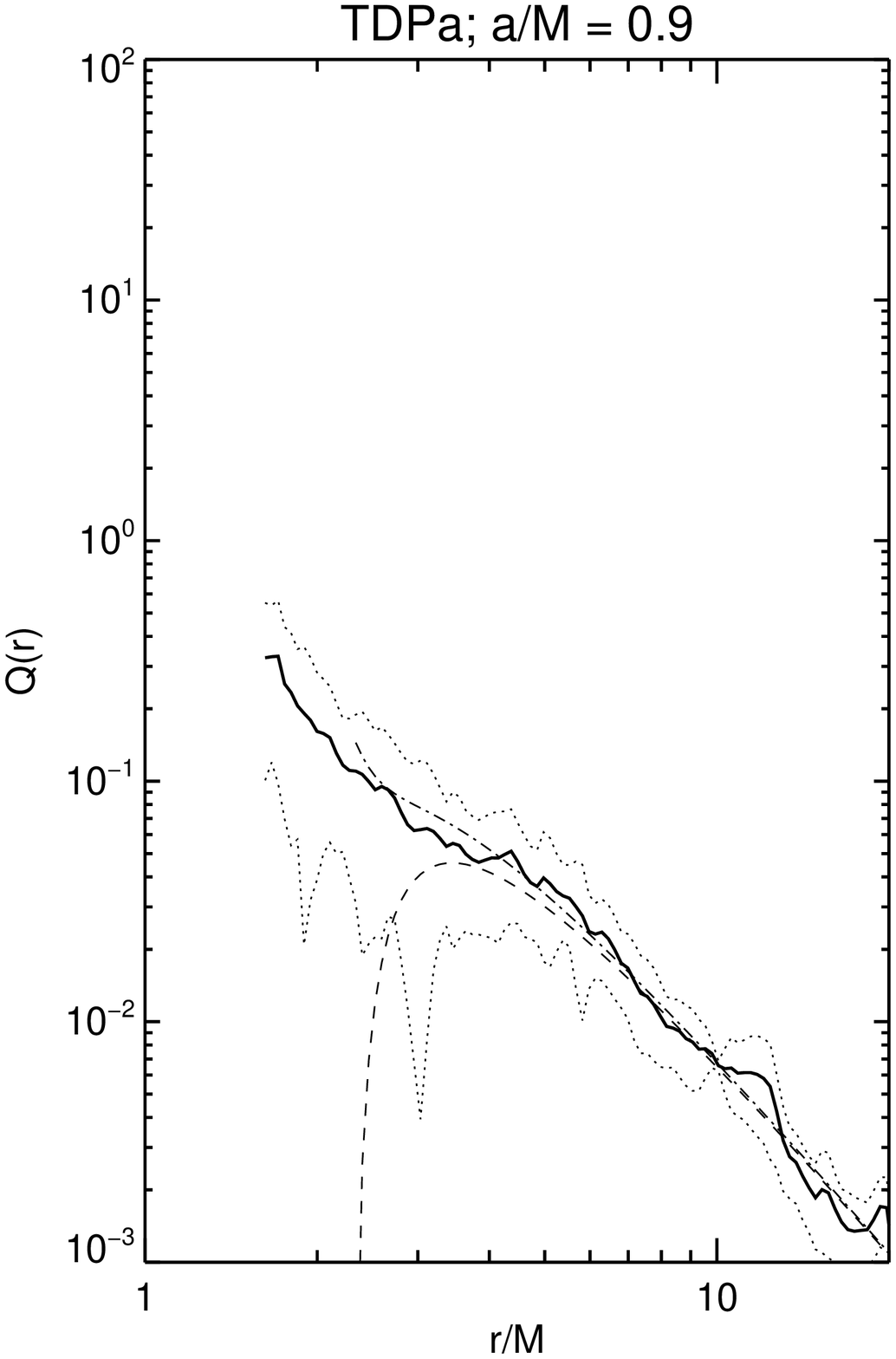}
\end{center}
\caption[]{Time-averaged radial profile of  $Q_{\mathrm{MW}}$
(solid lines) compared to $Q_{\mathrm{NT}}$  (dashed lines) and
$Q_{\mathrm{AK}}$ (dot-dash lines) for (from left to right) simulations
KDPg (dipole magnetic field topology), QDPa (quadrupole magnetic
field topology) and TDPa (toroidal magnetic field field topology). The
black hole spin was fixed at $a/M=0.9$. Dashed lines denoted $\pm1$
std. deviation from the mean $Q_{\mathrm{MW}}$. Simulation data was
time-averaged over the periods given in Table \ref{sims}}
\label{topdiss} 
\end{figure*}

To derive dissipation rates directly from instantaneous and local
simulation data, we return to the momentum-energy conservation
equation $\nabla_\nu T^{\mu\nu} = 0$.  If we define $T^{\mu\nu}$
in the way it was defined for the NT and AK models, projecting the
momentum-energy conservation equation onto $u_{\mu}$ yields:
\begin{equation}
\begin{split}
\nabla_{\mu} (\rho \epsilon u^{\mu}) + P \nabla_{\mu} u^{\mu} 
+ S^{\mu \nu} \nabla_{\nu} u_{\mu}\\
+ \nabla_{\nu} q^{\nu} + q^{\mu} u^{\nu} \nabla_{\nu} u_{\mu} = 0
\end{split}
\end{equation}
The GRMHD code evolves the internal energy equation 
\begin{equation}
\nabla_{\mu} (\rho \epsilon u^{\mu}) + P \nabla_{\mu} u^{\mu} = 0, 
\end{equation}
without either a
viscous stress (except for artificial bulk viscosity) or a radiative 
energy flux.  We can then use the orthogonality
of $q^{\mu}$  and $u_{\mu}$, 
along with the ansatz that energy liberated by the stress 
is promptly radiated, to write the energy balance equation as
\begin{equation}
S^{\mu \nu} \nabla_{\nu} u_{\mu} + \nabla_{\nu} q^{\nu} = 0.
\end{equation}
This description of the dissipation is consistent with the spirit of
the standard disc approximation
that the stress transporting angular momentum is the same stress
that locally heats the disc \cite[][]{Balbus:1999}.  However, if  $S^{\mu\nu}$
is calculated from the electromagnetic stress-energy tensor, 
$T^{\mu \nu}_{\mathrm{(EM)}}$, it is {\it not}
self-consistent for the reasons discussed in the context of $Q_{AK}$.

To solve this equation, we assume that the energy flux escapes 
perpendicular to the equatorial plane so that
$q^{\mu} = (0,0,0,q^{\theta})$ in the coordinate frame.  We also assume that
the flow is time-steady and axisymmetric (consistent with our use of
time- and toroidally-averaged values) and that $u^\theta = 0$ in
the coordinate frame.  With these assumptions,
we can expand the energy balance equation to yield:
\begin{equation}
\begin{split}
\partial_{\theta} (\alpha \sqrt{\gamma} q^{\theta}) = 
             - [S^{tr} \partial_{r} (\alpha \sqrt{\gamma} u_{t}) \\
+ S^{rr} \partial_{r} (\alpha \sqrt{\gamma} u_{r}) 
+ S^{\phi r} \partial_{r} (\alpha \sqrt{\gamma} u_{\phi})]
\end{split}
\end{equation}
where we have used the identity $\nabla_{\mu} (f v^{\mu}) =
(\alpha \sqrt{\gamma})^{-1} \partial_{\mu} (\alpha \sqrt{\gamma} f
v^{\mu})$. Solving for $q^{\theta}$ yields:
\begin{equation}
\begin{split}
q^{\theta} = - \frac{1}{\alpha \sqrt{\gamma}} \int_{\mathrm{disc}} [S^{t r} \partial_{r} (\alpha \sqrt{\gamma} u_{t}) \\
+ S^{rr} \partial_{r} (\alpha \sqrt{\gamma} u_{r}) 
+ S^{\phi r} \partial_{r} (\alpha \sqrt{\gamma} u_{\phi})] d\theta 
\end{split}
\end{equation}
where again the subscript ``disc'' denotes that we include only
contributions to the integral from bound material ($-hu_{t}<1$) that
lies within one density-scale height of the midplane. The fluid frame
dissipation rate computed from the electromagnetic stress-energy tensor
, $Q_{\mathrm{MW}}$,
is found from $Q_{MW} = e^{(\theta)}_{\mu} q^{\mu}$, which for $u^{\mu} =
(u^{t},u^{r},0,u^{\phi})$ yields:
\begin{equation}\label{eq:qmwdefn}
\begin{split}
Q_{\mathrm{MW}} = - \frac{\sqrt{g_{\theta \theta}}}{\alpha \sqrt{\gamma}}
\int_{\mathrm{disc}} [S^{t r} \partial_{r} (\alpha \sqrt{\gamma} u_{t}) \\
+ S^{rr} \partial_{r} (\alpha \sqrt{\gamma} u_{r}) 
+ S^{\phi r} \partial_{r} (\alpha \sqrt{\gamma} u_{\phi})] d\theta 
\end{split}
\end{equation}
We then set $S^{\mu \nu} = T^{\mu\nu}_{\mathrm{(EM)}}$ and calculate 
both this and the derivatives of $u_{\mu}$ directly from the simulation 
data.  This procedure is similar to that used by \cite{Armitage:2003} in 
their study of the observational implications of a pseudo-Newtonian 
cylindrical disc simulation.  In their procedure they set $S_{(r) 
(\phi)}$ equal to $B_{r} B_{\phi}/4\pi$ as calculated from the 
simulation.

Thus, this procedure generalizes the AK method in several ways.  It
allows for non-zero $u_r$ (but not non-zero $u_\theta$), and it permits
a smooth extension of the estimated dissipation across the ISCO.
However, it does so at the cost of some internal inconsistency
because $T^{\mu\nu}_{\mathrm{(EM)}}$ is not orthogonal to $u_\mu$,
a condition required by the model equations.  Fortunately, this
inconsistency may be of limited magnitude in the region of greatest
interest, the marginally stable region.  As shown by \cite{Krolik:2005},
$||b||^2 u^r u_\phi$ integrated over the volume occupied by bound material
is in most cases at least an order of magnitude less than $-b^r b_\phi$
integrated over the same region, and only deep inside the plunging region
or, at very high spin ($a/M = 0.998$) inside $r \simeq 2M$, does
the advected magnetic energy become close to being competitive.  Nowhere
does it exceed the conventional magnetic stress.

It should be noted, however, that by excluding those regions containing
bound matter but outside one density scale height from the plane we are
excluding any dissipation that might contribute to ``coronal" emission.
We do so in order to focus attention on the radiation coming from the
disc body, the component most likely to be thermalized.  Nonetheless,
it is possible that the ``coronal" portion we neglect here could be
significant in the total energy budget of the system.

In the simulations, only the portion of the flow lying inside of $r=20M$
is accreting in a quasi-steady state manner; outside this point matter
flows out as it absorbs angular momentum transported outward from the
inner disc.  For this reason, it is inappropriate to include regions
with $r\ge20M$ in the computation of $Q_{\mathrm MW}$.  However, standard
disc models dissipate 20--$60\%$ (for $a/M=0.998$--$0.0$) of the total
heat in the region $r>20M$; this is clearly non-negligible.  We assume,
therefore, that the total energy dissipated in the simulated disc outside
of $r=20M$ matches that specified by $Q_{\mathrm{AK}}$. 

\begin{figure}
\begin{center}
\includegraphics[width=\columnwidth]{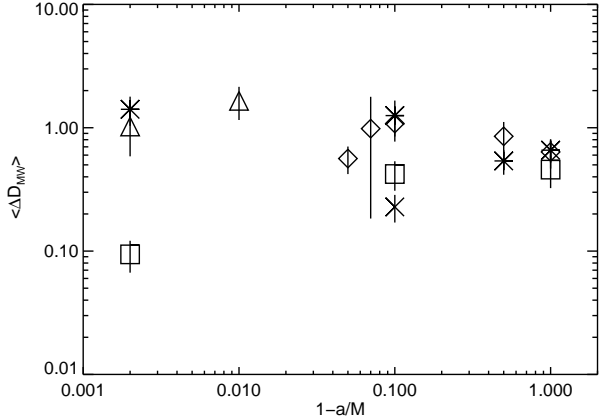}
\end{center}
\caption[]{As in Figure \ref{eta} for the time-averaged fractional 
increase in dissipation rate, $\langle \Delta D_{\mathrm{MW}} \rangle$ 
derived from $Q_{\mathrm{MW}}$.}
\label{eff} 
\end{figure}

Examination of the radial profile of $Q_{\mathrm{MW}}$ reveals that in 
the disc body it is often offset by a small amount, sometimes positive, 
sometimes negative, relative to $Q_{\mathrm{AK}}$.  We therefore 
renormalise $Q_{\mathrm{MW}}$ so that it exactly matches 
$Q_{\mathrm{AK}}$ at $r=10M$ in each case.  The shift is generally only 
a few to ten percent, and it ensures that both models predict the 
nearly the
same dissipation in the outer disc.  The error it induces in the 
dissipation rate in the inner accretion flow is small compared to the 
other uncertainties of our method.  Also note that this re-normalization 
is irrelevant to the location of the dissipation edge. For this 
quantity, all that is important is that the dissipation profile has a 
continuous extension beyond the $r=20M$ surface. Outside this radius, 
$Q_{\mathrm{AK}}$ follows an approximately $r^{-3}$ scaling, similar to 
that of $Q_{\mathrm{MW}}$ inside of $r=20M$ and as such is an 
appropriate choice for locating the radiation edge.

Figures~\ref{spindiss} and \ref{topdiss} show the radial profile
of $Q_{\mathrm{MW}}$ compared to $Q_{\mathrm{NT}}$ and $Q_{\mathrm{AK}}$
for (respectively) a variety of black hole spins and magnetic field
topologies.  Outside of the ISCO, $Q_{\mathrm{MW}}$, $Q_{\mathrm{NT}}$,
and $Q_{\mathrm{AK}}$ are all very similar.  However, as the ISCO is
approached from the outside, although $Q_{\mathrm{MW}}$ and $Q_{\mathrm{AK}}$
track each other well, both separate from $Q_{\mathrm{NT}}$ because
the artificial boundary condition imposed in the NT model forces the
stress to die there, whereas the simulations show that it in fact
continues to rise inward (as also seen in the data presented in
\cite{Krolik:2005} and \cite{Beckwith:2008a}).  Inside the ISCO,
only $Q_{\mathrm{MW}}$ is defined; in all cases, it rises steeply
with diminishing radius throughout the plunging region.

\begin{figure}
\begin{center}
\includegraphics[width=\columnwidth]{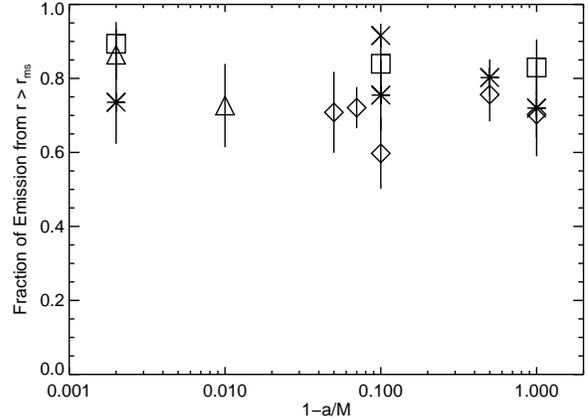}
\end{center}
\caption[]{The fraction of the dissipation $Q_{\mathrm{MW}}$ that
occurs outside the ISCO, $r>r_{ms}$, as derived
from time-averaged simulation data.  The symbols have the same
meaning as in Figure~\ref{eta}.}
\label{frms} 
\end{figure}

There is generally a close correspondence between $Q_{\mathrm{MW}}$
and $Q_{\mathrm{AK}}$ for all radii where both are defined.  This
is not surprising, given that both models share a similar origin
and the AK model is calibrated by the value of the stress at the ISCO
found in the same simulations used to determine $Q_{\mathrm{MW}}$.
In addition, we further constrain $Q_{\mathrm{MW}}$ to match
$Q_{\mathrm{AK}}$ at $r=10M$.
There are some notable differences in the curves, however.  In the high-spin KDE
model, for example, $Q_{\mathrm{AK}}$ is consistently greater than
$Q_{\mathrm{MW}}$ by factors of several for almost all radii
$r_{ms}<r< 4M$.
There are several possible reasons for a lack of agreement.  First there
is the aforementioned possibility that the simulations are not in a true
steady state.  Indeed, in most cases, even where the $Q_{\mathrm{AK}}$
curve differs noticeably from $Q_{\mathrm{MW}}$, it is nonetheless within
the lines indicating the one standard deviation fluctuations in the simulation
data.  The lack of radial dependence in the time-averaged accretion rate
argues that the simulations are, however, close to steady state in an
averaged sense.  Second, there is the lack of orthogonality between $u_{\mu}$
and $T^{\mu \nu}_{(EM)}$ (as described earlier in this section). However,
comparison of $Q_{\mathrm{MW}}$ and $Q_{\mathrm{AK}}$
calculated by replacing $T^{\mu \nu}_{\mathrm{EM}}$ with $b^{\mu} b^{\nu}$
(so that $u_{\mu} S^{\mu \nu} = 0$ is satisfied) reveals similar offsets
between the different dissipation profiles.  Third,
the velocities in the simulations are not purely Keplerian.  There is
a significant net radial component that becomes larger near and inside
the ISCO.  In particular, non-zero radial velocity introduces a new
term in the definition of $q^\mu$ not present in the AK or NT models
(see eqn.~\ref{eq:qmwdefn}).
Lastly, we have identified $S^{\mu\nu}$ with the Maxwell stress,
neglecting the Reynolds stress due to correlated radial and angular
velocity fluctuations, but it can also contribute to the total stress.
Although the magnitude of this contribution to the Reynolds stress is
difficult to determine within a global simulation, local shearing box 
simulations have shown that it is typically near or below
one third the Maxwell stress in the main disc body \cite[][]{Hawley:1995}.

\newpage

Figure~\ref{eff} shows the fractional increase in dissipation rate, $\langle
\Delta D_{\mathrm{MW}} \rangle$ derived from the Maxwell stress, using
a definition analogous to eqn.~\ref{fracin}, where $D_{\mathrm{MW}}
=  \int^{\infty}_{r_{in}} \int \, Q_{\mathrm{MW}} dx^{(t)}
dx^{(r)} dx^{(\phi)}$, and $dx^{(\mu)}$ are the coordinate elements
measured by an observer comoving with the fluid with $u^{\alpha} =
(u^{t},u^{r},0,u^{\phi})$.  We find values of $\langle \Delta
D_{\mathrm{MW}} \rangle$ from 10 to $200\%$, with most in the range
50--$100\%$.  In contrast with $\langle \Delta
D_{\mathrm{AK}} \rangle$, there is little dependence on black hole spin.
For topologies initialised with non-dipolar magnetic fields,
$\langle \Delta D_{\mathrm{MW}} \rangle$ is smaller than for those
with initially dipolar field.

In both the AK and the NT disc models all of the
dissipation in the disc occurs at or outside of the ISCO.  In the
simulations, however, the plunging region makes a contribution to the
total dissipation, as can clearly be seen from Figures \ref{spindiss}
and \ref{topdiss}.  Figure \ref{frms} shows the fraction of the total
dissipation for each of the simulations that comes from outside the ISCO.
The disc above the ISCO contributes from 60--90\% of the total energy
dissipation.  Comparing this figure with Figure \ref{eff} shows that, in
general, the greater the fractional increase in total dissipation, the
more significant the plunging region is compared to the rest of the disc.
As a group the dipole simulations tend to have larger stress in the
plunging region.

\begin{figure}
\begin{center}
\includegraphics[width=\columnwidth]{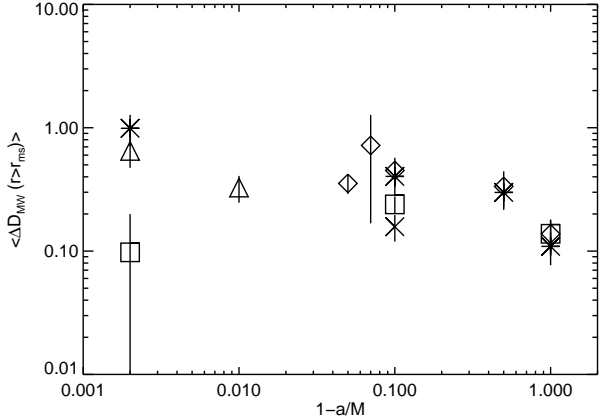}
\end{center}
\caption[]{The time-averaged fractional increase in dissipation
derived from $Q_{\mathrm{MW}}$ where the region under consideration
is limited to that lying outside the ISCO, $r>r{ms}$.  The symbols are
as defined in Figure~\ref{eta}.}
\label{fzeff_nrms} 
\end{figure}

In Figure \ref{fzeff_nrms} we return to the quantity $\langle \Delta
D_{\mathrm{MW}} \rangle$, but now include only that portion of the disc
outside of the ISCO.  Removing the contribution from the plunging region
reduces $\langle \Delta D_{\mathrm{MW}} \rangle$ to be on the order
of tens of percent.  There is perhaps a slight trend towards
larger numbers with larger spin.  Removal of the contribution from the
plunging region reduces the contrast between the different magnetic
field topologies.  This is mainly because in the dipole simulations
the profile $Q_{\mathrm{MW}}$ steepens significantly below the ISCO
(as can be seen from Figure \ref{topdiss}).

\newpage

\subsection{The Dissipation Edge}\label{diss}

\begin{figure}
\begin{center}
\includegraphics[width=\columnwidth]{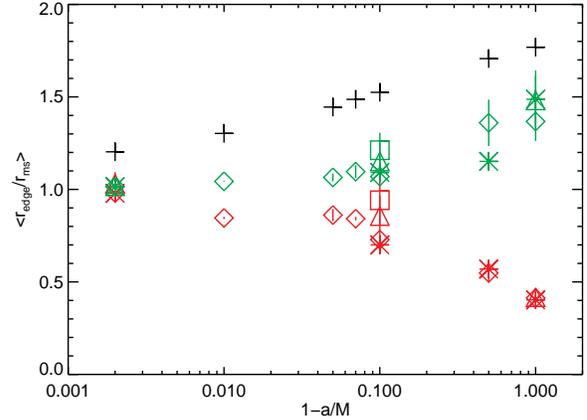}
\end{center}
\caption[]{The location of the dissipation edge in relation to the
location of the ISCO, $r_{ms}$, derived 
from $Q_{\mathrm{NT}}$ (black crosses), $Q_{\mathrm{AK}}$ (green symbols) 
and $Q_{\mathrm{MW}}$ (red symbols).  The symbols correspond to
different simulations as given in Figure \ref{eta}.}
\label{dissedge} 
\end{figure}

A way to characterize the relative distribution of $Q$ in all three
of the dissipation models is to compute the ``dissipation edge''
for each model, defined to be the radius outside of which $95\%$
of the total energy dissipated within the disc has been deposited.
Figure \ref{dissedge} displays the location of the dissipation edge for
$Q_{\mathrm{NT}}$, $Q_{\mathrm{AK}}$ and $Q_{\mathrm{MW}}$ in terms
of the radius of the ISCO, $r_{ms}$.  The dissipation edge for the NT
model lies between 1.2 (for $a/M=0.998$) and $1.8r_{ms}$ (for $a/M=0.0$).
Since the stress goes to zero at the ISCO in this model and reaches a
maximum value at
some location well outside of this point,  the majority of the dissipation
also occurs well outside the ISCO.  Consequently, radiation from these discs
does not probe (except by photon propagation) the deepest regions of
the black hole's gravitational potential.

A non-zero stress at the ISCO changes the picture.  In Figure~\ref{dissedge}
we see that the dissipation edge moves closer to the ISCO than in the standard
model, although in the AK model it obviously must
remain at a location $> r_{ms}$.  The AK dissipation edge is located between
$\sim 1.0 r_{ms}$ for $a/M=0.998$ and $\sim 1.5r_{ms}$ for $a/M=0.0$.
As would be expected from Figure \ref{eta}, changes in magnetic field
topology have little influence on the location of the dissipation edge
derived from $Q_{\mathrm{AK}}$.

Finally, we turn to the dissipation edge derived from $Q_{\mathrm{MW}}$. 
With stress permitted inside of the ISCO, the dissipation edge 
consistently lies either at or inside $r_{ms}$.  There is a visible 
dependence on black hole spin.  The dissipation edge ranges between 
$\simeq 0.4$ and $1.0r_{ms}$ as $a/M$ increases from 0.0 to 0.998.  
There is little apparent dependence on magnetic field topology.  Some of 
the dependence on spin may be an artifact of the simulations.  The GRMHD 
code uses Boyer-Lindquist coordinates, and there are fewer zones 
inside of $r_{ms}$ for high values of $a/M$.  As we shall see in the 
next section, however, even when the dissipation edge is deep inside the 
plunging region, the radiation edge is generally well outside, as few 
photons can escape from so near the horizon when the black hole spin is 
this rapid.

\section{Observed Quantities}\label{rad}

\begin{figure*}
\begin{center}
\includegraphics[width=0.32\textwidth]{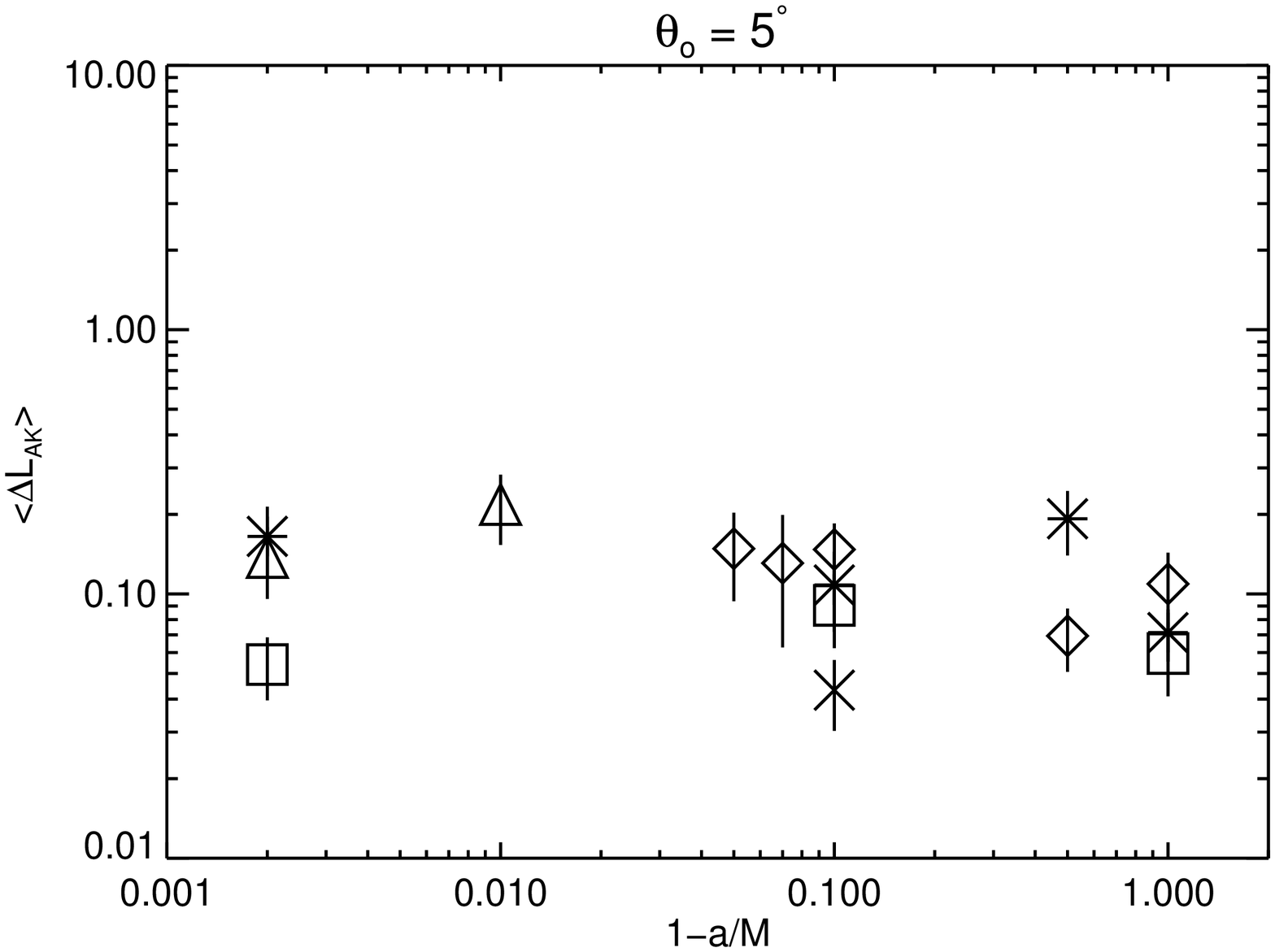}
\includegraphics[width=0.32\textwidth]{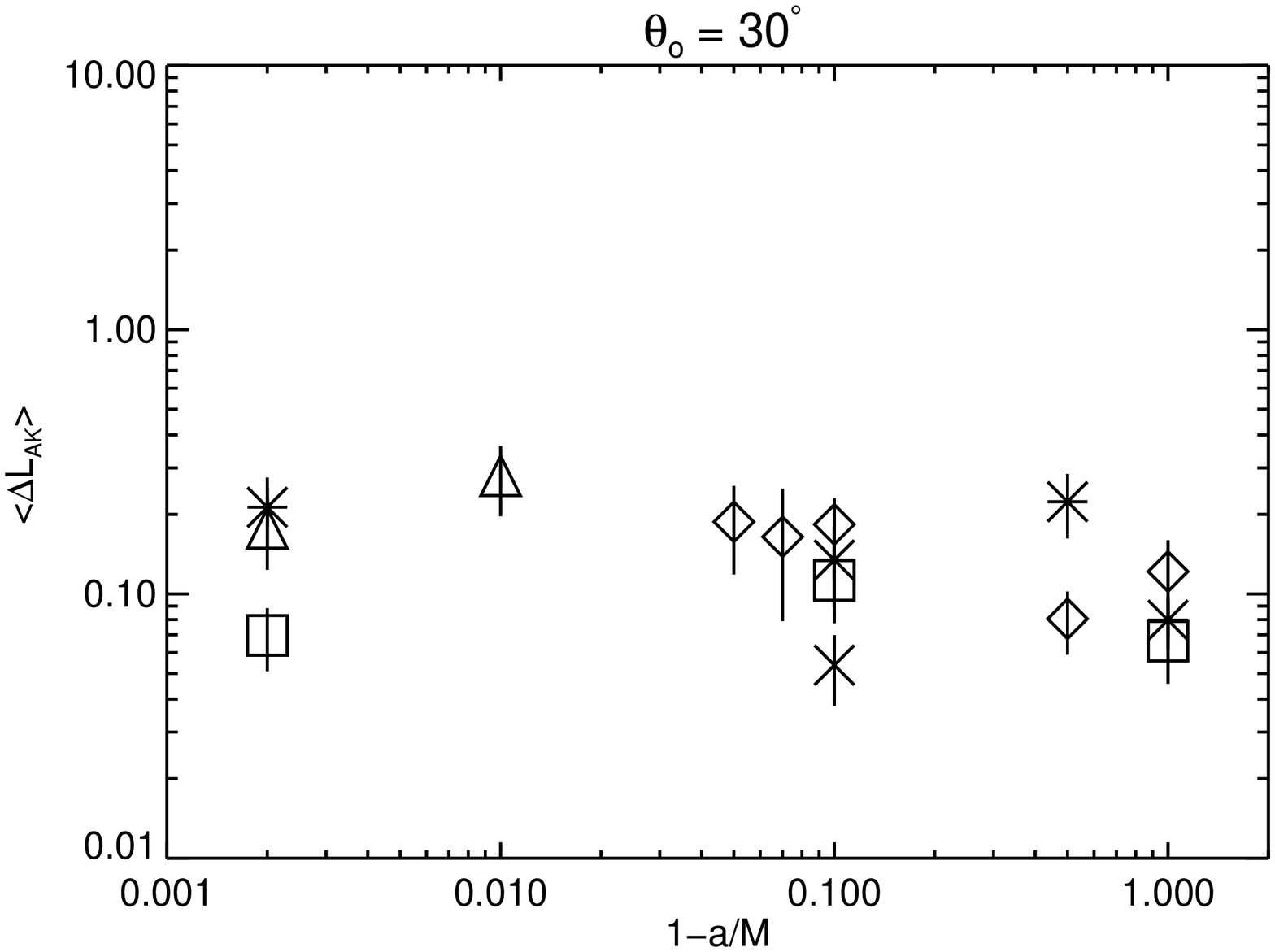}
\includegraphics[width=0.32\textwidth]{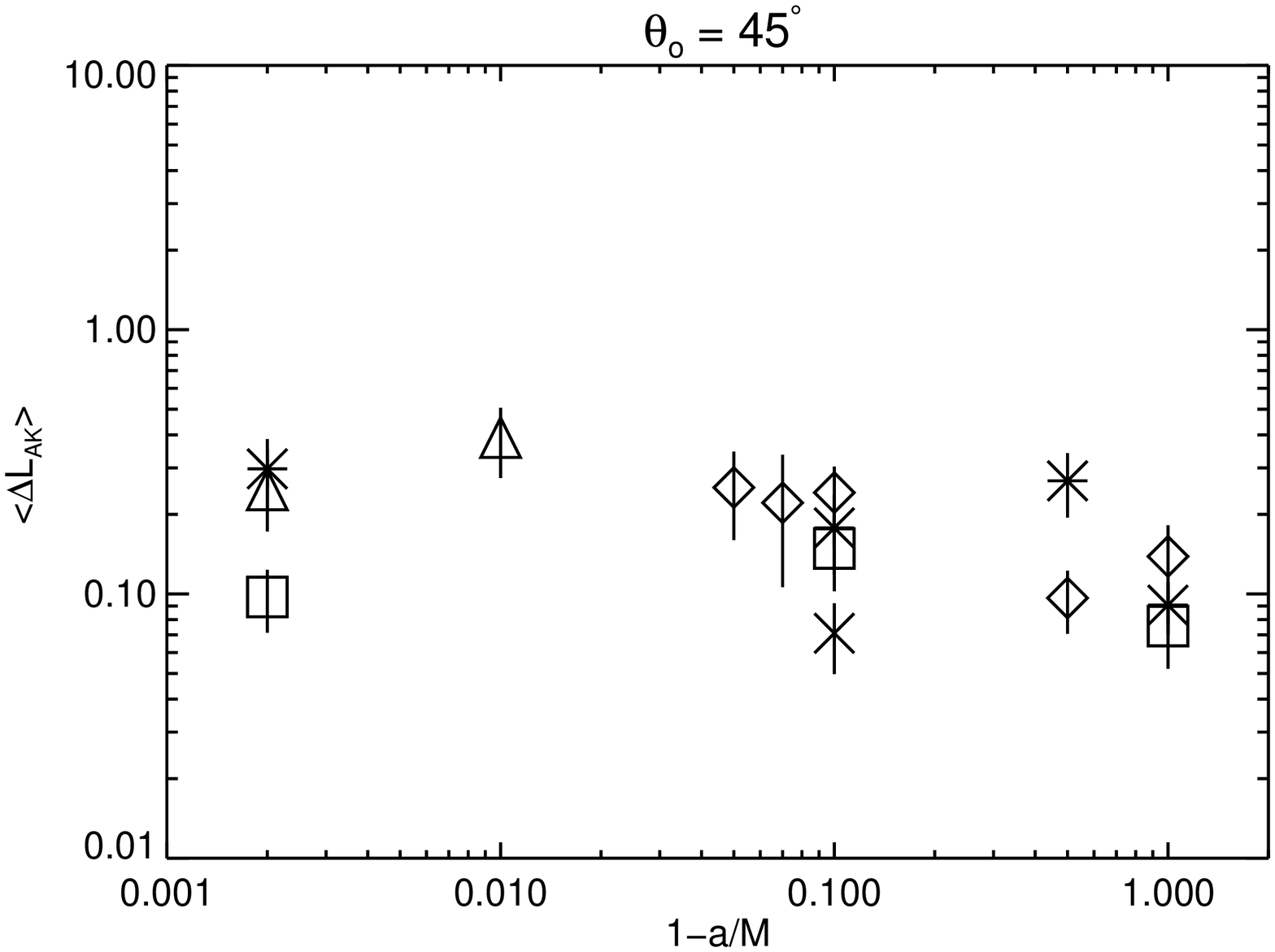}
\includegraphics[width=0.32\textwidth]{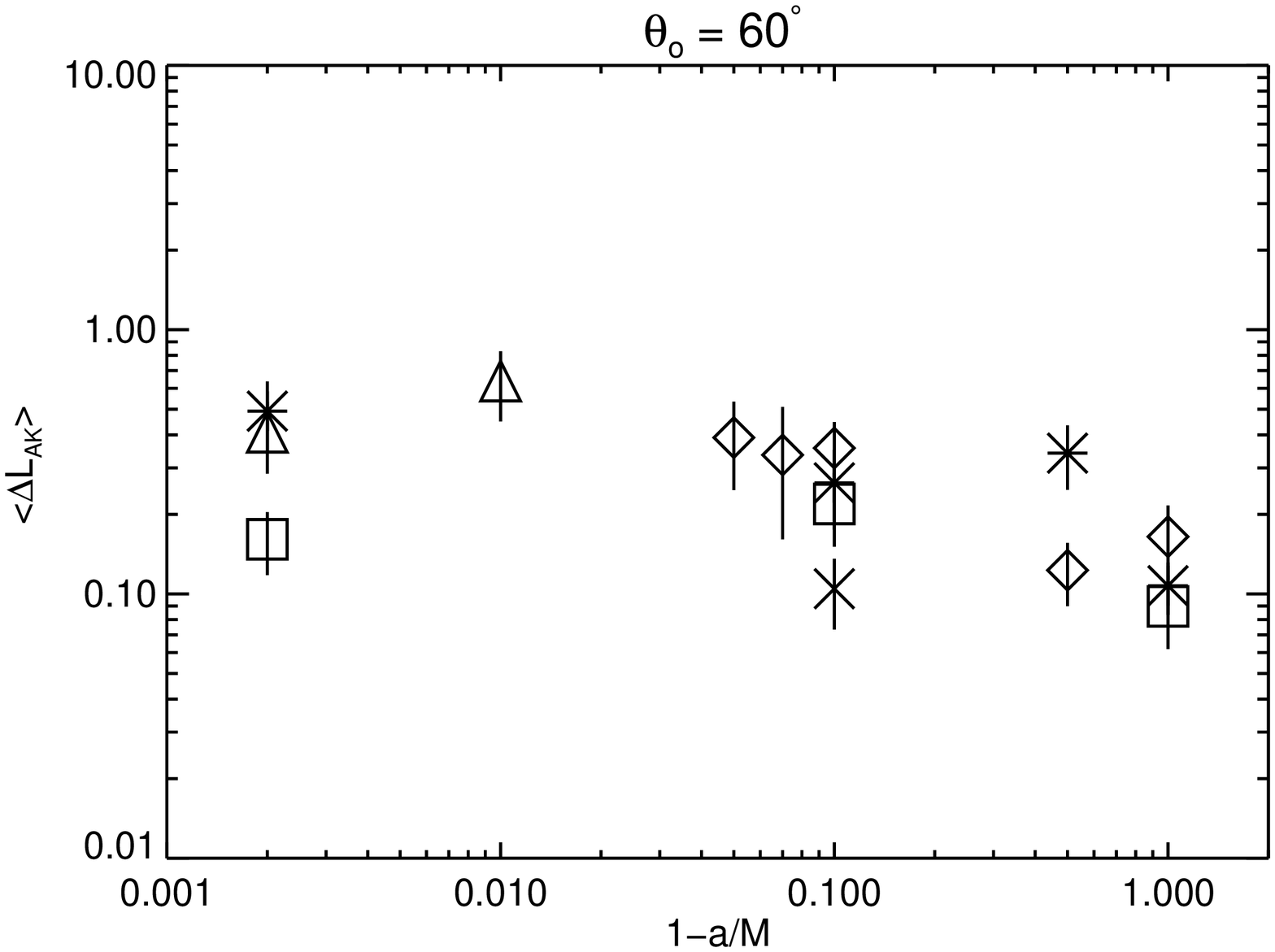}
\includegraphics[width=0.32\textwidth]{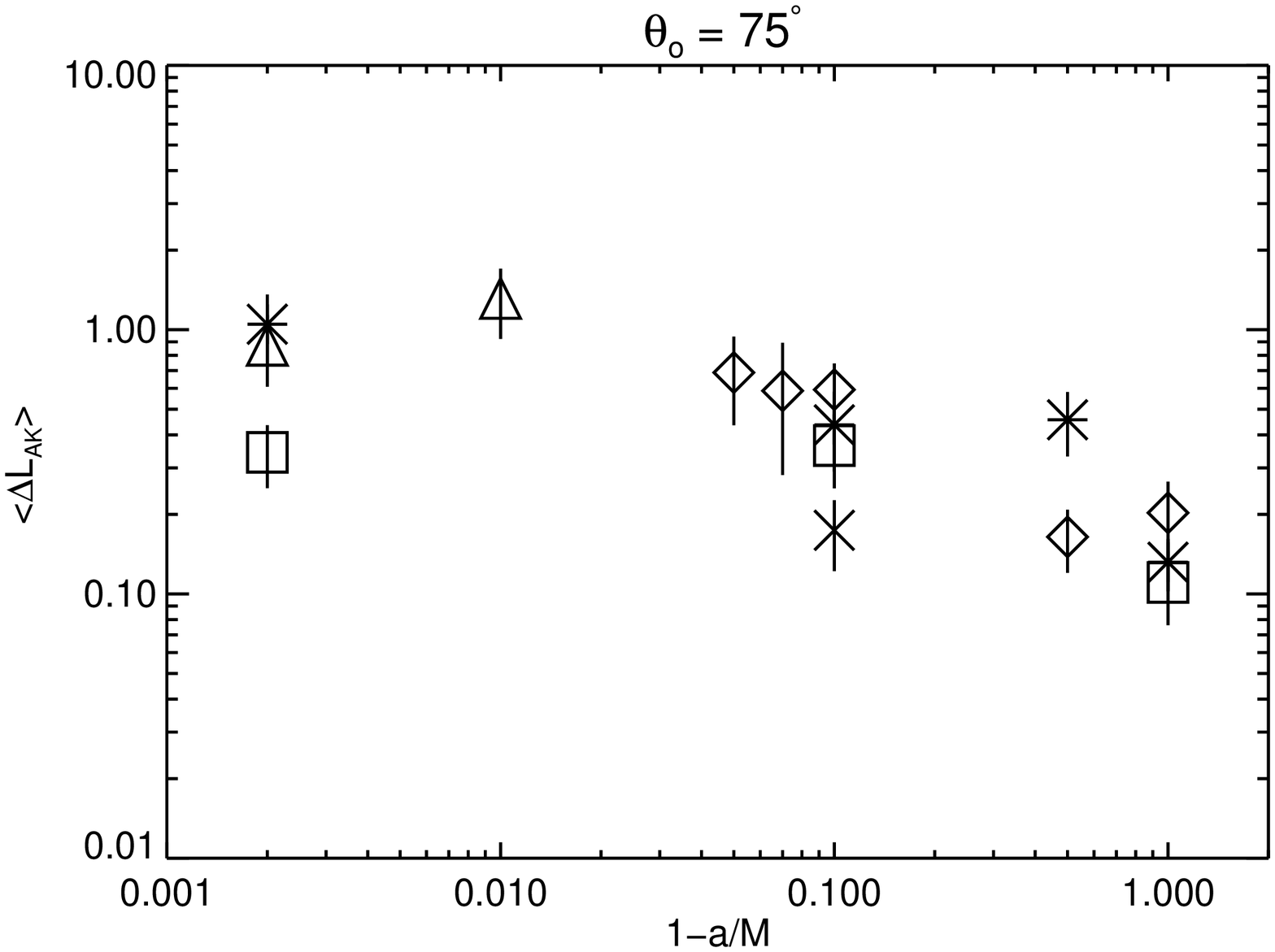}
\includegraphics[width=0.32\textwidth]{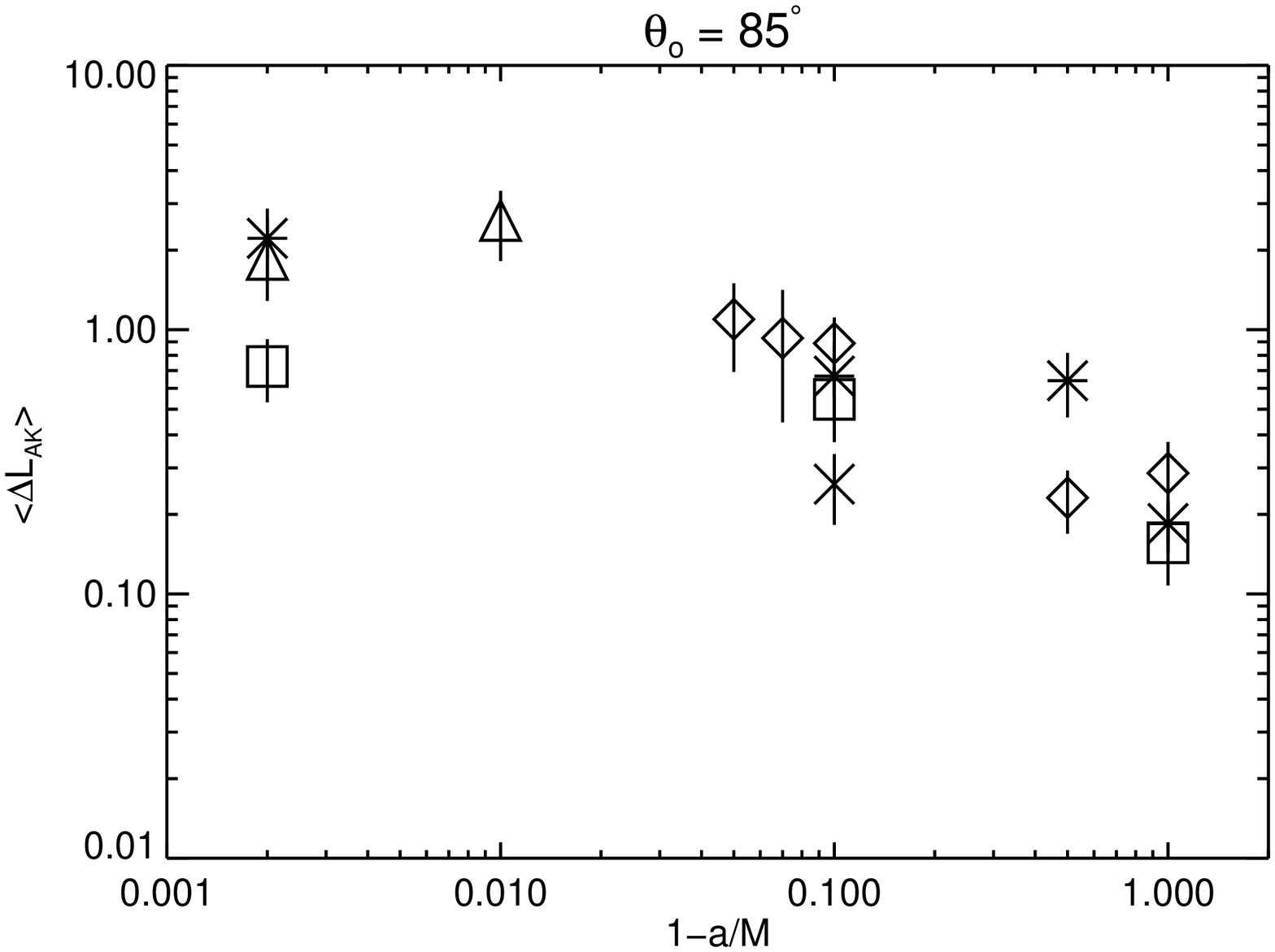}
\end{center}
\caption[]{The time-averaged fractional
increase in luminosity, $\langle \Delta L_{\mathrm{AK}} \rangle$
derived from $Q_{\mathrm{AK}}$ for a distant observer located at
$\theta_{o}=5^{\circ}$ (top left panel), $30^{\circ}$ (top centre
panel), $45^{\circ}$ (top centre panel), $60^{\circ}$ (bottom left
panel), $75^{\circ}$ (bottom centre panel) and $85^{\circ}$ (bottom
right panel).  Symbols correspond to different simulations as
defined in Figure~\ref{eta}.}
\label{aklum} 
\end{figure*}

In the preceding sections, we have seen how stress
near and inside the ISCO can significantly increase the total dissipation in the
accretion flow and alter the location of the dissipation edge.  In the
NT model, all but $5\%$ of
the total energy is dissipated outside of 1.2--1.8$r_{ms}$.  By allowing
for non-zero stress at the ISCO, the AK model moves the radial coordinate
of the dissipation edge
inward by factors of $\sim0.8$ compared to the NT model. Relaxing the
assumption that dissipation terminates at the ISCO causes the dissipation
edge to move inside the ISCO.  From an observational standpoint, these
distinctions can be important.  They affect the degree to which a given
accretion flow can probe a black hole's gravitational potential, and
also determine such observables as the peak in the disc thermal spectrum
and the width of the Fe~K$\alpha$ line profile.

While the fluid frame dissipation is important, the real test is whether
these differences carry over into the rest frame of a distant
observer.  In this section we examine this issue using the photon
transfer function, which incorporates effects due to gravitational
and orbital Doppler shifts, along with the influence of light-bending and
gravitational lensing.

We will examine the properties of these discs in two ways: first, when
observed from different inclination angles, and second by averaging 
over solid angle to obtain a total value.
The solid angle averaged value of some observed quantity $m_o$ is defined as
\begin{equation} 
\{ M \} = \frac{\int{ \int{ m_{o}
(\theta_{o} ) \sin \theta_{o} d\theta_{o}} d\phi}}{\int{ \int{ \sin \theta_{o}
d\theta_{o}} d\phi}} .
\end{equation} 
The solid angle average of a given quantity is not itself observable,
and for many potential observables the dependence on viewing angle is
considerable.  Nevertheless, the angle average provides a simple way to
summarise the impact of the effects we are studying.

\subsection{Fractional Increase in Luminosity}

For a given dissipation model, we can compute the
total luminosity carried to a distant observer using
the observed flux, eqn. ~\ref{obsflux}, and integrating:
\begin{equation}
\begin{split}
L =  \int \int g^{4}[x^{\mu}_{\mathrm{surf}}(\alpha,\beta)] 
        Q[x^{\mu}_{\mathrm{surf}}(\alpha,\beta)] \times \\
\delta (E_{o} - g E_{e}) d\alpha d\beta ;
\end{split}
\end{equation}
$Q$ corresponds to
$Q_{\mathrm{AK}}$, $Q_{\mathrm{NT}}$ or $Q_{\mathrm{MW}}$ as appropriate.
We can then compute the fractional enhancement of the luminosity over
the NT model using
\begin{equation}
\langle \Delta L_{\mathrm{AK,MW}} \rangle = \langle
\frac{L_{\mathrm{AK,MW}} - L_{\mathrm{NT}}}{L_{\mathrm{NT}}} \rangle.
\end{equation}

The results for the AK model are plotted in Figure \ref{aklum}
for a distant observer at inclinations ranging from 5--$85^{\circ}$.
In \S\ref{ak2000} we found $\langle \Delta D_{\mathrm{AK}} \rangle$
was strongly correlated with black hole spin, ranging from 10--500\%
as $a/M$ approached one. Figure~\ref{aklum}
shows that the corresponding fractional increases in luminosity are
more modest, ranging between 5--$40\%$ for $\theta_{o}\le45^{\circ}$
(low inclinations) to substantially above $40\%$ for the half of solid
angle with $\theta_{o}\ge60^{\circ}$ (high inclinations).   At low
inclinations, $\langle \Delta L_{\mathrm{AK}} \rangle$ exhibits little
dependence on black hole spin. At high inclinations, $\langle \Delta
L_{\mathrm{AK}} \rangle$ exhibits a similar (if slightly weaker)
dependence on black hole spin to that seen in $\langle \Delta
D_{\mathrm{AK}} \rangle$.

\begin{figure*}
\begin{center}
\includegraphics[width=0.32\textwidth]{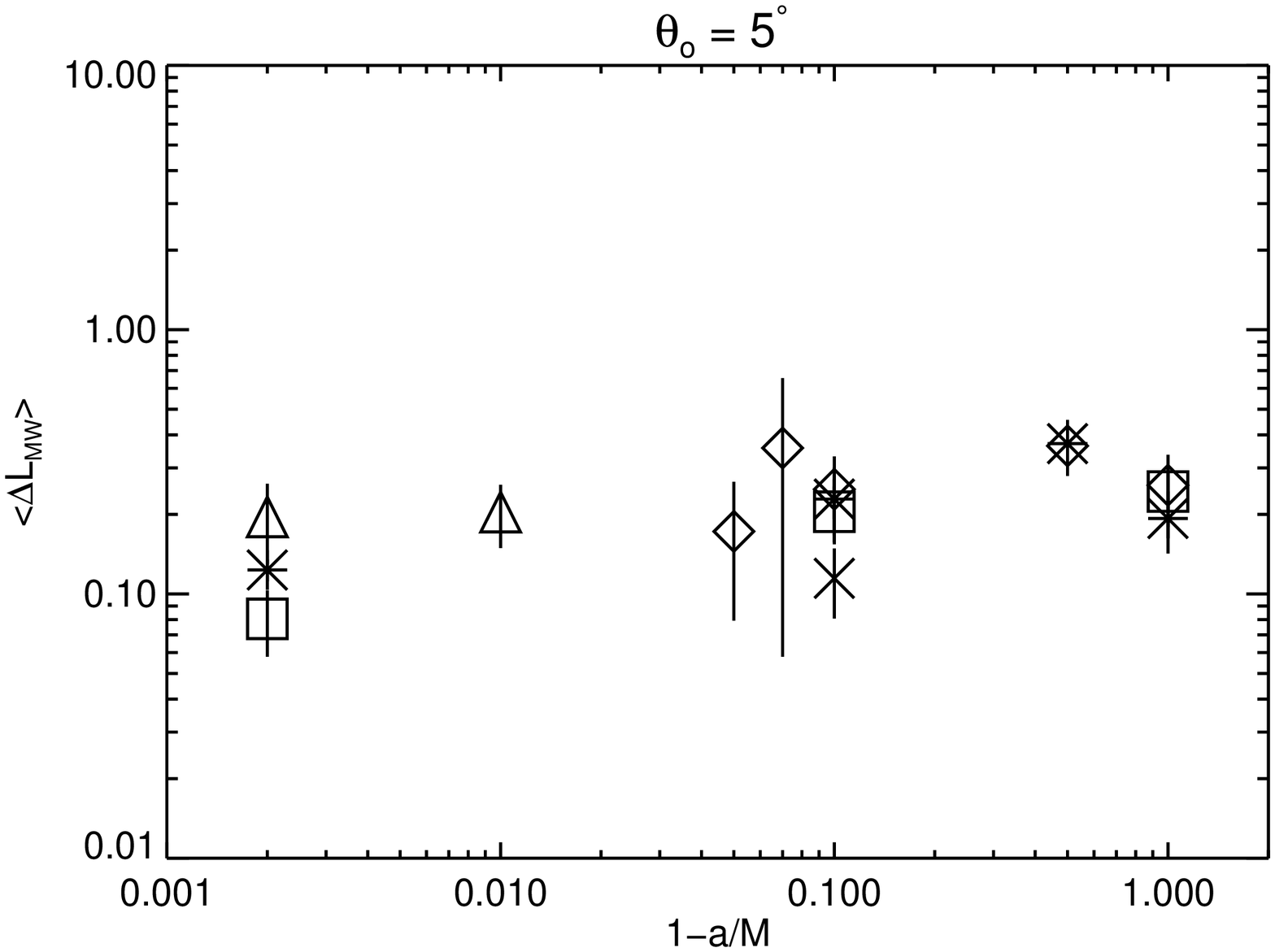}
\includegraphics[width=0.32\textwidth]{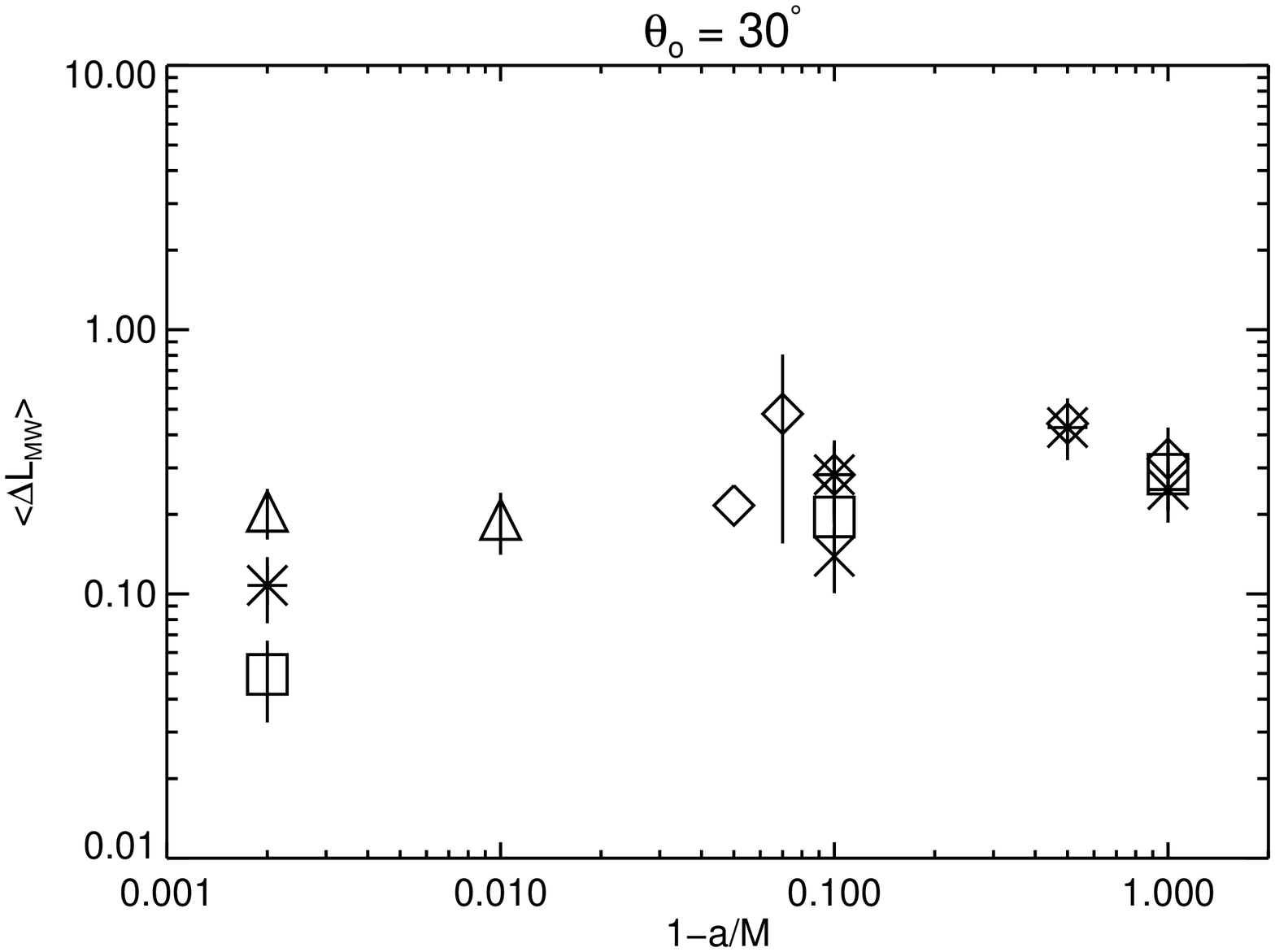}
\includegraphics[width=0.32\textwidth]{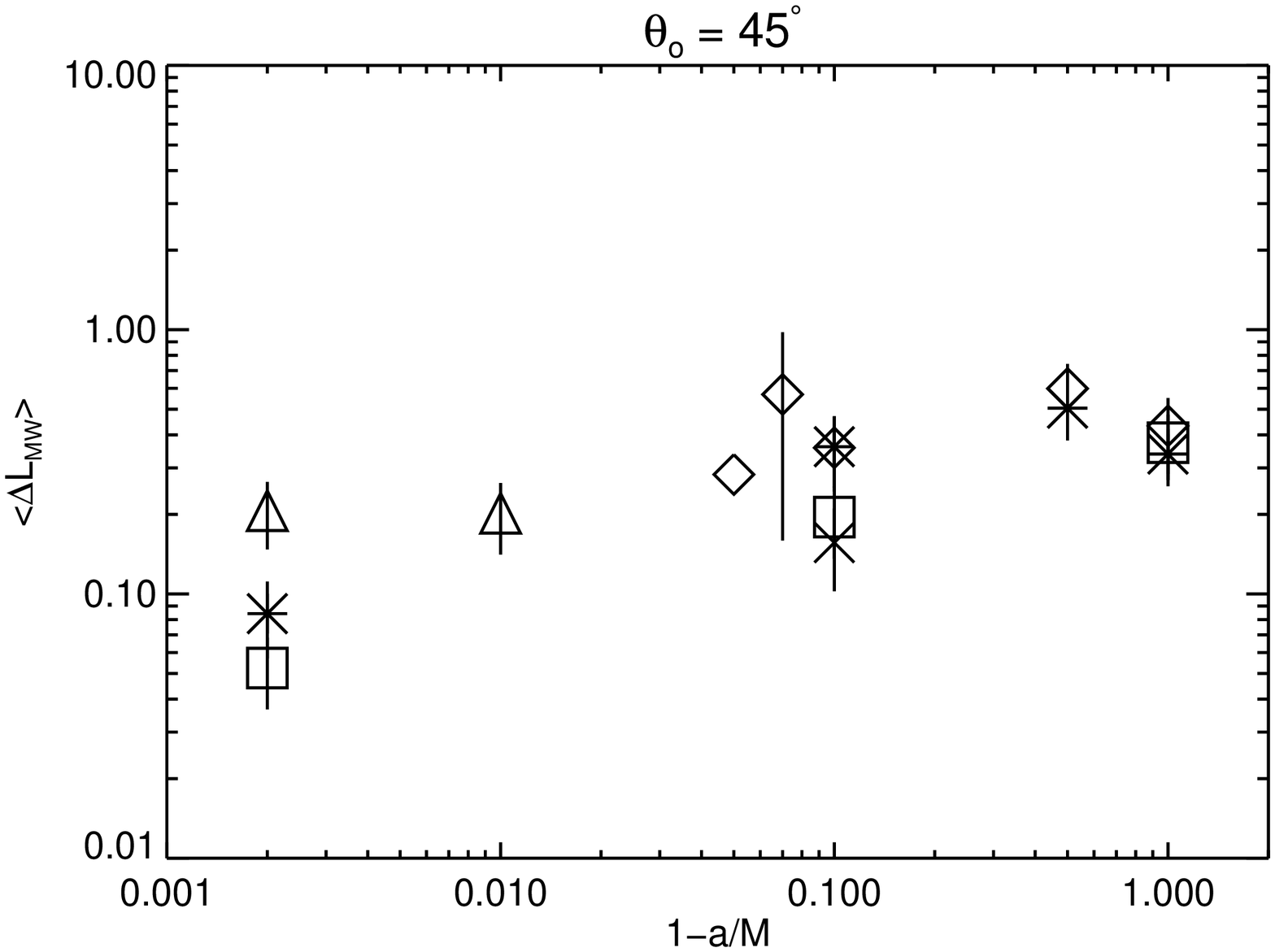}
\includegraphics[width=0.32\textwidth]{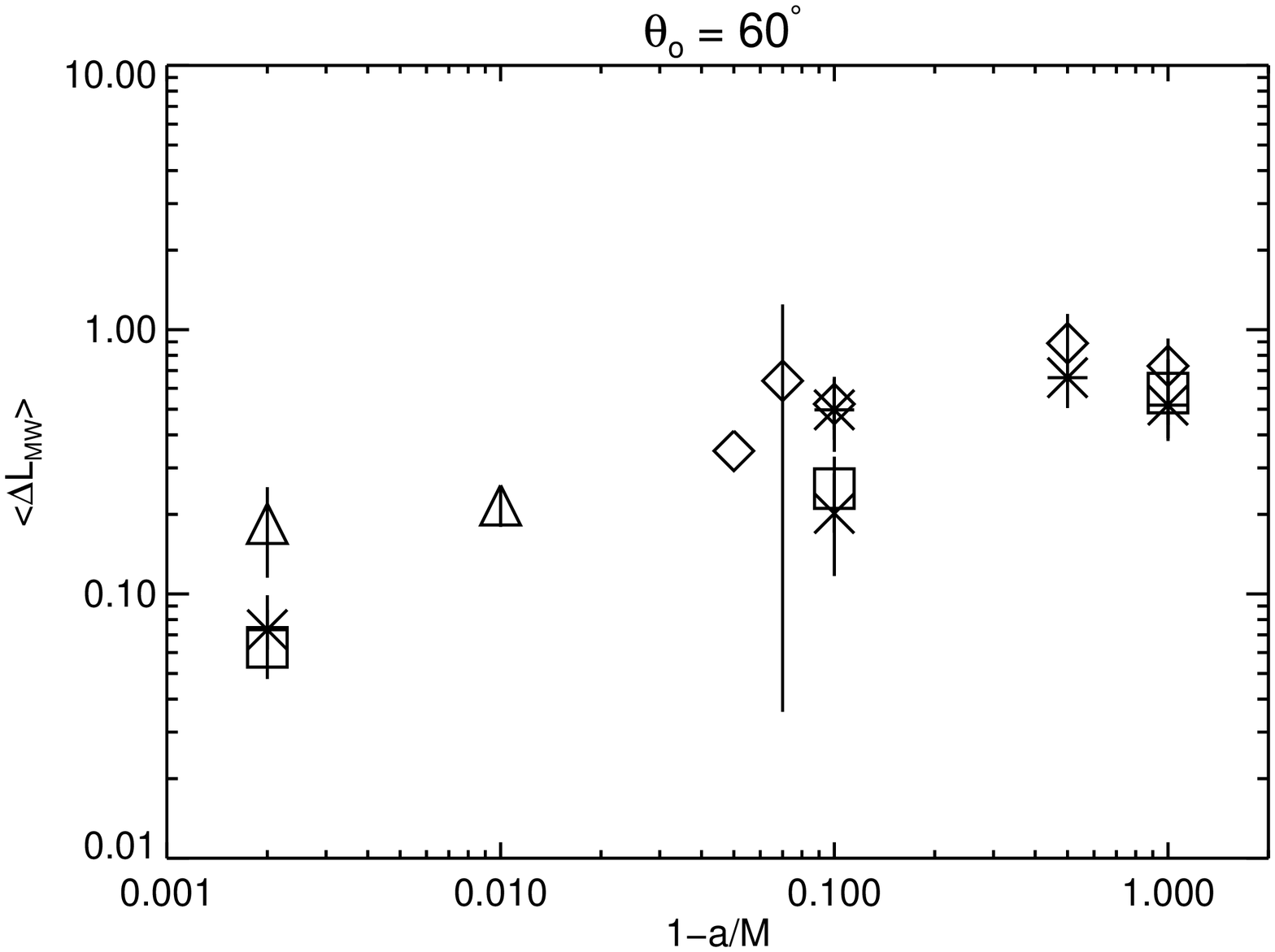}
\includegraphics[width=0.32\textwidth]{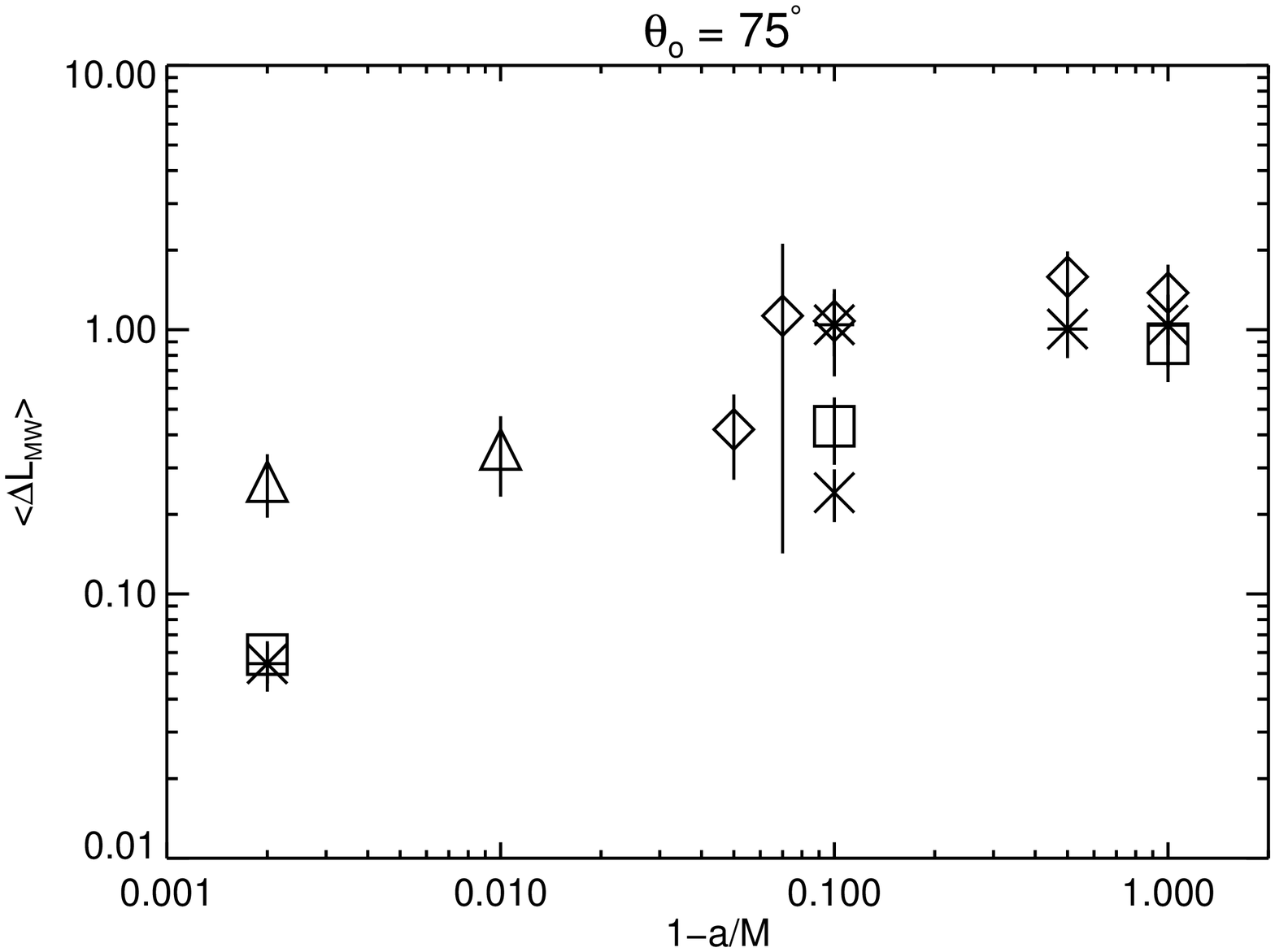}
\includegraphics[width=0.32\textwidth]{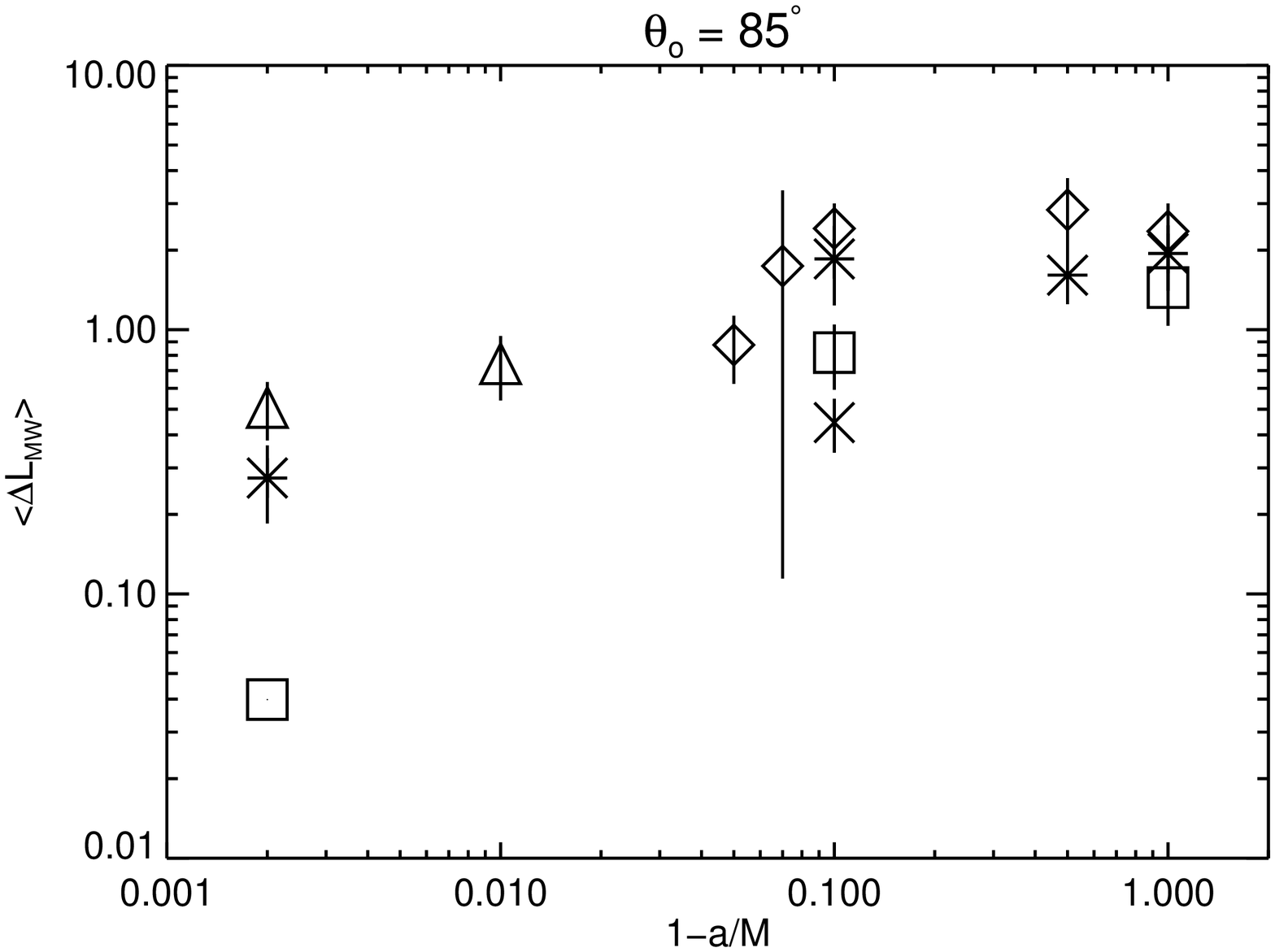}
\end{center}
\caption[]{As in Figure \ref{aklum} for the time-averaged fractional 
increase in luminosity, $\langle \Delta L_{\mathrm{MW}} \rangle$ derived 
from $Q_{\mathrm{MW}}$.}
\label{lum} 
\end{figure*}

Figure~\ref{lum} gives $\langle \Delta L_{\mathrm{MW}} \rangle$ for the 
same range of inclinations.  In \S\ref{pyntflx} (see Fig.~\ref{eff}), we 
found $\langle \Delta D_{\mathrm{MW}} \rangle$ was generally within a 
factor of 2 of $100\%$, with little consistent dependence on black hole 
spin.  Figure~\ref{lum} shows a different picture.  At all inclinations, 
$\langle \Delta L_{\mathrm{MW}}\rangle$ at the highest spin is a few 
tens of percent, varying by $\sim 2$ for the different magnetic 
topologies. At all inclinations, there is also a rise in $\langle \Delta 
L_{\mathrm{MW}}\rangle$ with diminishing spin, but the slope of this 
rise increases sharply with increasing inclination angle.  As a result, 
for angles of $75^{\circ}$ and more, $\langle \Delta L_{\mathrm{MW}} 
\rangle \gtrsim 100\%$ for all spins $a/M \lesssim 0.9$.  There is a 
weak tendency for $a/M = 0.5$ to yield the greatest fractional 
enhancement in luminosity, but it is not by a wide margin over the 
spinless case.

\begin{figure*}
\begin{center}
\includegraphics[width=0.32\textwidth]{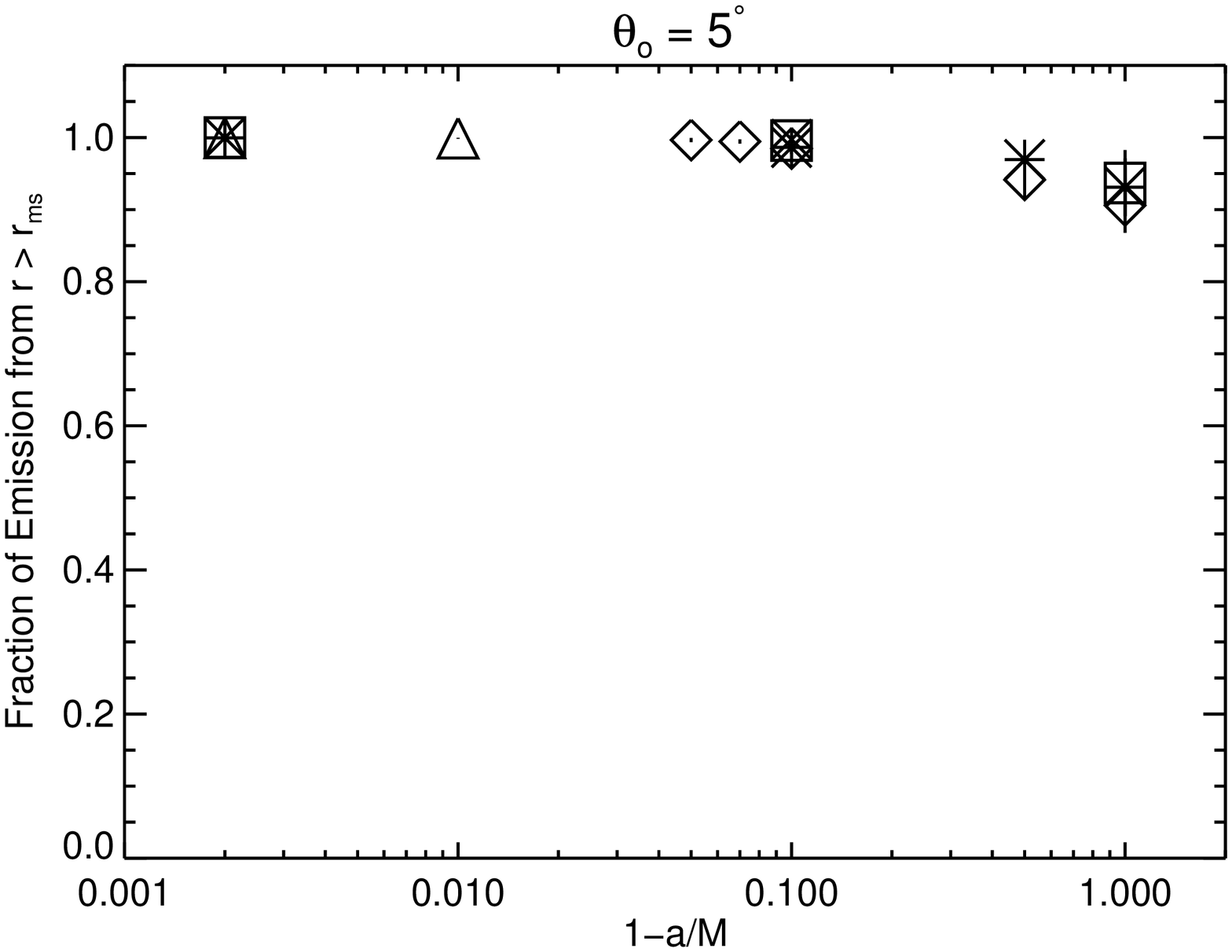}
\includegraphics[width=0.32\textwidth]{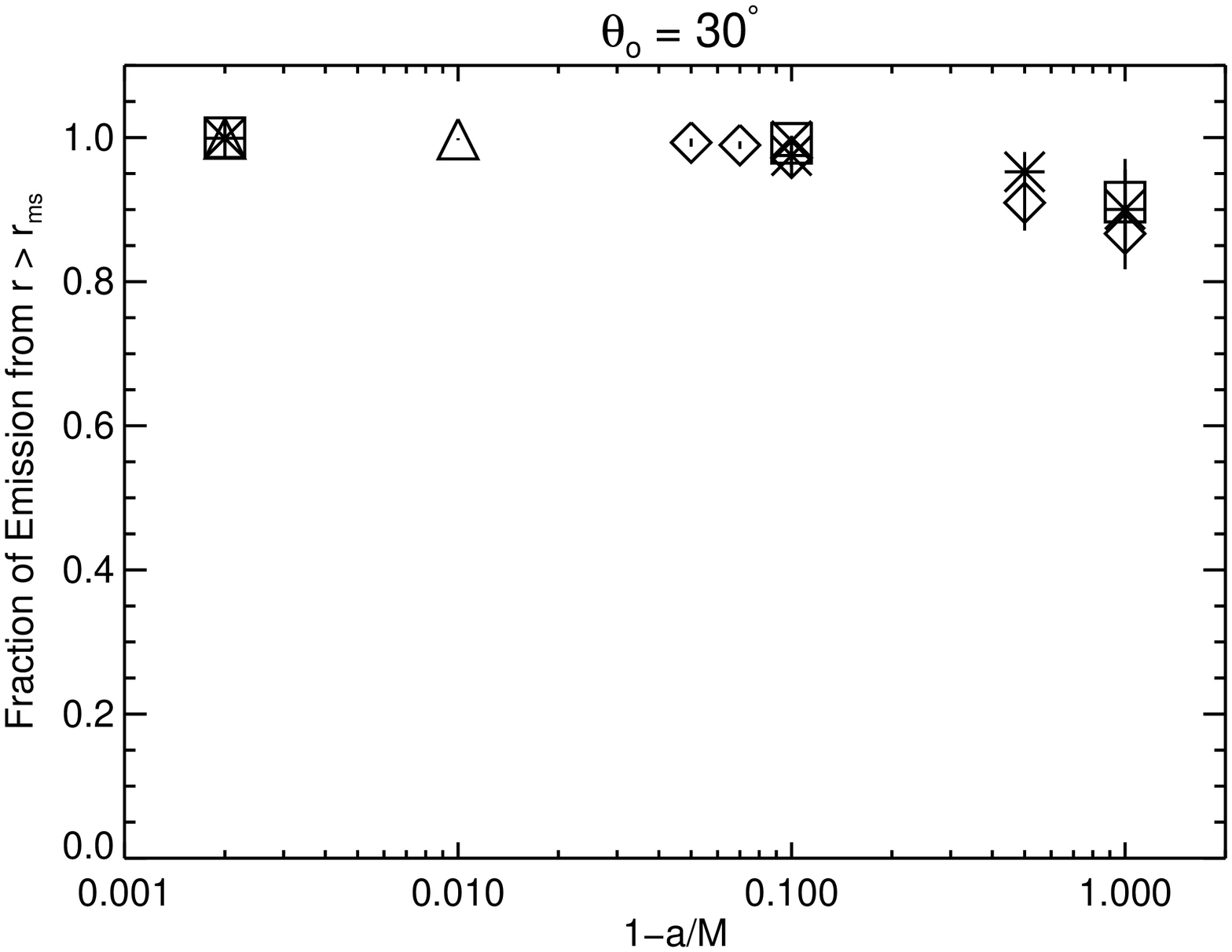}
\includegraphics[width=0.32\textwidth]{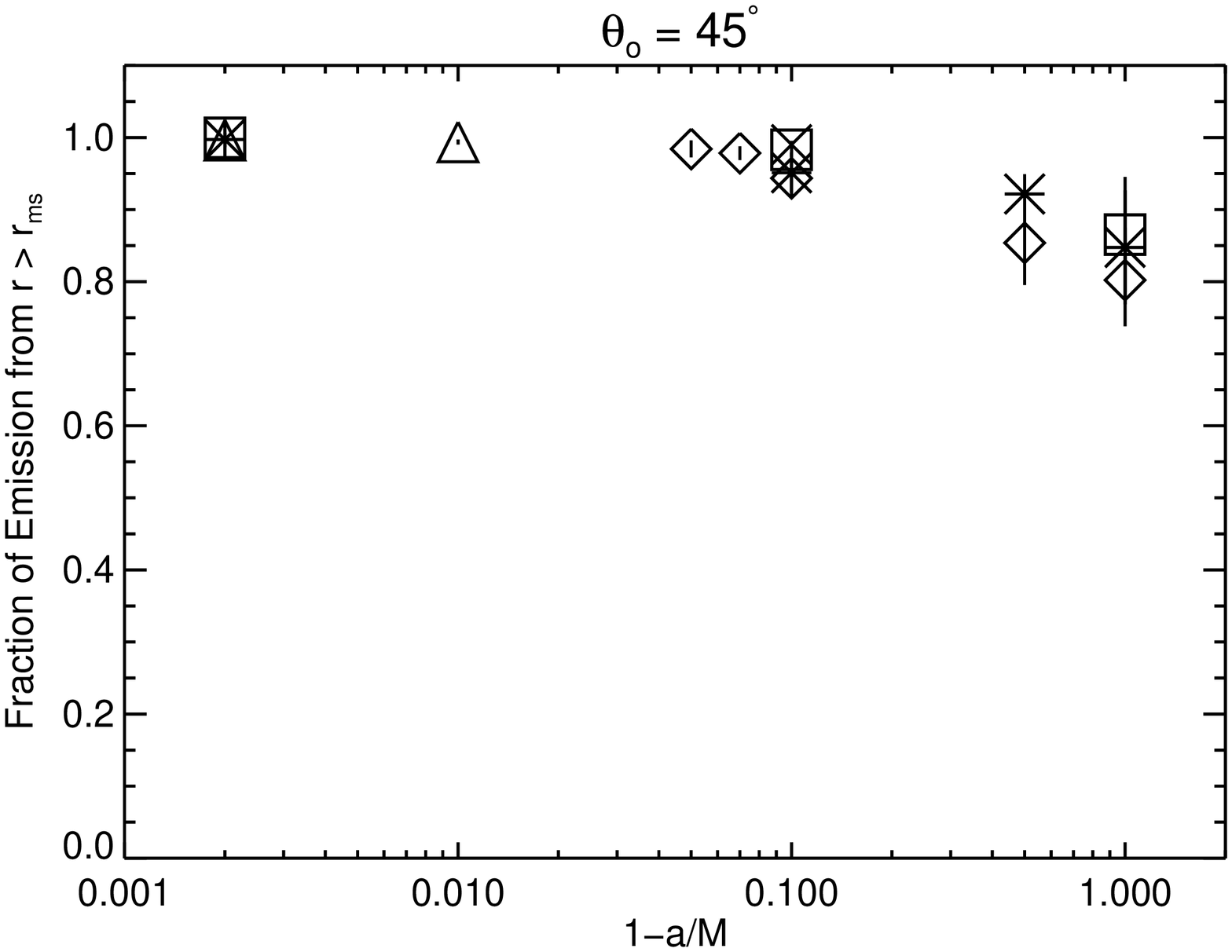}
\includegraphics[width=0.32\textwidth]{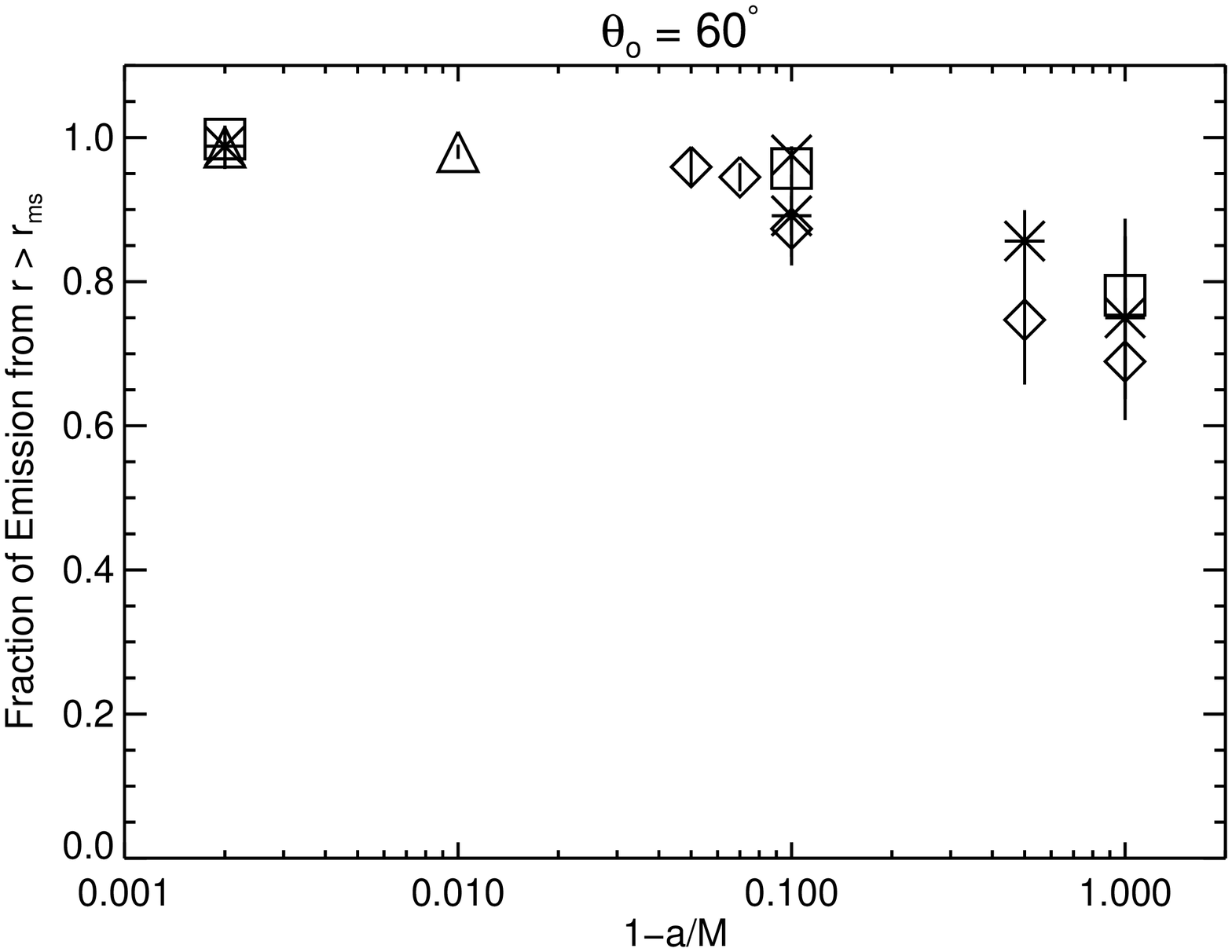}
\includegraphics[width=0.32\textwidth]{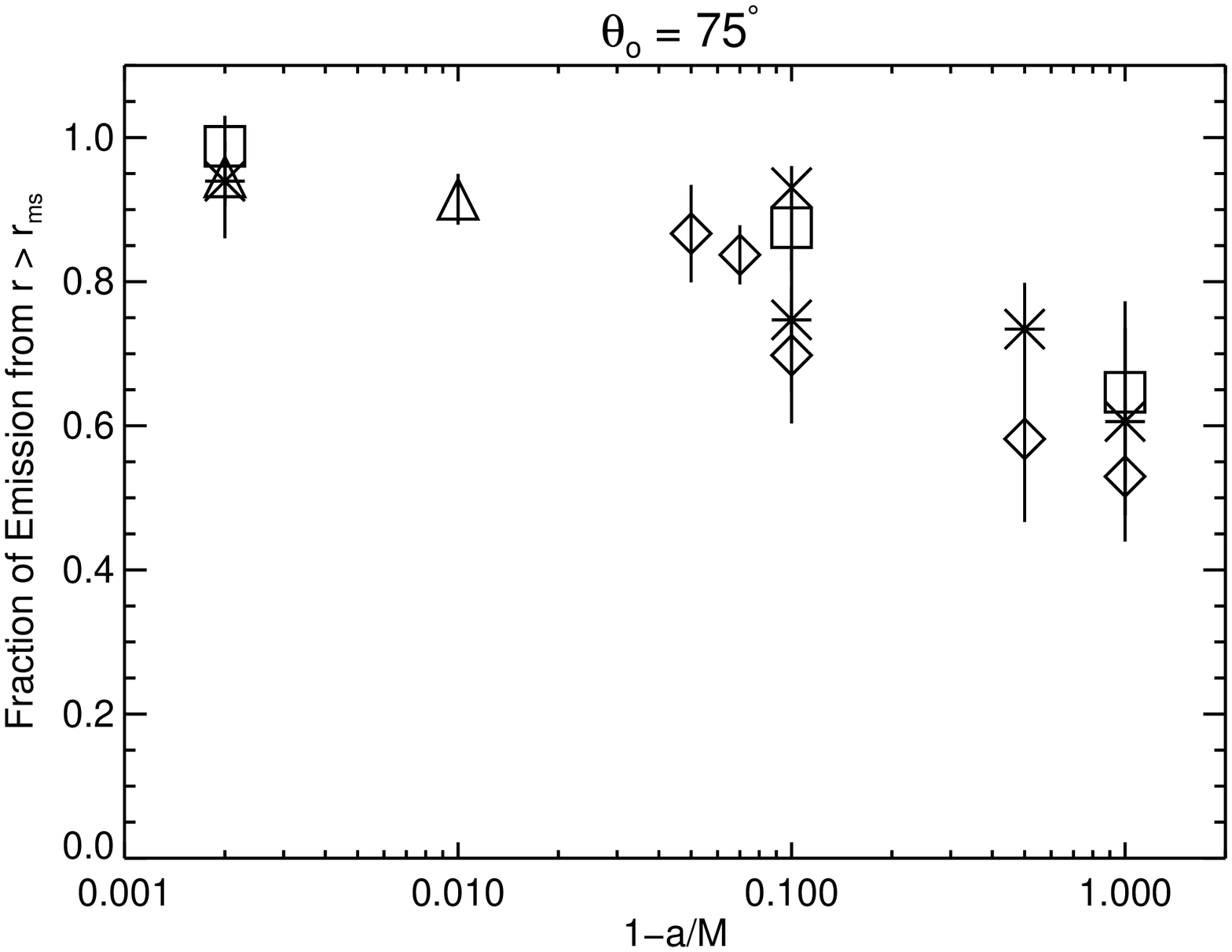}
\includegraphics[width=0.32\textwidth]{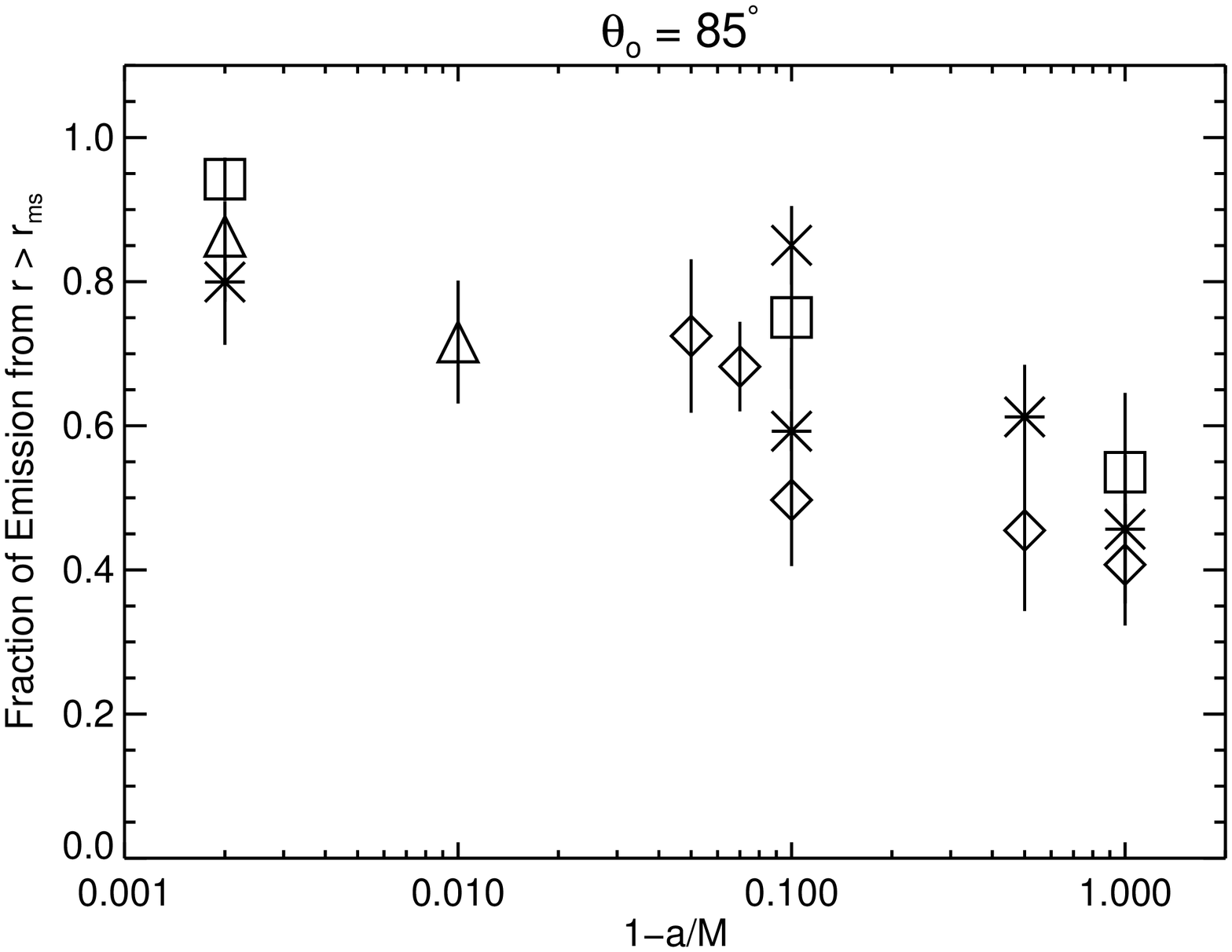}
\end{center}
\caption[]{As in Figure \ref{aklum} for the fraction of the total 
luminosity emitted outside of the ISCO derived from $Q_{\mathrm{MW}}$.}
\label{ofrms} 
\end{figure*}

In \S\ref{pyntflx} we found that the fraction of the total energy
dissipated outside the ISCO ranged from $\gtrsim60\%$ to $\gtrsim80\%$
for non-dipolar magnetic field topologies.  Figure \ref{ofrms} plots
the fraction of the total luminosity that is received from outside of
$r_{ms}$ for $Q_{\mathrm{MW}}$.  This fraction is near $100\%$ for low
inclinations and $a/M>0.9$.  As the spin of the hole is reduced below
$a/M=0.9$, the fraction falls to $\gtrsim90\%$ as more photons escape
from inside the ISCO.  At higher inclinations ($\theta_{o}\ge60^{\circ}$),
the contribution from the plunging region increases for all but the most
rapidly rotating black holes, in which $>80\%$ of the total observed
luminosity originates outside of the ISCO. Thus, the contribution of
the plunging region is greatest for low to moderate black hole spins:
at high inclination the plunging region contributes 20--$60\%$ of the
total observed luminosity of the disc for this range of spins.

\begin{figure*}
\begin{center}
\includegraphics[width=0.48\textwidth]{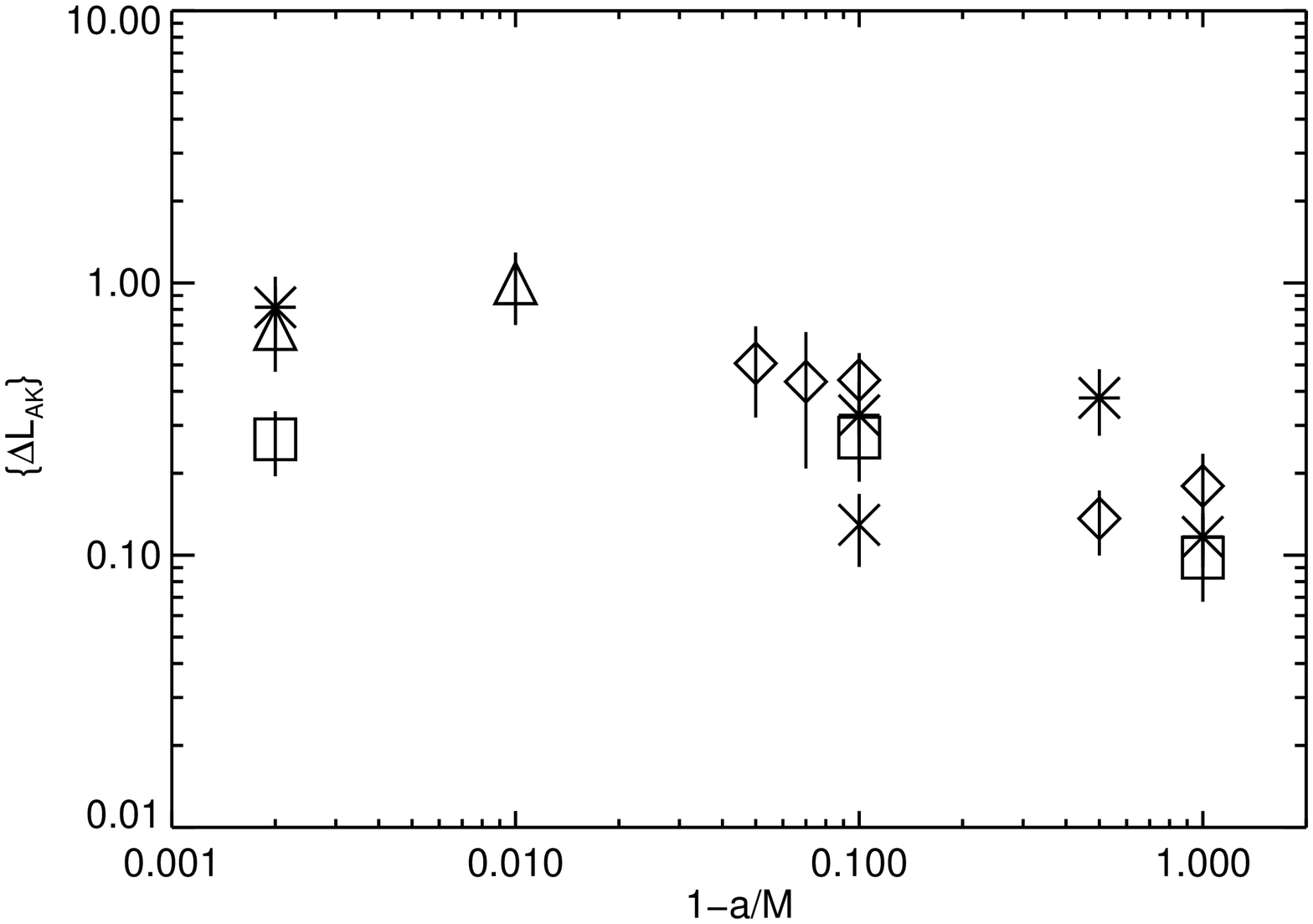}
\includegraphics[width=0.48\textwidth]{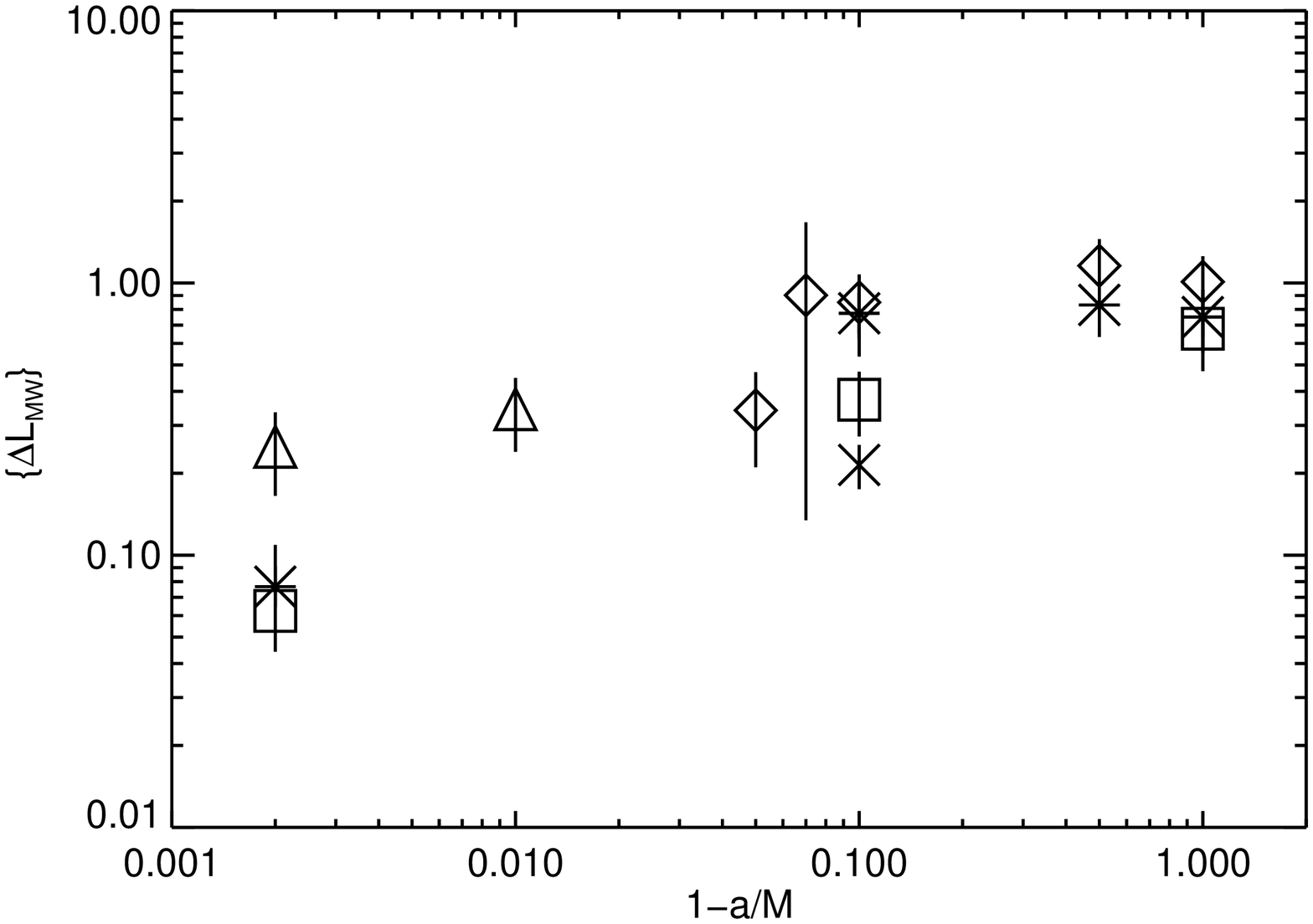}
\end{center}
\caption[]{Solid angle averaged $\{ \Delta L_{\mathrm{AK}}
\}$ (left panel) and $\{ \Delta L_{\mathrm{MW}} \}$ (right panel).
Symbols are as described in Fig. \ref{eta}.}
\label{saavg} 
\end{figure*}

The difference in solid angle averaged luminosity compared to the
standard NT value is shown in Figure \ref{saavg}.  Both $\{ \Delta
L_{\mathrm{AK}} \}$ and $\{ \Delta L_{\mathrm{MW}} \}$ are 10--100\%.
$\{ \Delta L_{\mathrm{AK}} \}$ generally increases with increasing $a/M$.
There is considerable variation between simulations with different
magnetic field topologies but with the same black hole spin.  The AK
model is sensitive to the precise value of the stress used at the
ISCO, but even for the weakest stress values used here there are still
significant increases in total luminosity.

In contrast, $\{ \Delta L_{\mathrm{MW}} \}$ \emph{decreases}
with increasing $a/M$ and is largest for $a/M=0.5$ where $\{ \Delta
L_{\mathrm{MW}} \} \sim 100\%$ In fact, $\{ \Delta L_{\mathrm{MW}} \}$
is larger than $\{ \Delta L_{\mathrm{AK}} \}$ at low spin, but smaller
at high spin, consistent with results obtained in the fluid frame.
Two effects are important here.  First, the plunging region is larger
for low spin holes, so emission from that region (absent by assumption
from the AK models) is more significant for those cases.  Second,
the stress tends to increase sharply at the ISCO for high spin holes,
boosting the emission in the AK model.  As we have already remarked in other
contexts, different magnetic
topologies produce significant variation among the simulations performed
at a given black hole spin.

\subsection{The Radiation Edge}\label{radedge}

The standard NT model is highly constrained. The assumption that
stress goes to zero at the ISCO, the location of which is determined by
the mass and spin of the hole, sets the location of the radiation edge.
Removing this boundary condition means that the radiation edge can move,
but by how much?  As emphasised in the Introduction, one principal goal
of this analysis is to locate the radiation edge for models that have
nonzero stresses at the ISCO.  While one might define the radiation
edge as that point outside of which 100\% of the light is emitted, such
a definition fails to distinguish between the AK and NT models because
for both this point is the ISCO.  To bring out the distinctions
in the distribution of the emission near the inner edge we define the
radiation edge as the radius outside of which $95\%$ of the light that
reaches distant observers is emitted.  To further
develop these contrasts, we also examine the cumulative distribution
function for the fraction of the solid angle-averaged luminosity
reaching infinity from outside a given radius.

\begin{figure*}
\begin{center}
\includegraphics[width=0.32\textwidth]{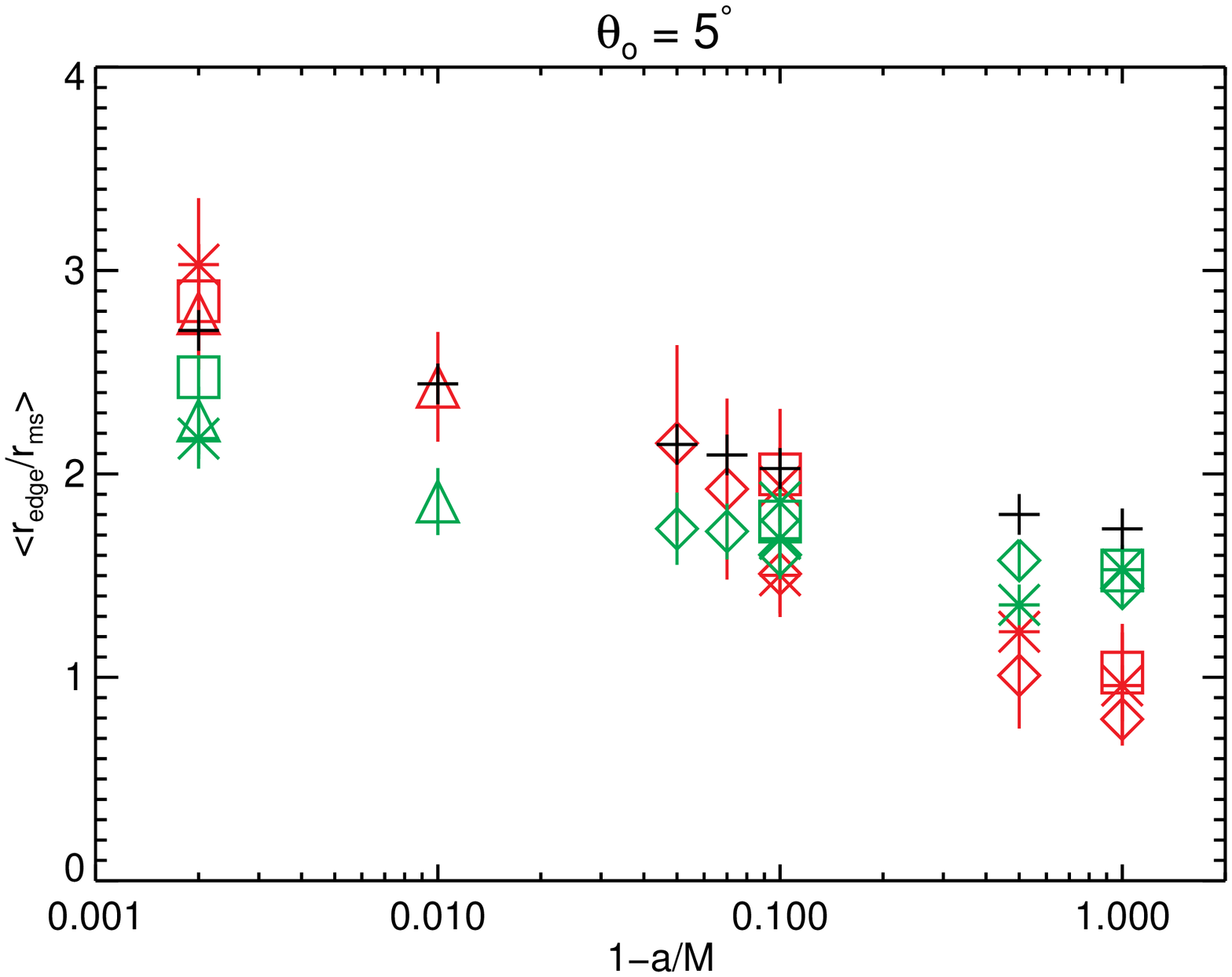}
\includegraphics[width=0.32\textwidth]{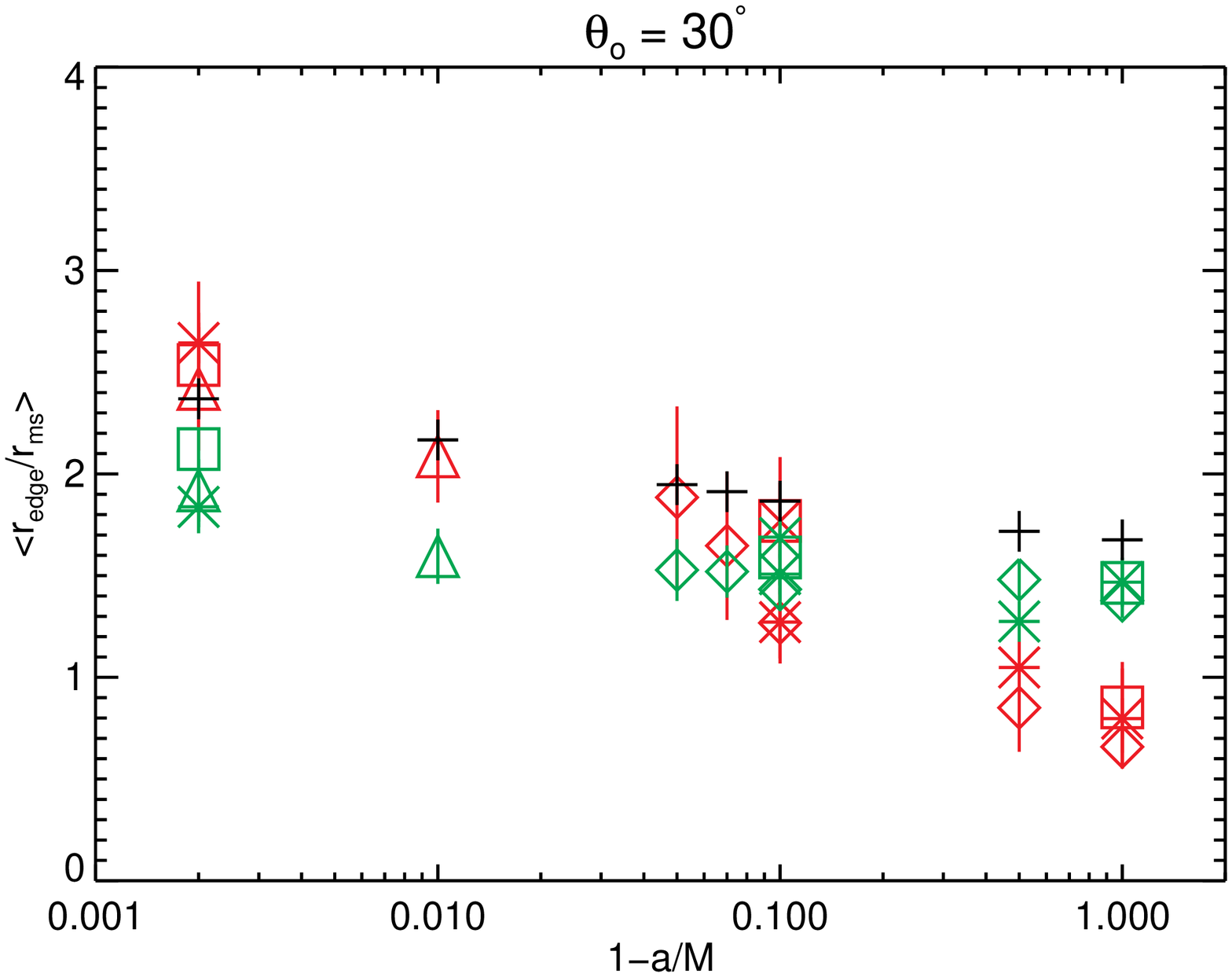}
\includegraphics[width=0.32\textwidth]{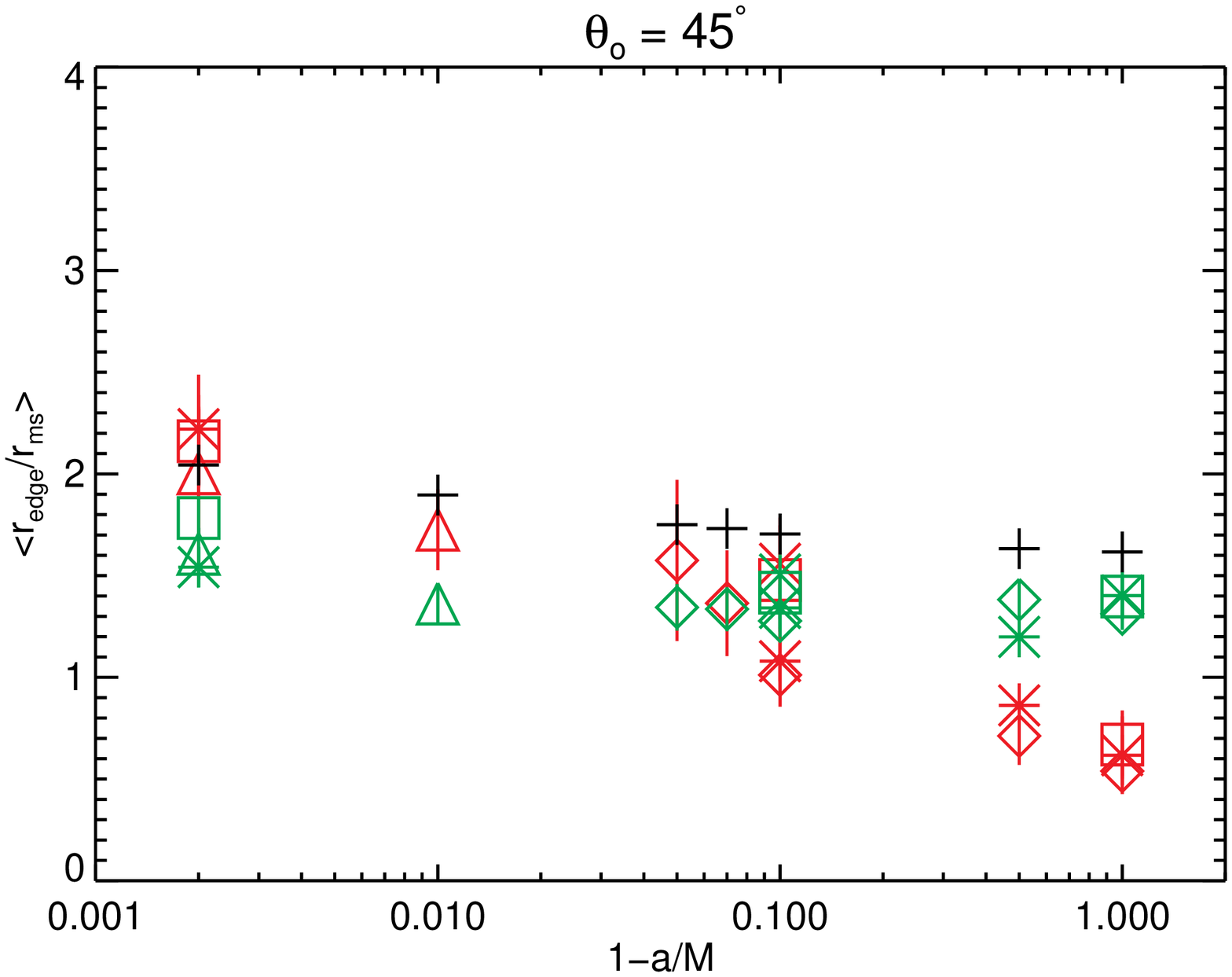}
\includegraphics[width=0.32\textwidth]{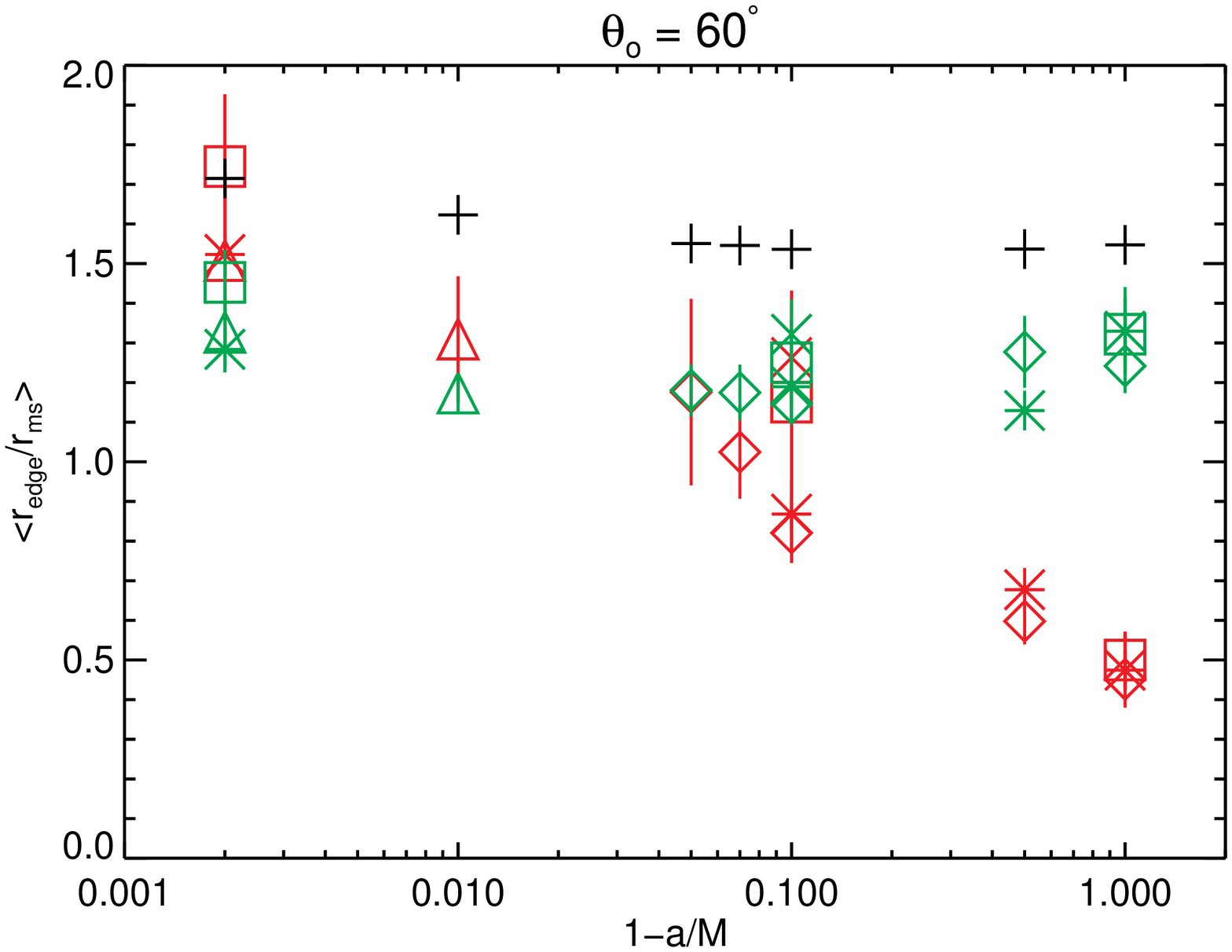}
\includegraphics[width=0.32\textwidth]{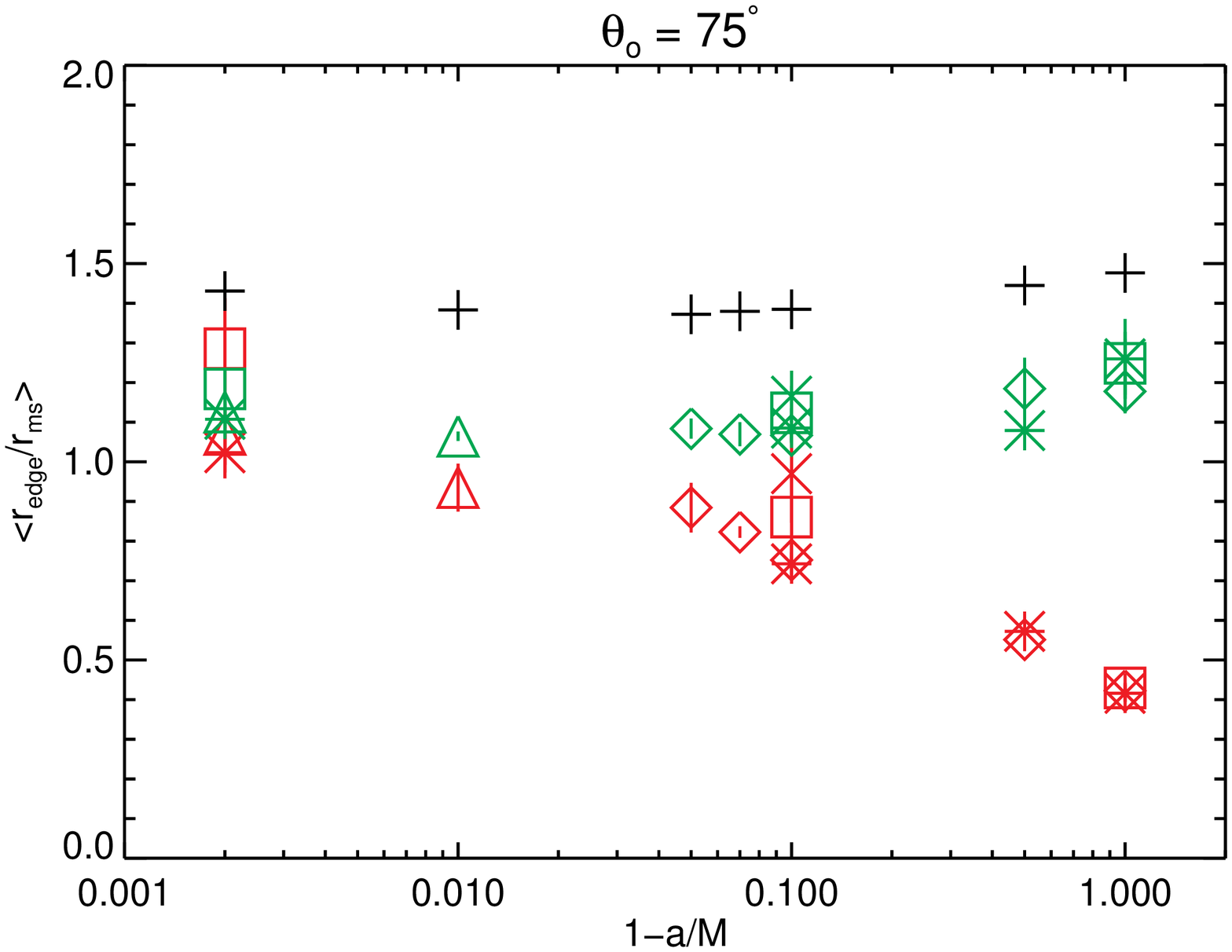}
\includegraphics[width=0.32\textwidth]{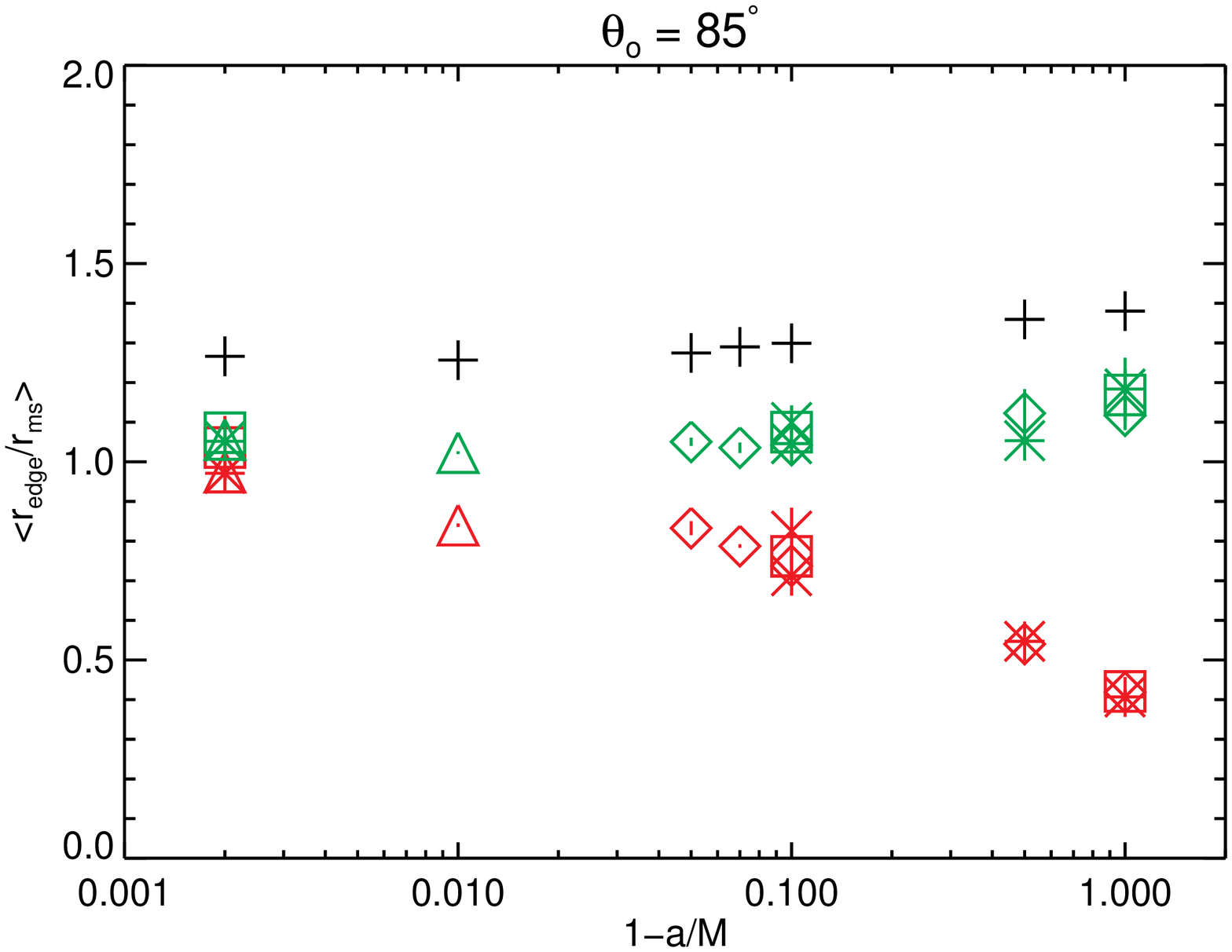}
\end{center}
\caption[]{Time-averaged ``radiation edge'' seen by distant observers in
in six different directions (arranged as in Fig.~\ref{aklum}:
$\theta_{o}=5^{\circ}$ (top left panel), $30^{\circ}$ (top centre
panel), $45^{\circ}$ (top centre panel), $60^{\circ}$ (bottom left
panel), $75^{\circ}$ (bottom centre panel) and $85^{\circ}$ (bottom
right panel).  The different models are coded by differently coloured
symbols: $Q_{\mathrm{NT}}$ (black), $Q_{\mathrm{AK}}$ (green)
and $Q_{\mathrm{MW}}$ (red).  Different symbol shapes denote
different simulations, as for Figure~\ref{eta}:
The dipolar field simulations
described by \cite{De-Villiers:2003b} are shown by stars,
those described by \cite{Hawley:2006} by diamonds and the
new high spin cases computed for this paper by triangles.
Simulations with quadrupolar topologies are shown by squares. The
toroidal field simulation ($a/M=0.9$) is denoted by a cross.}
\label{radedgeplt} 
\end{figure*}

In Figure~\ref{radedgeplt} we plot the radiation edge for each of the
dissipation models and for the same range of $\theta_{o}$ shown in
Figures~\ref{aklum}--\ref{ofrms}.  The first thing to note is that the
transport of photons to infinity introduces a significant separation
between the radiation edge and the previously computed dissipation edge.

For the NT model at angles $\theta_{o}\le45^{\circ}$, the radiation
edge lies between 1.7--$2.8r_{ms}$, increasing with increasing $a/M$,
whilst for $\theta_{o}\ge60^{\circ}$, the radiation edge becomes
approximately independent of black hole spin and can be found between
1.3 and $1.7r_{ms}$.  Comparing the location of the radiation edge
to the dissipation edge reveals that, for $\theta_{o}\le45^{\circ}$,
the radiation edge is either at (for spins $a/M<0.9$) or outside (for
$a/M\ge0.9$) the location of the dissipation edge.  When the inclination
is greater ($\theta_{o}\ge60^{\circ}$), the radiation edge lies either
inside ($a/M<0.9$) or near ($a/M\ge0.9$) the location of the dissipation
edge.

Similar behaviour is found for the AK radiation edge. The principal
difference is that, in this case, the radiation edge lies inside that
of the NT by a fraction that remains approximately constant as
$\theta_{o}$ is varied, with variations only on the $\sim5\%$ level
for a given simulation.   For example, for simulation KD0c, the AK
radiation edge lies between $0.83-0.80$ of the NT radiation edge as
$\theta_{o}$ varies between $5^{\circ}$ and $85^{\circ}$, whilst for
simulation KDPg, this fraction ranges between $0.80-0.75$ over this
same range of $\theta_{o}$.  The larger the black hole spin, the
greater the difference between the AK and the NT models.  In other
words, the relative visibility of the region near the ISCO remains the
same function of spin and inclination angle for both models;
the AK model simply has greater luminosity near the ISCO by an
amount that increases with black hole spin.

We next consider the radiation edge derived from $Q_{\mathrm{MW}}$. In 
\S\ref{diss}, we found that the location of its dissipation edge was 
inside that of the AK or NT model for all spins and field topologies.  
By contrast, the location of the radiation edge shows a wider range of 
behaviours.  At any given inclination, the radiation edge moves to larger 
$r/r_{ms}$ as the spin increases.  For a given spin, the radiation edge 
moves inward as the inclination grows.  Specifically, for discs that are 
viewed almost face-on ($\theta_{o}=5^{\circ}$), the radiation edge lies 
between 1--$3r_{ms}$, increasing with increasing black hole spin. At 
this inclination, the radiation edge lies approximately a factor of 
three outside the location of the dissipation edge, independent of black 
hole spin. At moderate inclinations 
($30^{\circ}\le\theta_{o}\le45^{\circ}$), the radiation edge moves 
inwards towards the ISCO, ranging between 0.5 and $2.5r_{ms}$, and again 
increasing with increasing black hole spin. For $\theta_{o}<45^{\circ}$ 
and $a/M>0.9$, the radiation edge lies either at or outside that derived 
from $Q_{\mathrm{NT}}$ and $Q_{\mathrm{AK}}$, whilst for $a/M<0.9$ the 
radiation edge lies either at or \emph{inside} that derived from 
$Q_{\mathrm{NT}}$ and $Q_{\mathrm{AK}}$.

These trends in the location of the radiation edge, and in particular,
the way it is offset from the dissipation edge (\S\ref{diss}) are driven
by the way photons travel through the relativistic potential.   When the
disc is viewed face-on, gravitational redshift dominates, reducing both
the energy of photons emitted close to the hole and the apparent rate
at which they are released; consequently, the radiation
edge tends to move outward as the inclination angle becomes smaller.  As
the viewing direction moves closer to edge-on, Doppler boosting of the
approaching side of the disc increases the energy of photons radiated there
and their rate of emission, bringing the radiation edge inward.  In the
NT and AK models, all velocities are azimuthal, so the peak approach
velocity occurs for matter passing through the sky plane when our view
is nearly in the disc equatorial plane; however, in the MW model,
the inward radial speed can be great enough, especially in the plunging
region, for Doppler boosting also to enhance the luminosity of matter
on the far side of the black hole.  In all models, emission from the
far side of the hole is also strengthened by gravitational lensing when
the line of sight is near the disc plane.  Thus, the radiation edge moves 
inward as the observer approaches the disc plane.

Trends in the position of the radiation edge as a function of black
hole spin at fixed inclination angle are the result of a different trade-off.
As the spin increases, the ISCO moves inward and all relativistic
effects are strengthened.  Those described in the previous paragraph
tend, on balance, to make the radiation brighter as the inclination
angle grows.  However, as the spin increases, the fraction of all
photons captured by the black hole also increases, likewise because
the ISCO moves inward.   In terms of how far inside the ISCO the
radiation edge falls, the latter effect is the strongest: the
ratio of the radial coordinate of the radiation edge to $r_{ms}$
is least for low-spin black holes viewed nearly edge-on.

One surprising result demands special discussion: at the highest spin
($a/M = 0.998$) and small inclination angle ($\theta_o \leq 30^{\circ}$),
the MW radiation edge either coincides with or lies a little bit outside
the NT edge, and both the MW and NT edges are well outside the AK edge.
At such high spin, photons escaping to infinity in the polar direction are
severely redshifted if their point of origin is near or inside the ISCO.
For this reason, all three radiation edges in this regime are at radii
several times $r_{ms}$.  As we have already discussed, however, when
the spin is this rapid, the AK model predicts a rather larger dissipation
rate in this range of radii than is predicted by either MW or NT.  The
disparity between the AK radiation edge on the one hand and the MW and NT
edges on the other follows directly from this contrast.  In addition, the
radiation edge for one high-spin simulation (KDEa) appears to fall at
particularly large radius, although this may well be an artefact of the lack of
inflow equilibrium in this particular simulation \citep{Krolik:2005}

\begin{figure}
\begin{center}
\includegraphics[width=0.48\textwidth]{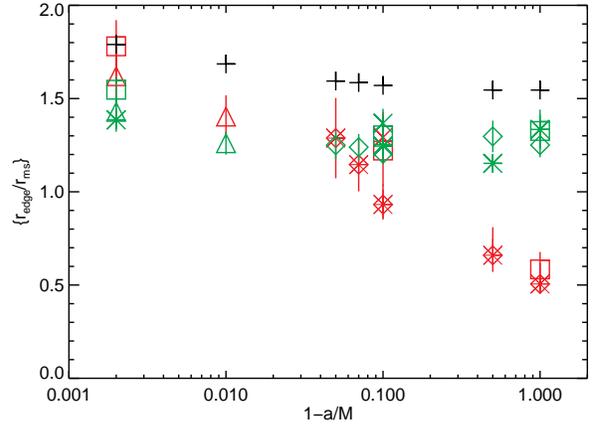}
\end{center}
\caption[]{Solid angle averaged $\{ r_{edge} / r_{ms} \}$. Symbols are as in
Figure \ref{dissedge}}
\label{saavgredge} 
\end{figure}

The radiation edge associated with the solid angle-averaged luminosity
is shown in Figure \ref{saavgredge}.
For $Q_{\mathrm{NT}}$, it lies between 1.5--1.8$r_{ms}$ and slowly
increases with increasing black hole spin.  For all models
as the spin increases, photons radiated deeper in the potential are
subject to larger gravitational redshifts and are more likely to be
captured.  For $Q_{\mathrm{AK}}$, the radiation edge moves inwards to
1.3--1.6$r_{ms}$, with the smallest values occurring for $a/M=0.9$ and
increasing as one moves away from this spin in either direction.  For this
model the enhanced dissipation at the ISCO compensates somewhat for the
greater likelihood for photon capture with increased black hole spin.
For $Q_{\mathrm{MW}}$, the radiation edge lies between 0.5--1.7$r_{ms}$,
increasing with increasing black hole spin. The enhanced dissipation
within the plunging region boosts the luminosity (and reduces the radius
of the radiation edge) for low spin models, but has little effect for
high spins.

Regardless of spin the solid angle-averaged radiation edge derived from
the simulation data lies within that of the NT model; enhanced stress
always moves this point inward.  Comparing $Q_{\mathrm{AK}}$ and
$Q_{\mathrm{MW}}$ we find a more complex picture.  The radiation edge
in the AK model must, by assumption, lie outside the ISCO, but for the
models with stress in the plunging region this need not be the case.
Indeed, for low spin holes the radiation edge can be inside the ISCO,
but the dissipation in the plunging region becomes less important
as black hole spin increases.  What matters most is the dissipation
level outside of the ISCO.  For the cases considered here, the stress
rises rapidly at the ISCO for high-spin models and the semi-analytic AK
formula predicts greater dissipation outside the ISCO than is implied
by the stress levels seen in the simulations.  Hence, for $a/M>0.9$
the solid angle-averaged radiation edge derived from $Q_{\mathrm{MW}}$
lies \emph{outside} that derived from $Q_{\mathrm{AK}}$.  This clearly
illustrates that the location of the radiation edge can be very sensitive
to the dissipation levels near the ISCO.

We have defined the radiation edge as the the radius
outside of which $95\%$ of the total observed luminosity is emitted.
This choice was made to provide a simple measure by which to contrast
the different luminosity profiles associated with
each of the dissipation models. In Figure \ref{saavg_frac}
we plot the solid-angle averaged fractional
cumulative distribution of observed
luminosity, $\{F(r/r_{ms})\}$ for each of the dissipation models
and individual datasets. We see that for $0.2 \leq F \leq 0.95$,
all these curves may be fitted to a reasonable degree of approximation
by the form $\{F(r/r_{ms}\}) \propto \exp[-(r/r_{ms})/r_*]$. Thus, there {\it is}
a characteristic radial scale for the luminosity profile that is always
related to, but not necessarily identical to, $r_{ms}$, and it may
be adequately parameterized by choosing a fiducial level of $F$.

The plots also provide a quantitative sense of the potential importance of
nonzero stress at the ISCO. First, note that the luminosity profile
in the NT model is practically independent of
spin (as can also be inferred from the solid-angle averaged radiation
edge shown in Figure \ref{saavgredge}). This somewhat surprising result
can be understood in terms of the dependence of the radiation edge on
$\theta_{o}$. At small inclination angle (face-on views),
the radiation edge derived from the
NT model for slowly spinning black holes lies inside that of the rapidly
spinning holes (relative to the location of the ISCO), whilst at large inclination angles the reverse is true. When
this data is averaged over solid angle, these changes cancel and so
the position of the radiation edge (and hence the fractional luminosity
distribution) relative to $r_{ms}$ is independent of $a/M$.  The NT model
is so tightly constrained by its assumptions that when the dependencies
due to observing angle are removed, there remains almost no contrast
other the location of the ISCO which is set by the spin of the hole.

\subsection{Characteristic Temperature}

We have seen how the apparent size of the disc varies with spin, inclination and
dissipation model.  However, observations do not directly measure
the radiation edge.  Rather, the concept of a radiation edge is
incorporated into the interpretation of the $L = A(a/M,\theta) T^{4}$
relationship \cite[][]{Gierlinski:2004}.  We have already seen how changes in the
dissipation function due to non-zero stress at and inside the ISCO have
the potential to change the apparent area and luminosity of the disc.
In this section, we attempt to gauge the impact of these changes on the
characteristic temperature of the accretion flow, $T_{char}$, as it might be
inferred from continuum fitting to the soft component of a spectrum.
Our analysis must necessarily be simple.  We define $T_{char}$ as
the maximum (observed) blackbody temperature found anywhere
in the disc.  It is determined by defining an effective temperature in
the fluid frame: $T_{eff}(r) = [Q(r)/\sigma ]^{1/4}$ and then transforming
this temperature to the rest frame of a distant observer via $T_{o}
(r) = g T_{eff}(r)$ \cite[][]{Cunningham:1975}. Then, since the characteristic
temperature of the associated blackbody spectrum will be very close
(but not identical to) the maximum black body temperature, we  make the
approximation $T_{char} \sim \mathrm{max} (T_{o})$.  We perform the same
procedure for all three dissipation models, ignoring color temperature corrections
and all observational effects except those due to inclination angle.

\begin{figure*}
\begin{center}
\includegraphics[width=0.32\textwidth]{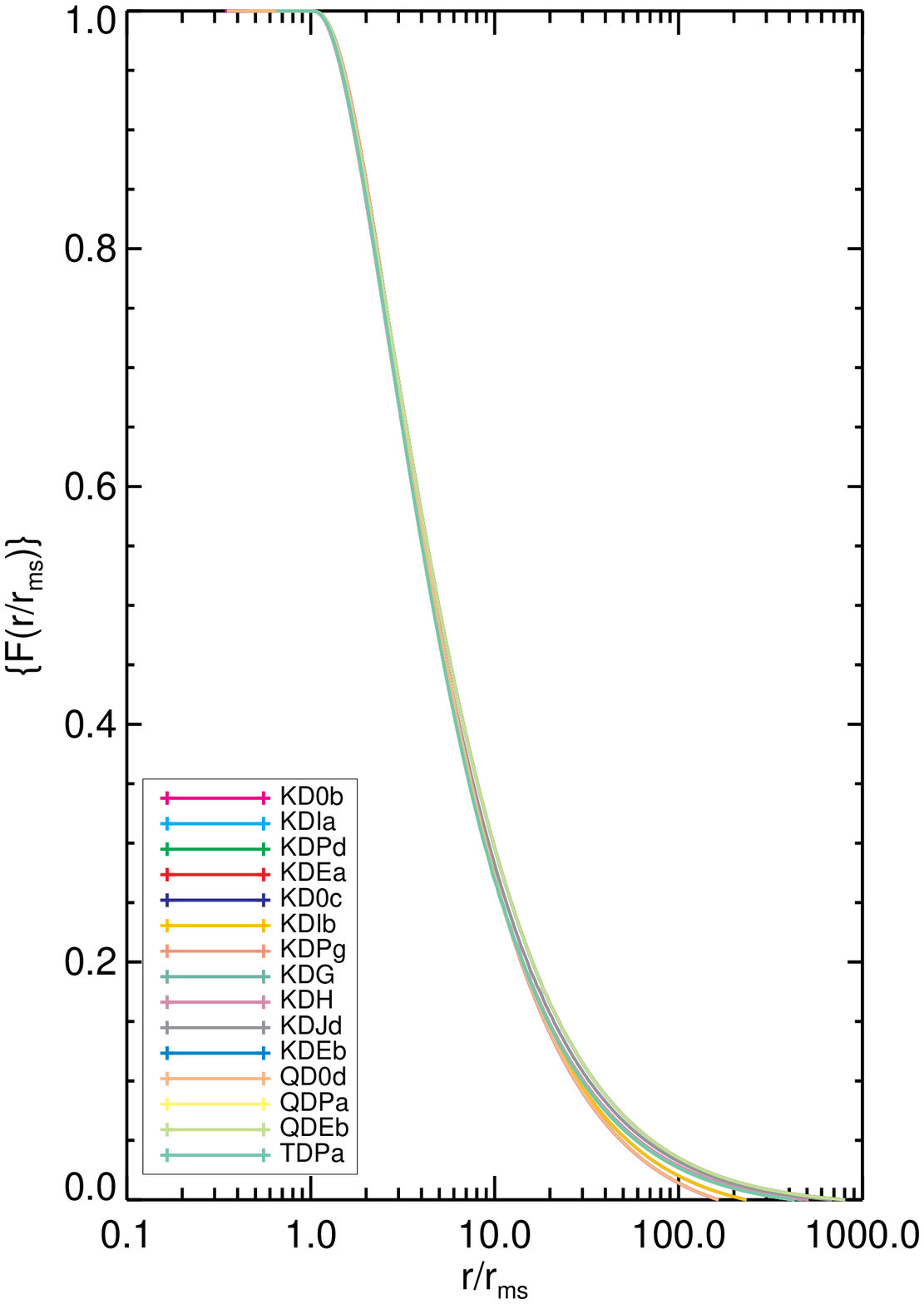}
\includegraphics[width=0.32\textwidth]{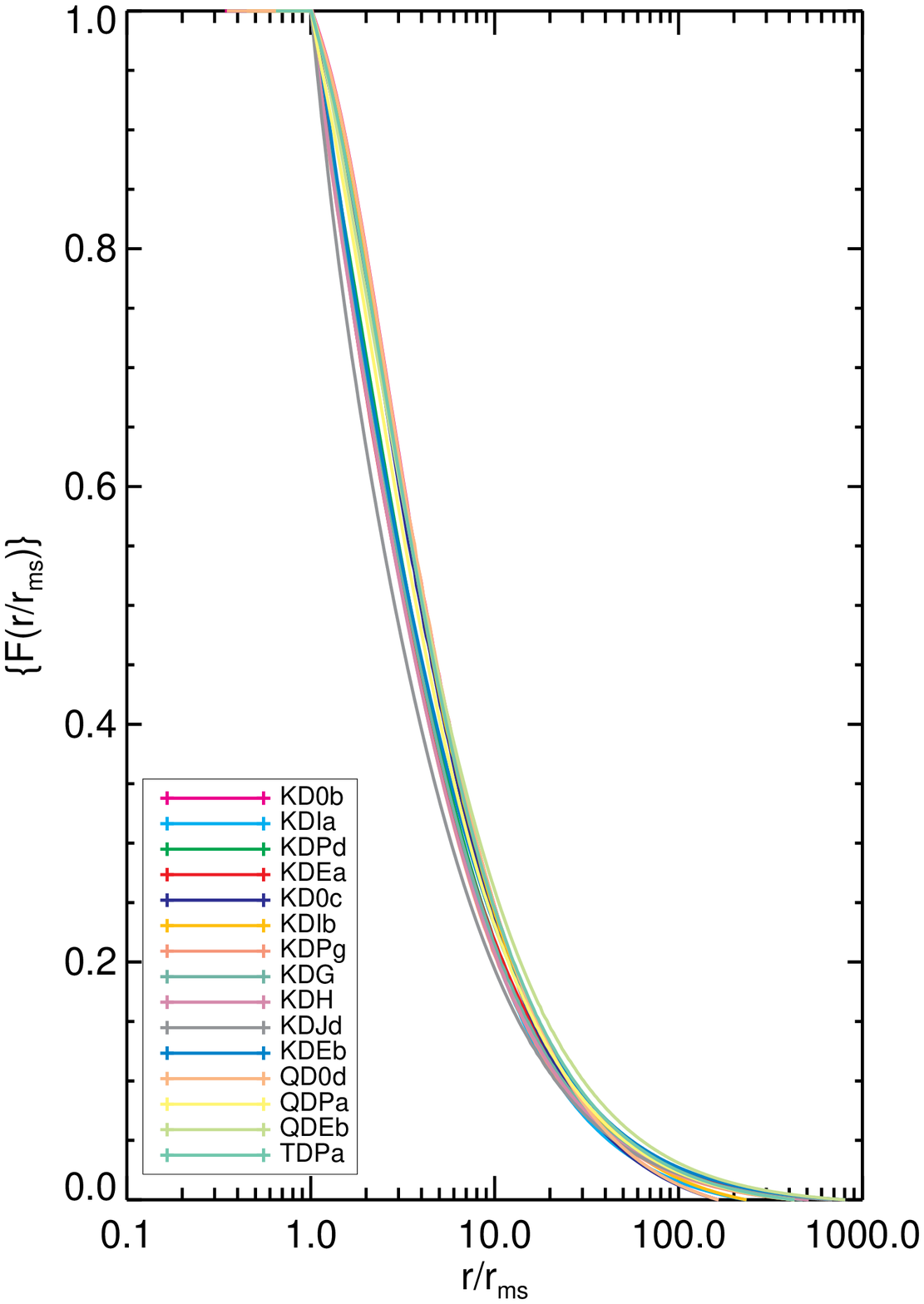}
\includegraphics[width=0.32\textwidth]{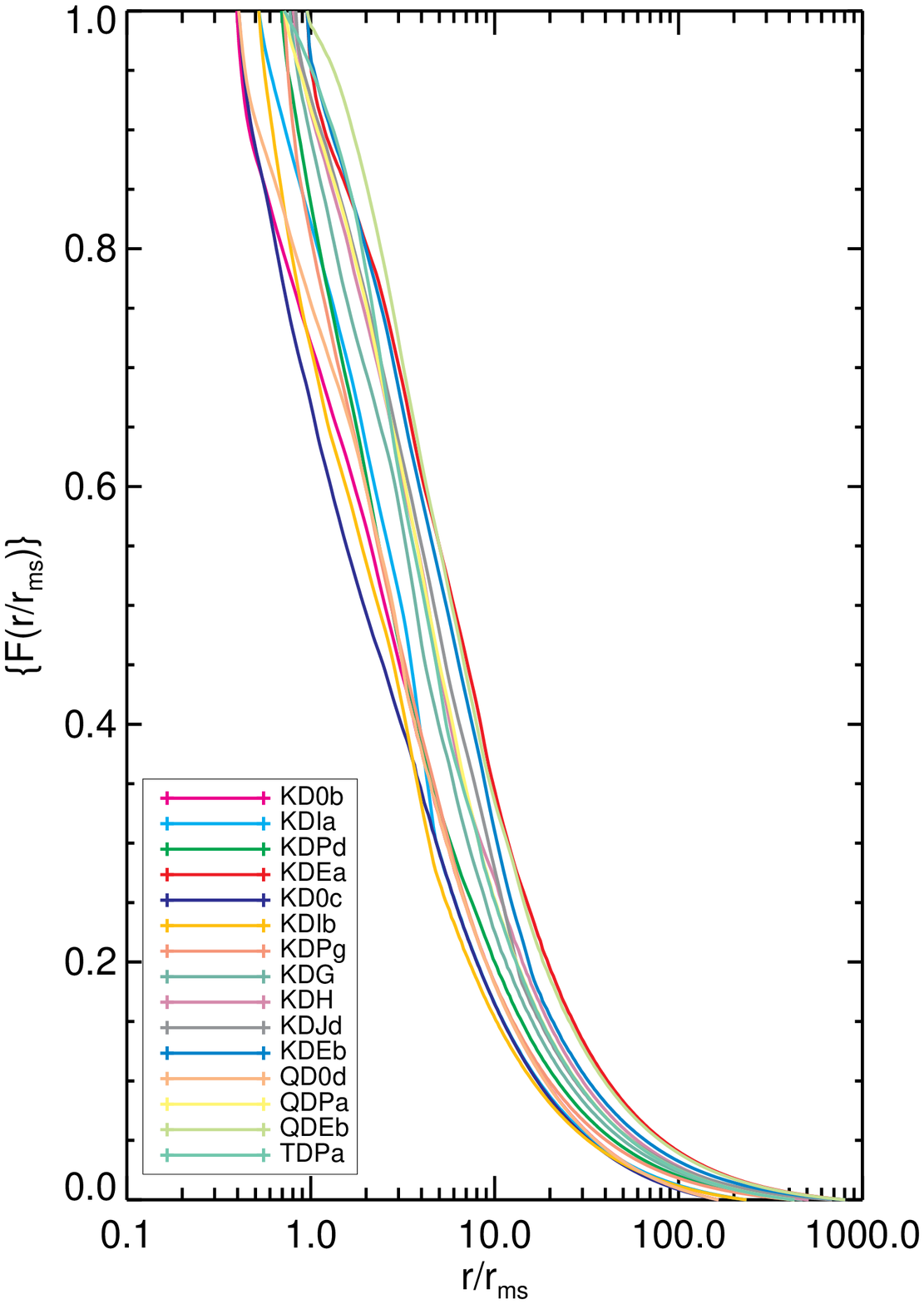}
\end{center}
\caption[]{Solid-angle averaged fractional distribution of observed luminosity, $\{F(r/r_{ms})\}$ derived from $Q_{\mathrm{NT}}$ (left-hand panel), $Q_{\mathrm{AK}}$ (centre panel) and $Q_{\mathrm{MW}}$ (right-hand panel).}
\label{saavg_frac} 
\end{figure*}

Although the geometric size of the disc in the AK model is
the same as in the NT model
(the curves are still constrained to reach $100\%$ at the ISCO), the middle
panel of Figure \ref{saavg_frac} shows how the addition of a finite
amount of stress at the ISCO breaks the NT degeneracy.  The most significant
differences come, naturally, from including contributions from inside
the ISCO.  The curves for the $Q_{MW}$ models (right hand panel) are
well separated for different spins.  For these models, the relative
importance of the plunging region is a function of both the amplitude of
the stress there and the spin of the hole.  The farther the ISCO from
the horizon, the greater the chance for radiation within the plunging
region to escape.

\begin{figure*}
\begin{center}
\includegraphics[width=0.32\textwidth]{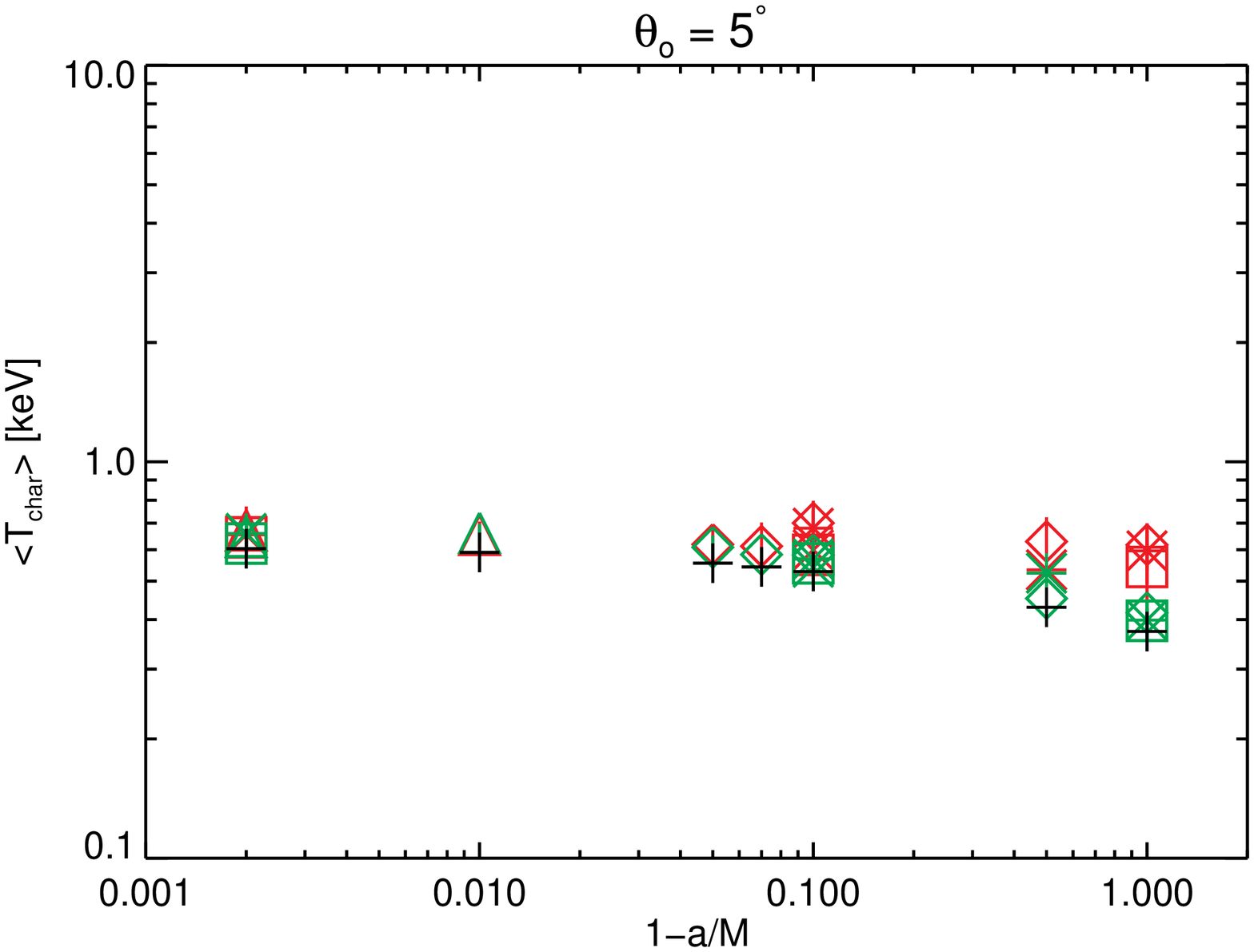}
\includegraphics[width=0.32\textwidth]{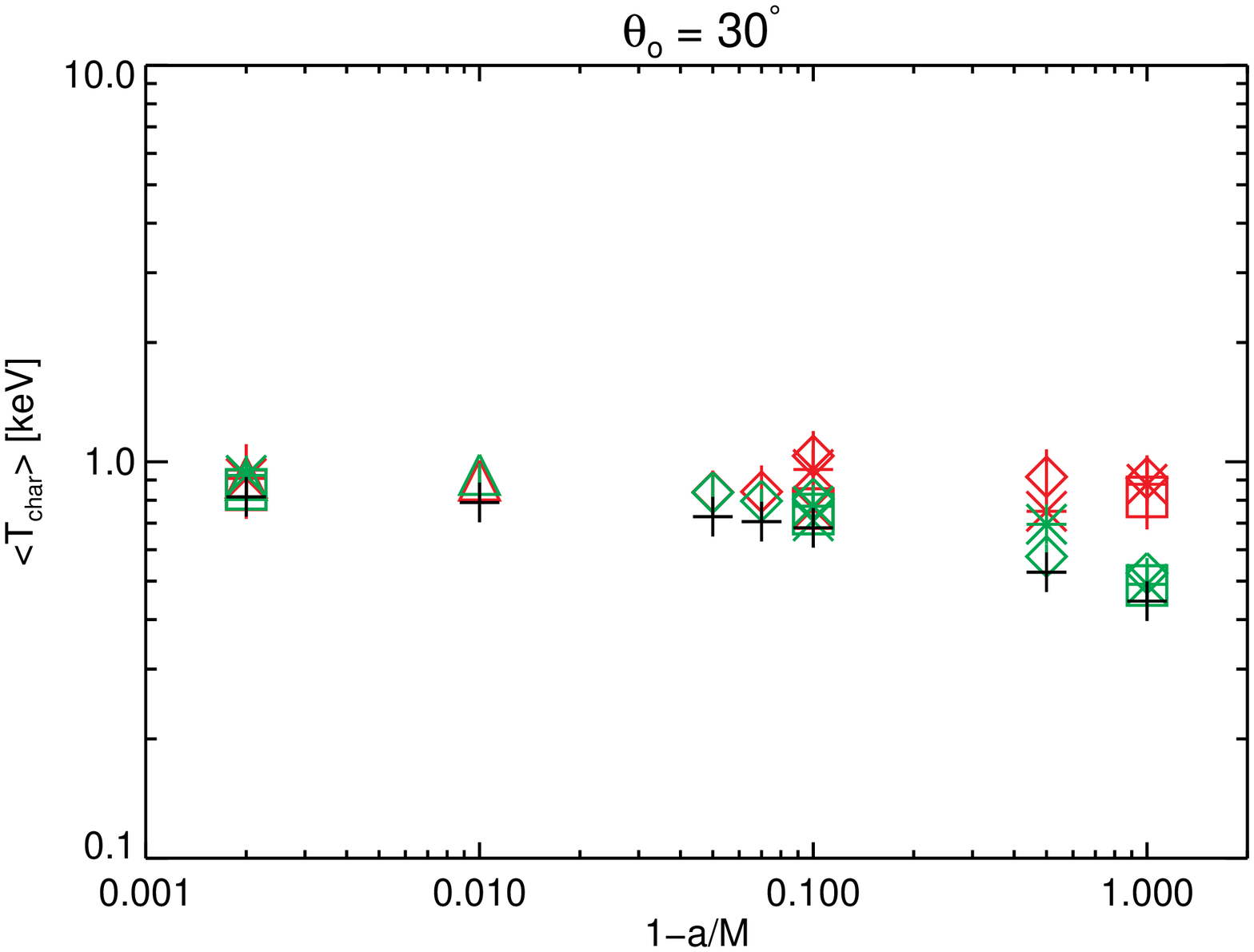}
\includegraphics[width=0.32\textwidth]{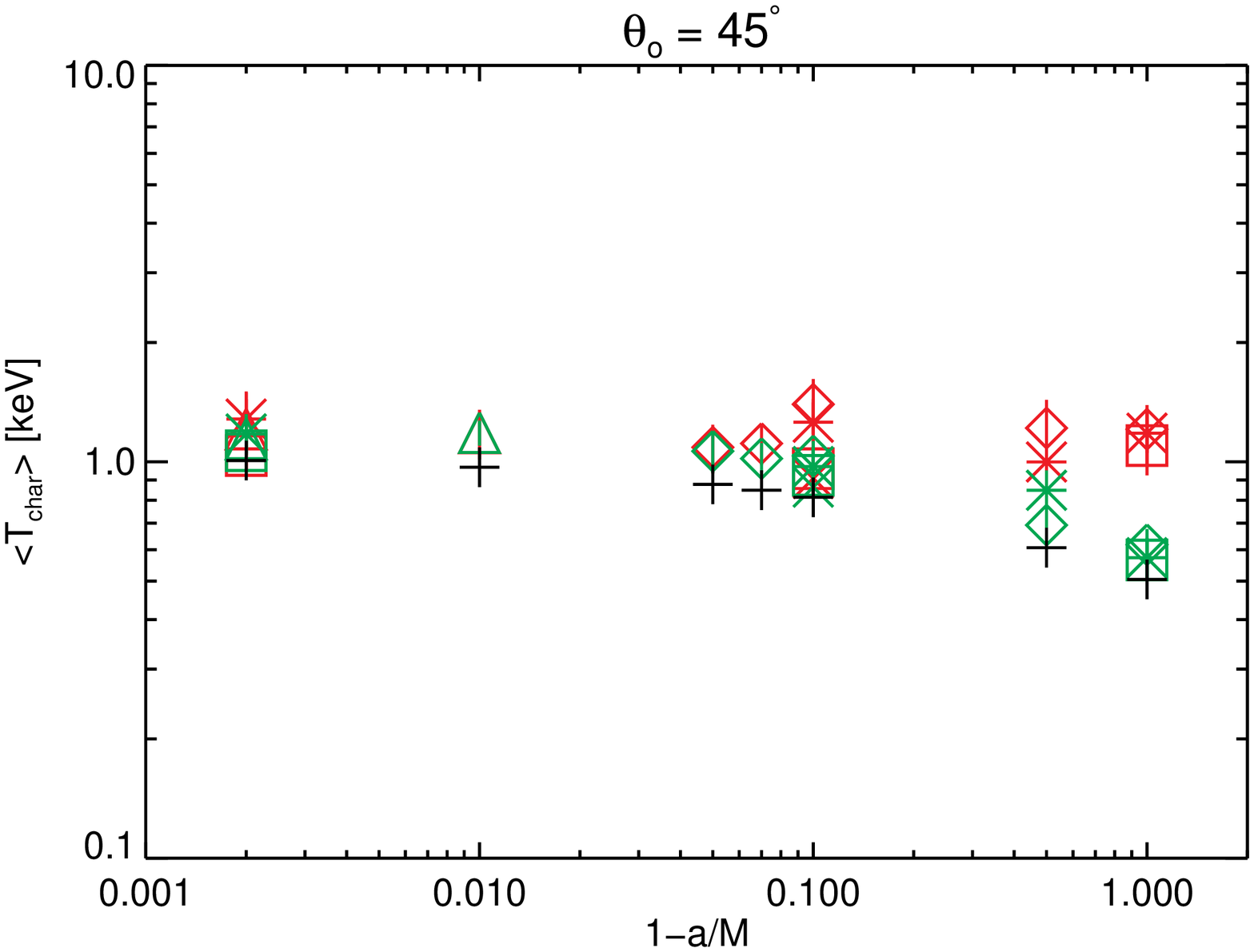}
\includegraphics[width=0.32\textwidth]{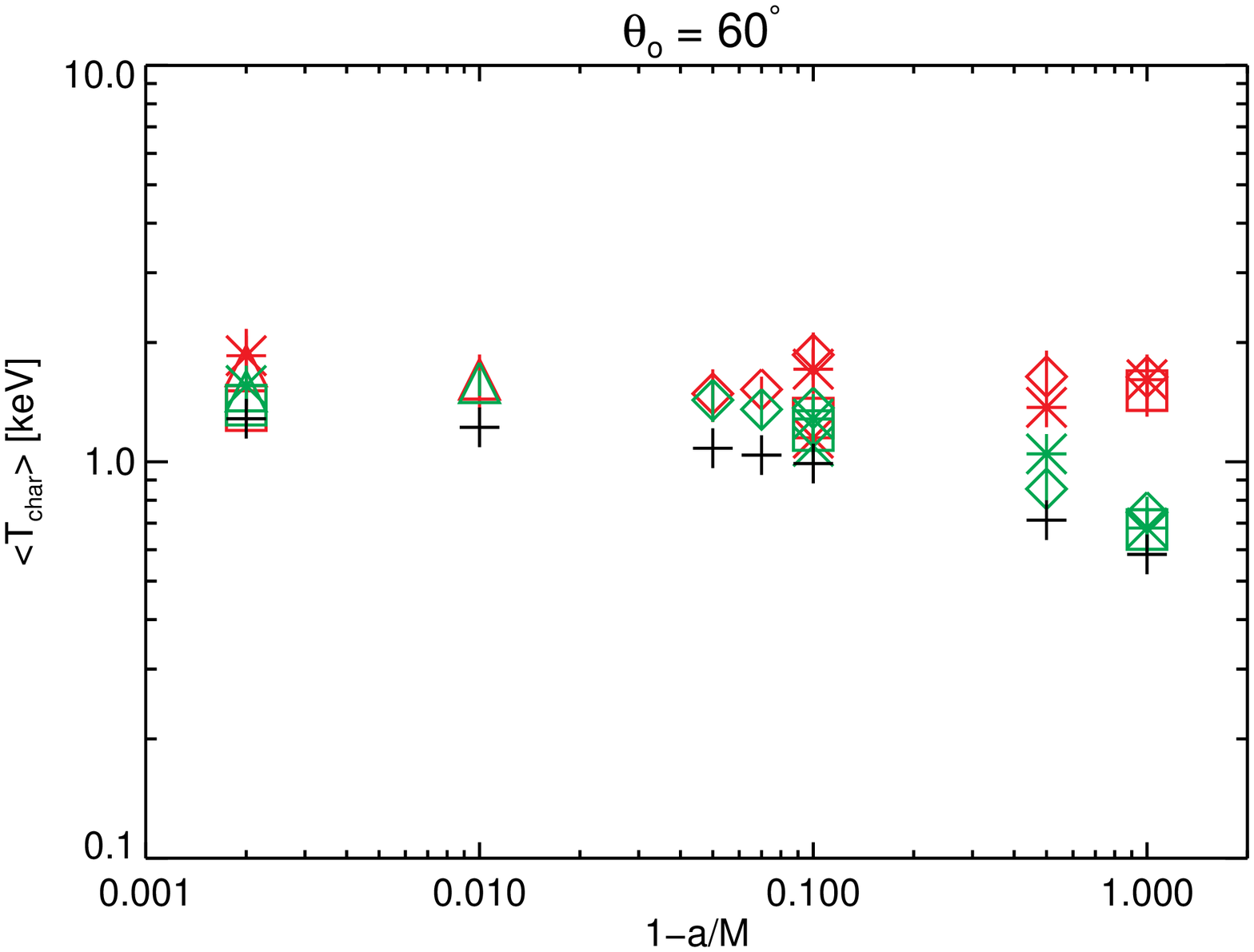}
\includegraphics[width=0.32\textwidth]{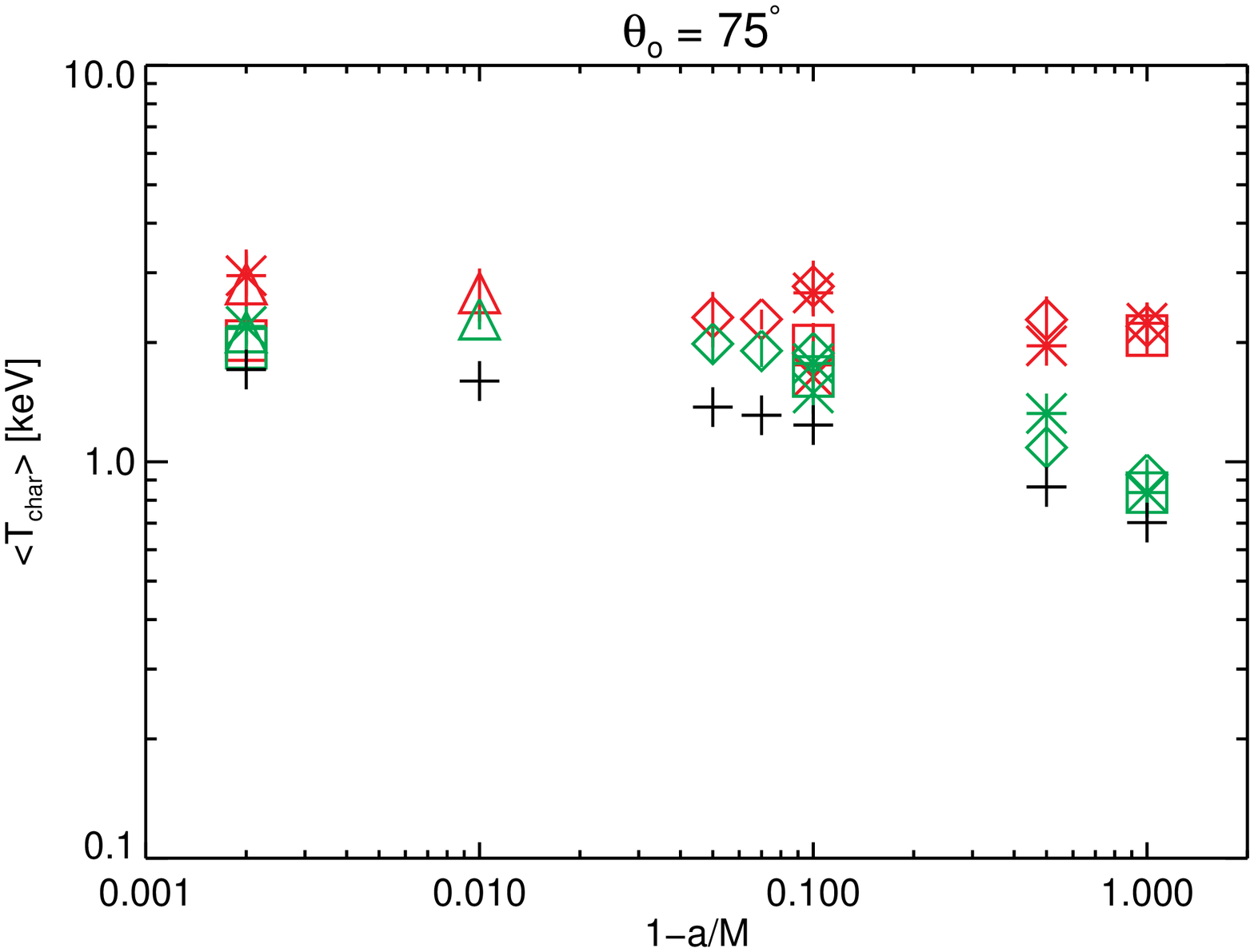}
\includegraphics[width=0.32\textwidth]{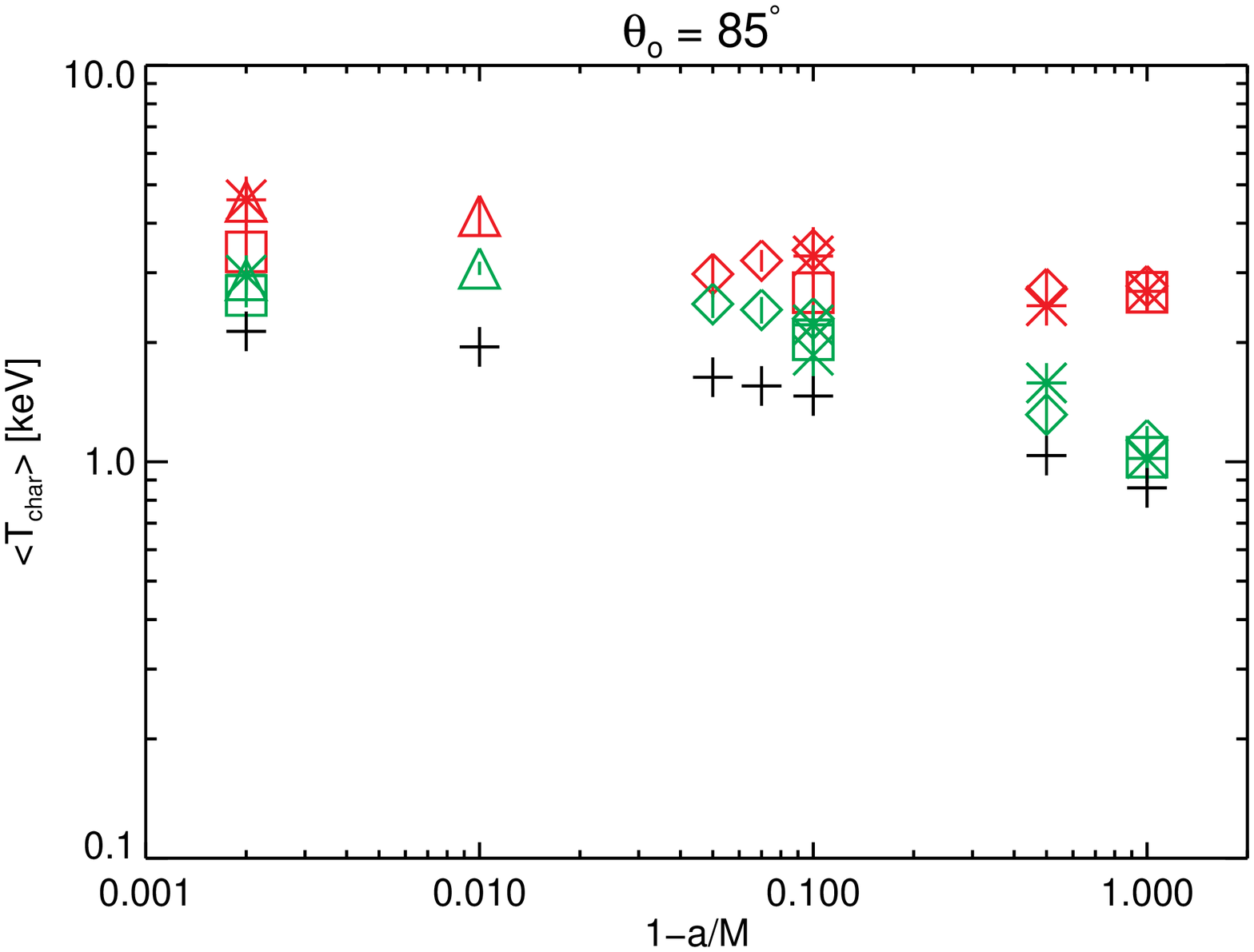}
\end{center}
\caption[]{Time-averaged characteristic black body temperature for
a $10 M_{\odot}$ black hole accreting at $0.1L_{edd}$ measured
by distant observers in in six different directions (arranged as in
Fig.~\ref{radedgeplt}, with the same key as that figure).}
\label{tcharplt} 
\end{figure*}

$T_{char}$ is plotted in Figure \ref{tcharplt} for the
different dissipation functions and observer inclinations shown in
Figure \ref{radedgeplt}. Each of the accretion flows has been scaled
such that its accretion rate would produce a flow with luminosity
$0.1 L_{edd}$. Several features are immediately apparent.  At all
inclinations $T^{NT}_{char}$, the characteristic temperature associated
with $Q_{NT}$ increases with increasing $a/M$.  This well known result
is the foundation of attempts to measure black hole spin with spectral
fitting techniques. $T^{AK}_{char}$ shows a similar trend, with an
increase over $T^{NT}_{char}$ by a factor $1.1-1.4$, an amount that is
comparable to typical color corrections  \cite[see e.g.][]{Done:2008}.
The AK model has the same (fixed) inner boundary, but has enhanced
luminosity for the same accretion rate.  The effect is most significant
for large inclination angles.

$T^{MW}_{char}$ exhibits a different behaviour. At low spins,
$T^{MW}_{char}$ is significantly greater than $T^{NT}_{char}$, whilst at
high spins the contrast is reduced. Overall, this has the effect of making
$T^{MW}_{char} (a/M)$ almost constant for a given inclination.  Overall,
$T^{MW}_{char}/T^{NT}_{char}$ varies between a
factor of $1.8-2.3$
for accretion flows accreting at the same luminosity, with the greatest
increases occurring for slowly spinning black holes viewed nearly edge
on.  Because energy extraction and radiation occur within the plunging
region, the distinctions between holes with different spins is greatly
reduced.  In other words, the effective inner boundary for all discs
lies close to the horizon.

\begin{figure}
\begin{center}
\includegraphics[width=0.48\textwidth]{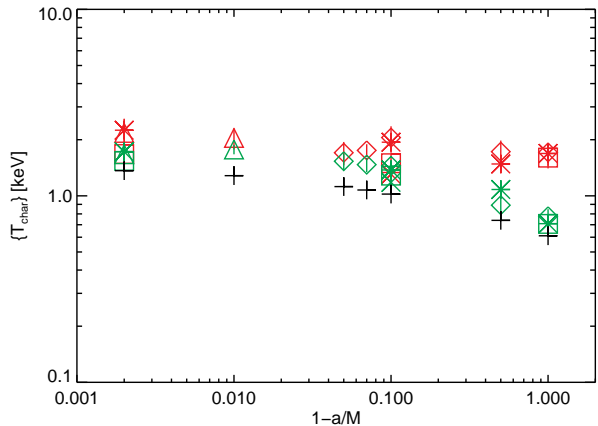}
\end{center}
\caption[]{Plot showing solid angle averaged $\{ T_{char} \}$
. Symbols are as in Figure \ref{dissedge}}
\label{saavgtchar} 
\end{figure}

These effects are summarised in 
Figure \ref{saavgtchar}, where we plot the characteristic temperature
associated with the solid-angle averaged radiation edge for a black
hole accreting at $0.1L_{edd}$.  Both the NT and the AK model share
the ISCO as their inner boundary, and as a result both $T^{NT}_{char}$
and $T^{AK}_{char}$ increase by around a factor of two over the range of
$a/M$ shown here (both increasing with increasing $a/M$).  On the other
hand $T^{MW}_{char}$ is approximately constant over the same range of
spin and is greater than $T^{NT}_{char}$ by a factor
that falls from $\simeq 2.3$ to $\simeq 1.2$ as $a/M$
increases from 0 to 0.998.  Unlike the radiation edge, the
contrast between $T^{MW}_{char}$ and both $T^{AK}_{char}$ and $T^{NT}_{char}$
always has the same sense: the MW model always predicts a higher characteristic
temperature.

\subsection{Impact on Measurements of $a/M$}

In the previous section, we examined the consequences of stress at the
ISCO on the characteristic temperature of the accretion flow and the dependence
of this quantity on both black hole spin and inclination.  We now examine this
question from the opposite direction, i.e., supposing one has a measurement of
the characteristic temperature of an accretion disc in a particular system
(along with the black hole mass, inclination of the binary orbit and disc
luminosity), how much uncertainty in the determination of $a/M$ is induced by
uncertainty about the correct dissipation profile in the inner disc?

\begin{figure}
\begin{center}
\includegraphics[width=0.48\textwidth]{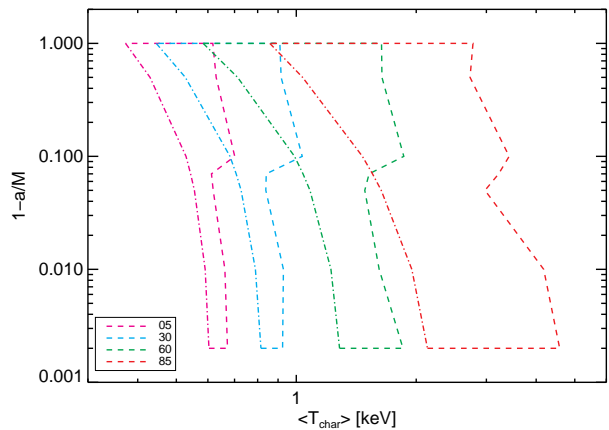}
\end{center}
\caption[]{Plot showing the region of the $(T_{char},a/M)$-plane consistent with
a given value of $\theta_{o}$ for a $10 M_{\odot}$ black hole accreting at $0.1L_{edd}$.
Different colors correspond to different $\theta_{o}$ (see key in bottom left
corner of plot), dot-dash lines show the $T_{char}$ curve predicted by $Q_{NT}$
at a given $\theta_{o}$, and the region enclosed by the dot-dash and dashed lines
of a given color show the region
of the $(T_{char},a/M)$-plane consistent with that value of $\theta_{o}$.}
\label{tcshape}
\end{figure}

To address this question, we use the data shown in Figure \ref{tcharplt}
to construct Figure \ref{tcshape}, in which four polygons, one for each of four
different inclinations, illustrates the range of $T_{char}$ and $a/M$
consistent with the different disc models.  To produce this figure, we first
fixed the inclination, spin, and black hole mass, and then collected the values
of $T_{char}$ predicted on the basis of any of our three candidate models (NT,
AK, and MW) from any of the several simulations we conducted with that particular
spin parameter.   The left-hand edge of each region traces the prediction of
the NT model because it always gives the lowest temperature.  The horizontal width
of each region is defined by the range of temperatures predicted by the complete
complement of models and simulations.  Because we have conducted simulations
at seven different spins, the right-hand edge is defined at seven points.

\newpage

Several qualitative features emerge from study of this figure.  First, for
each inclination, the maximum temperature change predicted by the NT model
as the spin runs from $a/M = 0$ to $a/M = 0.998$ (generally about a factor of 2)
is always less than the maximum temperature change at zero spin due to
model-dependent uncertainty in the dissipation profile (typically a factor of 3).
Thus, if we knew the inclination angle and mass for a given black hole, but
not its spin paramater, our uncertainty in predicting $T_{char}$ has a greater
contribution from our uncertainty about disc physics than it does from our
ignorance of its value of $a/M$.

Second, suppose that we know the inclination angle of a particular black hole
and obtain $T_{char}$ from X-ray spectra.  The uncertainty in $a/M$ due to uncertainty
in the dissipation profile may be estimated by imagining a vertical
line at that value of $T_{char}$ running through the region for the appropriate
inclination angle.  The uncertainty is then given by the difference in spins
between the points where the line intersects the boundary of the region.
The size of this uncertainty tends to be larger when our
view is more nearly edge-on.  However, for almost any value of $T_{char}$ that is
consistent with {\it any} of the area of a region, the magnitude of the
uncertainty in $1 - a/M$ is generally at least an order of magnitude.

Still another way to look at this diagram is to ask, ``If we estimate spin by
using the NT model, by how much may it be wrong if a different model is truer
to the disc's radiation profile?"  The answer is
the span of a vertical line whose lowest point is the NT prediction and runs
to the upper boundary of the associated region. Because the NT model always
gives the lowest prediction for $T_{char}$, the spin inferred from it is always
the greatest possible spin, and the error bar always stretches
toward lower rotation rates.  Given the characteristic curvature
of the permitted regions in this plane, the magnitude of the possible error
in $1 - a/M$ tends to be larger (in logarithmic terms) when the NT inference
of $a/M$ is closer to unity, but can still be substantial even when $a/M \lesssim 0.9$.
Note, however, that we only present data for $a/M>0$, so we cannot provide an explicit lower
bound on the range of $a/M$, beyond stating that $a/M=0$ in the magnetized case is always
consistent with the full range of $a/M$ in the NT case for fixed $\theta_{o}$.

\section{Summary, Discussion and Conclusions}\label{summ}

That magnetic forces might cause substantial stress at the ISCO was
foreseen very shortly after the invention of the standard model
\cite[][]{Page:1974}.
This possibility now appears to be an immediate corollary of the
well-established result that MHD stresses account for most of the
angular momentum transport in the bodies of accretion discs.  Indeed,
such stresses were seen in the first generation of three dimensional
GRMHD disc simulations.  The goal of this paper has been to
begin the linkage of these numerical MHD simulations to the observable
properties of accreting black holes even before the simulations are fully
equipped to make predictions about how these systems radiate.   To do so,
we have followed a path of cautious extrapolation from older methods.
We first used simulation data to fix the single parameter of a model
(called AK here) that changes the previous standard (the Novikov-Thorne
model) only by admitting the possibility of a non-zero stress at the ISCO.
Because the AK model is defined in a way that prevents its extrapolation
within the ISCO, we used the formalism underlying both it and the NT
model (i.e., vertically-integrating and azimuthally- and time-averaging
the equation of momentum-energy conservation under the assumption that
the four-velocity and the stress tensor are orthogonal) to create an
expression for the dissipation  (called $Q_{\mathrm{MW}}$ here) valid
both inside and outside the ISCO.

   Happily, in the body of the accretion disc well outside the ISCO,
both the AK and the MW method agree fairly well with NT more or less
independent of black hole spin and magnetic field topology, although
the irregularity in the curves of Figure~\ref{spindiss} reminds one
that the simulations are dynamic and time-varying, and that
the 26 samples of simulation data we used define only somewhat
imperfectly the long-term time-average.  In addition, with the exception
of the extreme high-spin example, where the AK and MW methods depart
from NT just outside the ISCO, they do so together.  This is noteworthy
because, although they are based on closely-related formalisms, they
are not identical: perhaps their most significant contrast is that
the AK method assumes that $u_r = 0$, while the MW method does not.
Lastly, inside the ISCO, where only the MW method is defined, it
follows a smooth extrapolation from larger radius.  When the black
hole spins slowly ($a/M \lesssim 0.9$), $Q_{\mathrm MW}$ extends
with hardly any change in logarithmic derivative with respect to
radius.  For higher spin, the extension gradually steepens toward
smaller radius, but the next step in our formalism shows that this
makes little difference to observed radiation: relativistic ray-tracing
shows that the volume deep inside the plunging region, particularly
at high spin, contributes little energy to the luminosity reaching
observers at infinity.  Thus, we are relatively confident that, despite
the uncertainties involved, our estimate of the location of the
radiation edge is comparatively insensitive to the exact relation
between the flow's detailed dynamical properties and the dissipation
rate.

   The dependence of the radiation edge on spin may be summarised
succinctly: At the highest spin, there is relatively little difference
between the different methods of estimating its position because the
ISCO is so close to the horizon that the great majority of photons
released in the plunging region never reach infinity, or if they do, are
severely redshifted.  It moves from 2--3 times the radius of the ISCO
when the disc is viewed face-on to almost exactly at the ISCO when the
disc is viewed nearly edge-on.   This inward movement of the radiation
edge with increasing inclination angle is quite model-independent, as
it stems from relativistic photon propagation effects: when photons
from the plunging region do reach infinity with substantial energy,
it is because they are emitted in the direction of the orbital motion.
Probability of escape is then enhanced by a combination of special
relativistic beaming and gravitational lensing; energy at infinity is
enhanced by special relativistic Doppler boosting.  As the black hole
spin decreases, the diminishing depth of the potential immediately inside
the ISCO makes it progressively easier for photons to escape from that
region and reduces the gravitational redshift they suffer when they do.
The result is that the radiation edge moves farther inside the ISCO
as {\it either} the spin diminishes (at fixed viewing angle) or the
inclination angle moves toward the equatorial plane (at fixed spin).
At its most extreme, the case of $a/M = 0$ and $\theta_o = 85^{\circ}$,
$r_{re}$ can be $\simeq 0.5 r_{ms}$.

   Figure~\ref{ofrms} gives additional cause to believe that these
results are comparatively insensitive to dissipation model.  Although
the radiation edge can move well inside the ISCO at low spin and high
inclination, most of the light received by distant observers
is generally emitted in the region near and outside the ISCO.  Only at
the highest inclinations ($\theta_o \gtrsim 75^{\circ}$) and lowest
spins ($a/M \lesssim 0.5$) does the contribution of the plunging region
to the luminosity approach $50\%$.   Thus, most of the light seen
at infinity likely comes from a region where the predictions of the
AK and the MW models differ little.

    Nonetheless, because it is also true that most of the light
is emitted within a radius at most a few times the ISCO (except for
the highest spin viewed more or less face-on), the contrast in
total luminosity between the AK and MW models on the one hand, and
the NT on the other, are order unity for all spins $\leq 0.9$.
For higher spins, the effect may be smaller, but the uncertainties
are also greater.

These conclusions have immediate implications for a number of
phenomenological issues.  Firstly, as suggested by \cite{Falcke:2000},
it may be possible to image the nearest supermassive black hole, the one
in Sgr A*.  Because its accretion flow, unlike those of intrinsically
brighter systems, could well be radiatively inefficient, a simulation
scheme that conserves total energy is more appropriate to analysing its
emission.  \cite{Noble:2007}, using such a code (albeit an axisymmetric
version), have produced predicted images that illustrate several of the
effects emphasised here, although in their work so far they have not
reported quantitative descriptions of characteristic emission radii.

Secondly, the enhanced total radiative efficiency due to dissipation
in the marginally stable region may affect estimates of population-mean spin
parameters \cite[e.g., as for AGN by][] {Elvis:2002,Yu:2002}.  Because
the efficiency rises with increasing prograde spin in the NT model, the
spin inferred by this method may overestimate the actual spin of accreting
black holes if this enhancement is ignored.

The additional luminosity from enhanced stress in the innermost part of the accretion 
flow could significantly alter the emergent spectrum. Employing a simple thermal 
model, we have found that the characteristic temperature of the flow increases by a 
factor of 1.2 --1.4 over that predicted by the NT model. As a consequence, the thermal 
peak of the disk spectrum (at $\sim 1$~keV in Galactic black holes, $\sim 10$~eV in 
AGN) may be pushed to somewhat higher energies.

Several caveats must be mentioned, however, in regard to this prediction.
First, this number supposes an emergent 
spectrum that is Planckian, but most estimates of the disc atmosphere's structure 
suggest that it is scattering-dominated, so that the color temperature of the 
spectrum is shifted upward from the effective temperature. The magnitude of 
this shift depends on details of the disc's vertical structure that are not 
as yet well known \cite[see e.g.][]{Davis:2005}. Furthermore,
\cite{Blaes:2006} \citep[using the vertically
stratified shearing box simulations of][]{Hirose:2005} show that magnetic
pressure support changes the vertical structure of the disk resulting in a noticeable hardening of the emergent disk spectrum compared to the standard Novikov-Thorne picture due to non-LTE
effects. Second, it is possible that some 
of the enhanced dissipation will occur where the density and optical depth are 
too low to accomplish thermalisation. Strengthening of the ``coronal", 
i.e., hard X-ray, emission, rather than hardening the thermal disc spectrum 
would then be the likely consequence. Third, our treatment ignores those photons 
emitted deep in the potential that neither escape directly to infinity nor are 
captured by the black hole, but instead strike the disc. As shown by \cite{Agol:2000}, 
this ``returning radiation" can be a substantial fraction of all photons emitted 
when $r \lesssim 5r_g$. Depending on their spectrum and the structure of 
the disc atmosphere where they strike, these photons may be either reflected 
(with Doppler shifts) or absorbed and their energy reradiated at a different 
(in general, lower) temperature. Quantitatively evaluating all three of these 
effects is well beyond the scope of this paper, but can be done in future work.

There are also implications for attempts to determine black
hole spin from spectral fitting.  In all three models, the characteristic
radius of emission is always {\it near} the ISCO, but does not, in general,
coincide with it.  Generally speaking, this characteristic radius is largest
for the NT model, smaller (but still outside the ISCO) for the AK model, and
smaller still, possibly moving into the plunging region inside the ISCO,
for the MW model.  Because the ISCO moves to smaller radial coordinate as
$a/M$ increases, these characteristic radii always become smaller for faster
spin.  However, the fractional amount by which the characteristic emission
radius moves inward in the MW model is greatest for the lowest spins, so that
in the end, the MW model predicts a relatively slow variation of radiation
edge with black hole spin.
The AK model, like the NT model, does not radiate from inside the ISCO,
but the additional stress at and just outside the ISCO in this model (relative
to the NT prediction) produces
a systematic increase in the characteristic temperature.  The magnitude
of this shift in characteristic temperature rises, of course, with increasing
additional stress.  When there is emission from the plunging region, as
in the MW model, the characteristic temperature can rise still higher, but
the highly relativistic motions there make observed properties more strongly
dependent on inclination angle.  In addition, a larger fraction of the emitted
photons can be captured by the black hole.

When all these considerations are combined,  we find that, for fixed
black hole mass, luminosity, and inclination angle, the uncertainty in the
characteristic temperature of the radiation reaching distant observers due
to uncertainty in the dissipation profile is {\it greater} than the that due
to a complete lack of knowledge of the black hole's spin. Clearly, our incomplete understanding of accretion disc physics (here specifically the magnitude of the stress at and inside the ISCO) makes it difficult to determine a black hole spin based on continuum model-fitting.  The best one can say is that estimates based on the traditional Novikov-Thorne
model can be expected to yield the most rapid spin possible, but the
actual spin may be significantly slower.

Our results demonstrate the potential importance of nonzero
stresses at and inside the ISCO.
But how representative are the specific values obtained in these simulated
discs?  There are two considerations: those arising from
purely numerical effects, and those limitations arising from
the assumptions and parameters of the model used.

First, the results of numerical simulations can be influenced by finite
resolution and the limitations of the numerical technique.  All of the
simulations presented in this work were performed at a resolution of
$192\times192\times64$ $(r,\theta,\phi)$ grid zones using ideal MHD and an
internal energy equation.  The equation of state and the numerical energy
dissipation are unlikely to have a direct effect on our conclusions
as $Q_{MW}$ is derived directly from the \emph{physical} Maxwell
stresses within the disc, rather than by measuring some \emph{numerical}
dissipation rate.  Low resolution usually causes the Maxwell
stress to be {\it undervalued}; if so, the implications
of this paper would
be strengthed by improved resolution.  Until available
computer power makes better-resolved three-dimensional simulations possible, the
best way we have to test the effects of finite resolution is to
compute axisymmetric simulations with higher resolution.
A variety of such simulations were presented in
\cite{Beckwith:2008a} with resolutions up to $1024^2$.  We observe that
greater resolution reduces the rate of numerical reconnection and improves
the ability of the simulation to maintain certain field configurations
and small-scale field structures.  Overall the amplitude of the turbulent
Maxwell stresses in the disc remained largely unchanged as resolution
was increased.  We have also calculated ${\cal W}^{(r)}_{(\phi)}$ and
$Q_{MW}$, and find no significant qualitative differences from the results
presented in this work.

\newpage

Beyond the purely numerical issues, the value of the Maxwell
stress at the ISCO may depend on a number of disc properties.  In the
ensemble of simulations presented here, the stress levels are determined
in part by the initial field topology (dipole versus quadrupolar,
poloidal versus toroidal, and the presence or absence of a net vertical field).
Indeed, local shearing-box simulations suggest that the
saturated field strength can increase substantially when large-scale vertical
field threads the disc \cite[][]{Balbus:1998}.

It is also possible that the saturation stress depends
on disc thickness.  To quantify disc thickness, we define the scale-height
$H$ as the proper height above the plane at which the time- and
azimuthally-averaged density falls by $1/e$ from its similarly averaged value
on the equatorial plane (see \S\ref{dissmodels}):
$H= \int^{\theta_h}_{\pi/2} \sqrt{g_{\theta \theta} (r=3M,\theta)} d\theta$.
We similarly define $R$ as the proper (as opposed to coordinate) radial
distance from the horizon to $3M$ plus the coordinate distance from the
origin to the horizon:
$R = r_{in} + \int^{r=3M}_{r_{in}} \sqrt{g_{rr} (r,\theta=\pi/2)} dr$,
with $r_{in}$ as given in Table \ref{sims}).  We then define the disc thickness
as the ratio $H/R$.  Even though $r=3M$ lies well
within the plunging region for the lower spin cases, we find that there is
a slow enough radial variation in this quantity to make $H/R$ a reasonably
well-defined parameter.  Measured in this way, our discs are modestly thick,
with a characteristic aspect ratio $H/R = 0.06$--0.2 at
$r=3M$.  Most of the range in $H/R$ results from the fact
that the maximum pressure in the initial condition for these simulations
varies somewhat between different $a/M$.

It is difficult to say, however, just what sort of dependence there may
be on disc thickness.  There are some arguments suggesting
that magnetic effects may increase with increasing $H/R$.  For example,
local shearing-box simulations find that the Maxwell stress is proportional to
magnetic pressure \cite[][]{Balbus:1998,Sano:2004}, but there have not yet been
any systematic studies of what regulates the magnetic pressure in global,
vertically-stratified discs.  \cite{Afshordi:2003} have argued
that inner disc stresses and dissipation may depend on disc thickness as
well as on accretion rate, an argument reiterated by \cite{Shafee:2007},
but their arguments are framed in an essentially hydrodynamic context,
and therefore eliminate any possibility of predicting
magnetic stresses.  They are also non-relativistic, and therefore eliminate
any effects due to frame-dragging.  There are also arguments that any dependence on
$H/R$ may be weak.  In the simplest analytic or quasi-analytic MHD models,
magnetic torques at the ISCO remain significant even in the limit of a zero
pressure disc \citep{Krolik:1999a,Gammie:1999}.  As discussed
in \cite{Krolik:2005}, processes analogous to the Blandford-Znajek mechanism can
readily transport energy and angular momentum from rotating black holes
to the accretion flow, and there is no particular reason to think that
these processes should be tightly connected to the pressure in the disc.
In the end, only direct simulations with the resolution to describe thin
discs adequately will answer this question in a satisfactory way.
The results of the present investigation provide yet one more reason
why it will be important to do so.

The final question that we address in this work is how the results
presented here relate to current state of the art measurements of 
black hole spin via spectral fitting of
the disk continuum. The 6 systems with the best data \cite[see
e.g.][]{Davis:2006,Shafee:2006a,Middleton:2006,Liu:2008} all have spins
in the range $a/M\sim0.1-0.8$ based on disk models that assume the
stress-free inner boundary condition. From the perspective of the disk
stress models these spins are upper limits.  This might indicate that
the hole is counter-rotating, but also opens the possibility that spin
determinations might themselves constrain the physics near the ISCO.
Firstly, the stress levels at the ISCO could be near the value assumed
by the stress-free inner boundary condition; secondly, the classical
relationship between stress and dissipation might not hold for enhanced
magnetic stresses ear the ISCO; thirdly, the density levels at and
inside the ISCO could be insufficient to thermalize the dissipated
heat; fourthly, the time-scales for thermalization and radiation of the
dissipated heat could be longer than the inflow time-scale.  Another
uncertainty which we have not examined is that the plane of the disk and
the equatorial plane of the hole could be misaligned and so the disk is
subject to the Bardeen-Peterson effect \cite[see e.g.][]{Fragile:2007}.
Understanding the roles played by the these effects will be crucial in
providing robust estimates of black hole spins via spectral fitting of
the disk continuum.

\section*{Acknowledgements}
This work was supported by NSF grant PHY-0205155 and NASA grant NNG04GK77G
(JFH), and by NSF grant AST-0507455 (JHK). KB thanks Sean Matt, Chris Done and Shane Davis for useful discussions; S.M. for reading of the manuscript and CD. for explaining the current state and interpretation of black hole accretion disk temperature measurements. We thank an anonymous referee whose suggestions greatly improved an earlier version of this manuscript. We acknowledge Jean-Pierre De~Villiers for continuing collaboration in the development of the
algorithms used in the GRMHD code.  The simulations were carried out on
the DataStar system at SDSC.

\newpage

\appendix
\section{Basis Vectors of the Local Rest Frame}

\label{sec:appa}

The calculation of the photon transfer functions requires the
introduction of a set of basis vectors describing the local rest frame
of the fluid (the ``fluid frame"). Such a tetrad set was presented by
\cite{Krolik:2005}, who used a Gram-Schmidt orthonormalization
procedure to construct it. Unfortunately, some of the expressions
given in that work were
incorrectly transcribed. The correct version is as follows:

\begin{align*}
e^{\mu}_{(t)} = u^{\mu} \\
e^{\mu}_{(\phi)} =
-\frac{1}{k_{\phi}\sqrt{|-g^2_{t\phi}+g_{\phi\phi} g_{tt}}|} \times \\
\left[g_{\phi\phi} u^{\phi}+g_{t\phi} u^{t},0,0,-\left(g_{t\phi} u^{\phi}+g_{tt} u^{t}\right) \right] \\
e^{\mu}_{(r)} = -\frac{1}{k_{r}k_{\phi}}
\left[ \sqrt{g_{rr}} u^{r} u^{t},
\frac{k^{2}_{\phi}}{\sqrt{g_{rr}}},
0,
\sqrt{g_{rr}} u^{r} u^{\phi} \right] \\
e^{\mu}_{(\theta)} = -\frac{1}{k_{\theta}k_{r}}
\left[
\sqrt{g_{\theta \theta}} u^{\theta} u^{t},
\sqrt{g_{\theta \theta}} u^{\theta} u^{r},
\frac{k^{2}_{r}}{\sqrt{g_{\theta\theta}}},
\sqrt{g_{\theta \theta}} u^{\theta} u^{\phi} \right]
\end{align*}
where:
\begin{align*}
k_{\phi} = \sqrt{|g_{\phi\phi} \left(u^{\phi}\right)^2+u^{t} \left(2 g_{t\phi} u^{\phi}+g_{tt} u^{t} \right)|} \\
k_{r} = \sqrt{|g_{\phi\phi} \left(u^{\phi}\right)^2+g_{rr} \left(u^{r}\right)^2+u^{t} \left(2 g_{t\phi} u^{\phi}+g_{tt} u^{t} \right)|} \\
k_{\theta} = \sqrt{|g_{\phi\phi} \left(u^{\phi}\right)^2+g_{rr} \left(u^{r}\right)^2+g_{\theta\theta} \left(u^{\theta}\right)^2+u^{t} \left(2 g_{t\phi} u^{\phi}+g_{tt} u^{t} \right)|}
\end{align*}

\label{lastpage}

\end{document}